\documentclass[10pt,twocolumn,letterpaper]{article}
\usepackage{cvpr}  

\usepackage{booktabs} 
\usepackage[ruled]{algorithm2e} 

\usepackage{graphicx}
\usepackage{xcolor}
\usepackage{multirow}
\usepackage{array}
\usepackage{enumitem}
\usepackage{soul}
\usepackage{wrapfig}
\usepackage{subcaption}
\usepackage[ruled]{algorithm2e} 

\usepackage{soul} 
\usepackage{multirow}
\usepackage{array}
\usepackage{graphicx}
\usepackage{booktabs} 
\usepackage{enumerate}
\usepackage{times}
\usepackage{epsfig}
\usepackage{graphicx}
\usepackage{amsmath}
\usepackage{amssymb}
\usepackage{multirow}
\usepackage{booktabs}
\usepackage{wrapfig}
\usepackage{footnote}
\usepackage{subcaption}
\usepackage{algorithmicx}
\usepackage{algpseudocode}
\usepackage{enumitem}
\usepackage{xcolor}
\usepackage{graphicx}
\usepackage{amsmath}
\usepackage{amssymb}
\usepackage{booktabs}
\usepackage{enumitem}
\usepackage{hyperref}

\newcommand{\filluptopage}[1]{%
  \clearpage
  \loop\ifnum\value{page}<#1\relax
    \null\clearpage
  \repeat
  \loop\ifnum\value{page}=#1\relax
    \null\clearpage
  \repeat
}

\SetAlFnt{\tiny}
\SetAlCapFnt{\small}
\SetAlCapNameFnt{\small}
\SetAlCapHSkip{0pt}

\begin{document}
\title{Creating Language-driven Spatial Variations of Icon Images}
\author{
\normalsize Xianghao Xu\\
\normalsize Brown University\\
\normalsize USA\\
\and
\normalsize Aditya Ganeshan\\
\normalsize Brown University\\
\normalsize USA\\
\and
\normalsize Karl D.D. Willis\\
\normalsize Autodesk\\
\normalsize USA\\
\and
\normalsize Yewen Pu\\
\normalsize Autodesk\\
\normalsize USA\\
\and
\normalsize Daniel Ritchie\\
\normalsize Brown University\\
\normalsize USA\\
}

\twocolumn[{%
\renewcommand\twocolumn[1]{#1}%
\maketitle
\begin{center}
    \centering
    \includegraphics[width=0.9\linewidth]{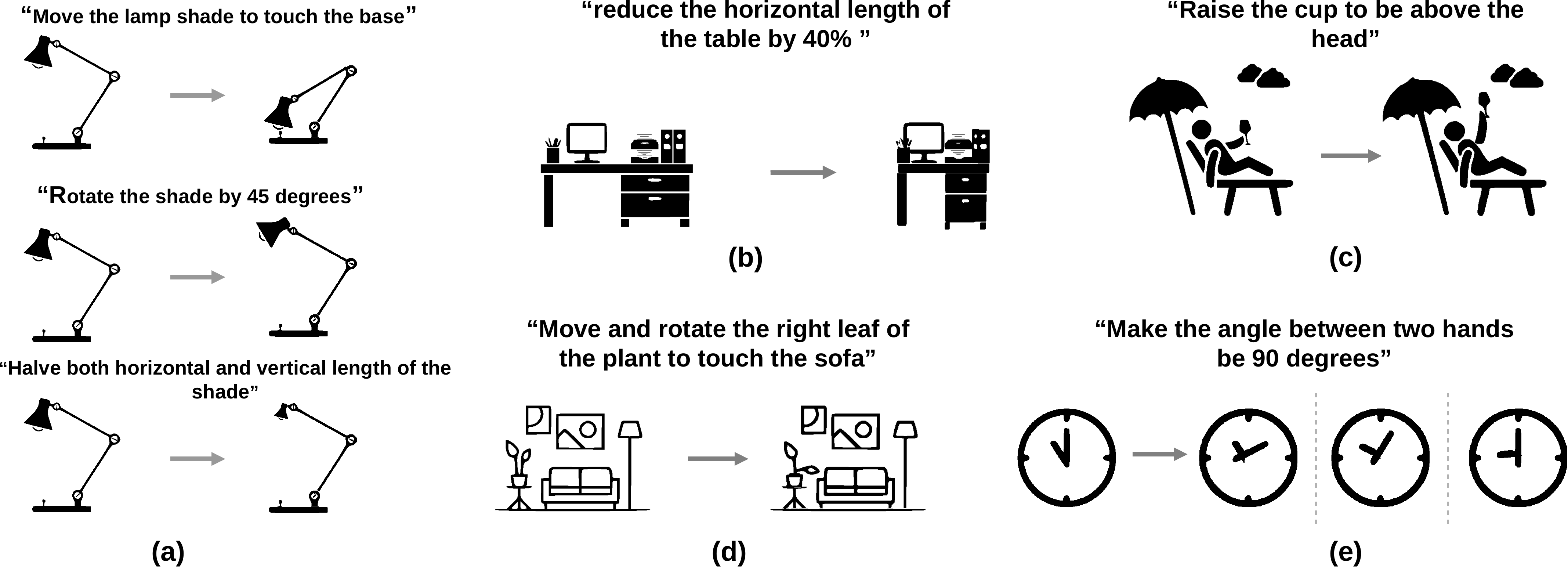}
    \captionof{figure}{Language driven icon image variations created by our method}
    \label{fig:teaser}
\end{center}%
}]

\begin{abstract}
Editing 2D icon images can require significant manual effort from designers.
It involves manipulating multiple geometries while maintaining the logical/physical coherence of the objects depicted in the image. Previous language-driven image editing methods can change the texture and geometry of objects in the image but fail at producing spatial variations, i.e. modifying spatial relations between objects while maintaining their identities. We present a language-driven editing method that can produce spatial variations of icon images. Our method takes in an icon image along with a user’s editing request text prompt and outputs an edited icon image reflecting the user's editing request. Our method is designed based on two key observations: (1) A user's editing requests can be translated by a large language model (LLM), with help from a domain specific language (DSL) library, into to a set of geometrical constraints defining the relationships between segments in an icon image. (2) Optimizing the affine transformations of the segments with respect to these geometrical constraints can produce icon images that fulfill the editing request and preserve overall physical and logical coherence. Quantitative and qualitative results show that our system outperforms multiple baselines, enabling natural editing of icon images.

\end{abstract}
\setlength{\abovedisplayskip}{3pt}
\setlength{\belowdisplayskip}{3pt}

\section{Introduction}
\label{sec:intro}
2D icon images are widely used in advertising, logos, street signs, and more to convey important information in a simplified and understandable way. 2D icons are typically created as vector graphics by manually creating and editing vector curves. Creating icons in this manner can be a time-consuming process requiring designers with specialized skills. 
To reduce this effort, new icon images can be created by producing \emph{spatial variations} of existing icons; that is, modifying the spatial relations between and within objects in the image (e.g. re-sizing or re-arranging).

Natural language provides a useful interface for specifying many such variations; in recent years, many methods have been developed for editing images given a text-based editing request~\cite{hertz2022prompt,brooks2022instructpix2pix,kawar2023imagic}.
Although these methods are effective at modifying the texture and geometry of objects in an image (e.g. ``change the apple to an orange''), they fail at producing spatial variations (e.g. ``raise the cup to be above the head'' in Fig.~\ref{fig:teaser}c).
In addition, since these methods are based on models that have been trained on detailed natural images, they perform poorly when applied to icon images.

Despite the simplicity of icon images, producing spatial variations of them can be challenging --- involving transforming multiple geometries while maintaining logical and physical coherence of the objects depicted in the image.
For example, to ``move the lamp shade to touch the base'' in Fig.~\ref{fig:teaser}a, the regions in the image corresponding to ``shade'' and ``base'' need to be recognized; when bringing the shade down to the base, the chain of arm connectors should also be adjusted to keep the physical structure of the lamp intact.

We approach this non-trivial icon image modification problem by dividing it into two sub-problems: (1) How to associate the concepts referred to in the editing request to segments of the image, and (2) How to realize the editing request by modifying the segments in the image. Although our main contribution focuses on the second problem, for the first problem we assume a labelled segmentation is manually created, or a semi-automatic segmentation pipeline using existing open-vocabulary concept grounding models~\cite{ren2024grounded} is used.
For the second and main problem we address in this work, we frame it as finding affine transformations for each segment such that the editing request is satisfied.
One obvious approach to solving this problem is to use a pretrained large language model (LLM) to translate the editing request into affine transformation parameters.
However, we observed that LLMs perform quite poorly at this task, struggling to produce appropriate continuous parameter values.
We observed that they are much better at describing, in words, the key constraints that the transformed scene must satisfy. We speculate that this is due to the fact that humans also use descriptive words to specify the desired constraints when specifying an edit ~\cite{acquaviva2021communicating}. 
Based on these observations, we design a domain specific language (DSL) for describing these constraints (position, pose, size, distance, angle, interaction, etc.) and task an LLM with translating the input edit request into a program in this language.

We further find that LLMs struggle to accurately reason about all the constraints that must be satisfied for every object in the image (e.g. ``secondary'' effects that must happen in response to the input editing request, such as the lamp arms need to remain connected in Fig.~\ref{fig:teaser}a).
Thus, our system only relies on the LLM to specify the primary constraints that relate directly to the user's input editing request.
We then rely on a graph based search method that progressively searches for and optimizes additional constraints that preserve the physical and logical coherence of all the objects in the image. We present the following contributions:

\begin{itemize}
\item A novel system for converting natural language editing requests into spatial variations of icon images.
\item An extensible DSL of differentiable operators that can be used to specify geometrical constraints for visual elements.
\item A graph-based constraint search method that progressively identifies appropriate additional constraints for achieving an editing request while preserving the logic and physical coherence of the image scene.
\end{itemize}

Additionally, we will open source our code along with the evaluation benchmark dataset we crafted upon publication. 

\section{Related Work}
In this section we review related work on language driven diffusion models and icon generation and manipulation.
\label{sec:related_work}
\subsection{Language Driven Image Editing Models}
Language driven image editing were widely explored, \cite{Text2live}, \cite{gal2022textual}, \cite{kwon2021clipstyler}, \cite{Photorealistic_Text-to-Image}, \cite{lin2023text}, \cite{de-net}, etc. Diffusion models in particular have been widely utilized for language driven natural image editing tasks, showcasing remarkable results. \cite{rombach2021highresolution} enable text guided generation via a general cross-attention module. \cite{hertz2022prompt} inject text control into a diffusion generation process by manipulating the attention layers. \cite{dall-e-2} utilizes the clip latent space for text conditioned image generation. \cite{Avrahami_2022_CVPR} allow users to specify a region in the image and replace the masked content with a diffusion process. \cite{ruiz2022dreambooth} enables text guided generation by fine-tuning a diffusion model with a handful of examples and special tokens. \cite{kawar2023imagic} approaches the task by optimizing a latent code towards a target latent code generated by a diffusion model conditioned on the editing text. \cite{brooks2022instructpix2pix} train an edit diffusion model on editing paired data. More related work in this line of research include but not limited to \cite{Kim_2022_CVPR},  \cite{wang2023mdp}. However, a shortcoming of these methods is they can typically only change the texture and geometry of objects in the image. They typically fail at fine-grained manipulation, such as editing the geometry or spatial relationships between objects whilst maintaining the identity of the elements in the image.

\subsection{Icon Generation and Manipulation}
Related to the modification of icon images is the adjacent problem of conditioned generation of icons images~\cite{arsvg, wu2023iconshop, sketchrnn, carlier2020deepsvg, wang2023deepvecfont}. These works utilize neural networks and learn to generate a sequence of Scalable Vector Graphics (SVG) operations (e.g. points, lines, circles) that can be executed to produce an image. Recent work~\cite{clipdraw, jain2023vectorfusion, xing2023diffsketcher} removes the need for direct SVG supervision by learning to generate icon images directly. The important problem of modifying existing icons to produce spatial variations is not addressed by these methods, as their foremost focus is on generation rather than editing. What's more, these methods suffer the same problem as natural image editing methods: during generation the identity preservation of the original visual elements in the image is not guaranteed. More closely aligned with our work is Lillicon~\cite{lillicon}, that tries to solve the icon image editing problem. Lillicon uses transient widgets to select and manipulate features of icons to create their scale variations. However, their method does not take language as input, their system requires considerable user interaction, and the only variations created are limited to scale variations. In contrast, our method doesn't require much user interaction and can create a wide range of different icon variations driven by language input.  

\section{Overview}
\label{sec:overview}
\begin{figure*}[t!]
    \centering
    \includegraphics[width=0.9\linewidth]{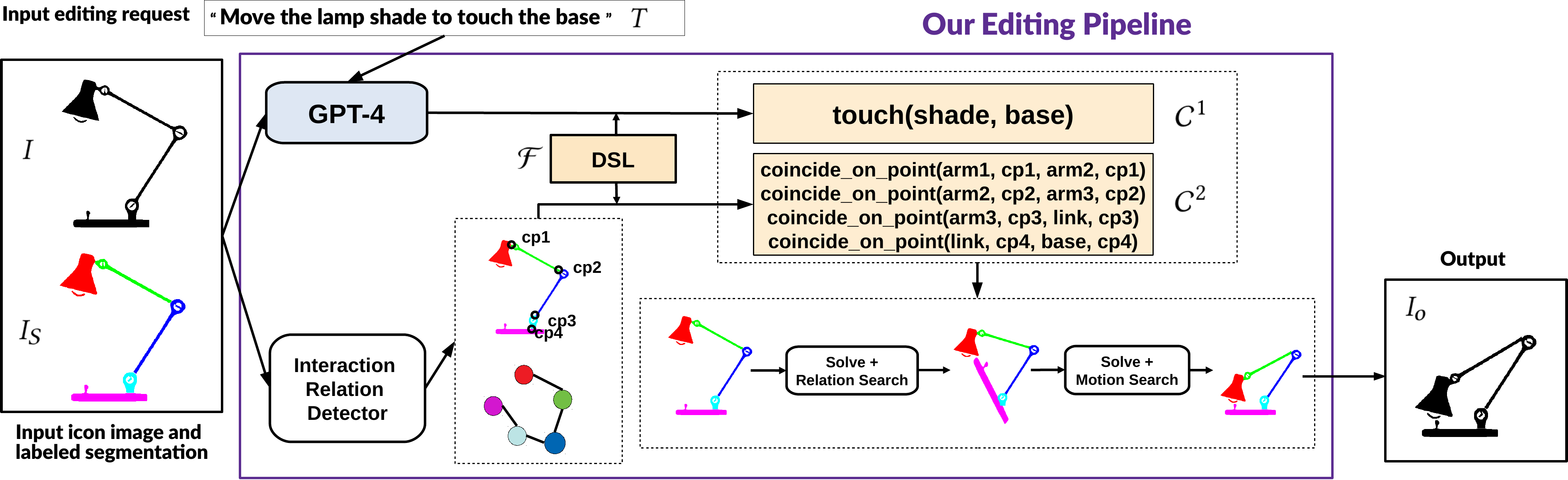}
    \caption{Overview of our editing pipeline. The pipeline takes in an icon image $I$, a segmentation map $I_S, \mathcal{S}$ and an editing request text prompt $T$, then outputs an edited image $I_o$ that is aligned with $T$. With the help of the DSL library $\mathcal{F}$, primary constraints $\mathcal{C}^1$ are generated by GPT-4 and candidate secondary constraints $\mathcal{C}^2$ are generated by a spatial relation detector. A combination of discrete search and continuous optimization is used to find the affine transformations for the segments $\mathcal{S}$ under the constraints, the transformed segments are re-rendered into an image $I_{o}$ as the output.}
    \label{fig:edit_pipeline}
\vspace{-2mm}
\end{figure*}

In the following sections, several components of our editing system will be introduced, followed by the introduction of our segmentation tool. Our editing system, shown in Fig.~\ref{fig:edit_pipeline}, takes in an icon image $I$ and its associated segmentation map $I_S$, an edit request text prompt $T$. And it outputs an edited image $I_o$ that reflects the user's editing request. The input image $I$ is a (512*512*3) rgb image, the associated segmentation map $I_S$ is an integer value (512*512) mask where each different integer value corresponds to a different segment. Each segment $S_i \in \mathcal{S}$ is also assigned a semantic label $l_i \in \mathcal{L}$ expressed as (ObjectName: PartName), e.g. (Table:Leg). This segmentation map $I_S$ can be created manually by the designer, optionally assisted by our semi-auto segmentation tool introduced in \ref{sec:seg_tool}. The output edited image $I_o$ is a (512*512*3) rgb image. Our editing system first convert the image $I$ into a set of labelled segments $S_i \in \mathcal{S}$, where each segment corresponds to a set of pixels in the input image $I$. Then the editing request $T$, which is originally expressed in natural language, is converted into a set of primary constraints $\mathcal{C}^1$ by a pre-trained Large Language Model(LLM) assisted by our design DSL constraint library $\mathcal{F}$. Next, a hybrid search and optimize algorithm is used to progressively add secondary constraints $\mathcal{C}^2$ while optimizing the affine transformation variables such as translation $t_i \in R^2$, rotation $r_i \in R$ and scaling $s_i \in R^2$ for every relevant segment $S_i \in \mathcal{S}$ with respect to the full constraints $\bigcup(\mathcal{C}^1, \mathcal{C}^2)$ to generate the edited segments. Affine transformation is chosen because our goal is to re-arrange the visual elements present in the image while still maintaining their identity; non-affine editing or deformation will not provide such a guarantee. Finally the edited segments are re-rendered into an image $I_{o}$ as the output. 

\section{Image as segments}
\label{sec:image_as_seg}
In order to manipulate the image at object/part level, we choose to discretize the input image $I$ as a set of labelled segments using the associated segmentation map $I_{S}$. For each segment, it is originally represented as a set of pixels in the image, however, this representation is not convenient for affine manipulation, we thus represent each segment $S_i$ by its boundary geometry defined by a set of 2D paths. A boundary path $B_i \in \mathcal{B}$ is represented as a closed loop of connected line segments, each line segment is defined by its two end vertices $[x_0, y_0]$ and $[x_1, y_1]$. In our boundary formulation, we ignore the inner holes of each segment, this is a choice compatible with our geometrical DSL design which we introduce in \ref{sec:constraint}. With this representation, the affine transformation of the segments can be easily achieved by manipulating the 2D positions of the vertices in the boundary loop. We provide further details in the supplement materials for how the occlusion between segments are handled .

\section{Constraint Formulation}
\label{sec:constraint}
With the input image discretized into a set of editable segments $\mathcal{S}$, the next step is to establish some desired geometrical constraints for these segments to achieve the goal expressed in the edit request. In this section, we first introduce the design of a DSL constraint specifier library, then describe how it interacts with a pretrained LLM to bridge the gap between natural language editing requests and primary geometrical constraints for segments. Followed by the introduction of a search procedure to add additional secondary constraints from the DSL library to preserve the logical/physical coherence of the scene. 

\subsection{Design of Domain Specific Language}
In order to express editing requests as constraints on segments our DSL library is designed to contain two types of functions: (1) Constraint Specifiers for specifying constraints and (2) Compute Operators for accessing segments. 
The constraint specifiers are functions that directly specify either a motion type constraint or a geometry violation value, they are formatted as:
\begin{equation}
  \begin{aligned}
F_{m}(S_i), \quad v_o = F_v(v_i, v_j), \quad v_o = F_v(S_i, S_j)\\
  \end{aligned}
\end{equation}
with $F_m \in \mathcal{F}$ represents motion constraint specifier and $F_v \in \mathcal{F}$ represents value constraint specifiers, $S_i \in \mathcal{S}$ and $S_j \in \mathcal{S}$ represents segment as input parameters, $v_i \in R$ and $v_j \in R$ represents scalar/vector as input parameters. The output scalar value $v_o \in R$ indicating how much the constraint is violated. 

The compute operators are functions that can get attributes of the segments, or get the spatial relations between segments, they are formatted as:
\begin{equation}
  \begin{aligned}
v_o = F_c(v_i), \quad v_o = F_c(v_i, v_j)\\
v_o = F_c(S_i), \quad v_o = F_c(S_i, S_j)\\
  \end{aligned}
\end{equation}
with $F_c \in \mathcal{F}$ represents the compute operators, and $F_v \in \mathcal{F}$ represents value constraint specifiers, $S_i$ and $S_j$ represents segment as input parameters, $v_i \in R$ and $v_j \in R$ represents scalar/vector as input parameters. In order to make the DSL general enough to handle segments with different geometry, attributes of segments can only be accessed by using the compute operators, so that all the segment specific geometry computations are handled inside compute operators. These computations are implemented to be fully differentiable with respect to the segment transformation variables to enable gradient based optimization. What's more, the functions are designed in a way so that they can be combined recursively to express constraints with complex structures. Please see table \ref{table:dsl} for a partial overview of the functions in the library.
\begin{table}
\begin{tabular}{|p{0.9\linewidth}|}\hline
\fontsize{11}{12}\selectfont \textbf{\underline{Constraint Specifier}}:\\
\textbf{Motion constraint specifier(Segment to None)}:\\ \indent $translate$ \textbar $rotate$ \textbar $scale$ \textbar $stay$\\
\textbf{Value constraint specifier(Values to Value):}\\ \indent $equal$ \textbar $smaller$ \textbar $larger$\\
\textbf{Derived value constraint specifier(Segments to Value):}\\ 
\indent $coincide\_on\_point$ \textbar $inside$ \textbar $touch$ \textbar $overlap$ \textbar $detach$ \textbar $on\_top$ \textbar $on\_bottom$ \textbar $on\_left$ \textbar $on\_right$\\
\hline
\fontsize{11}{12}\selectfont \textbf{\underline{Compute Operator}}:\\
\textbf{Arithmetic operator(Values to Value):}\\ \indent $plus$ | $minus$ \textbar $mul$ \textbar $div$ \textbar $min$ \textbar $max$\\
\textbf{Attribute operator(Segment to Value):}\\ 
\indent $vert\_len$ \textbar $hori\_len$ \textbar $center\_x$ \textbar $center\_y$ \textbar $long\_dir\_x$ \textbar $long\_dir\_y$ \textbar $short\_dir\_x$ \textbar $short\_dir\_y$ \textbar $min\_x$ \textbar $min\_y$ \textbar $max\_x$ \textbar $max\_y$ \\
\textbf{Attribute operator(Segment to Segment):}\\ \indent $old$ \textbar $top$ \textbar $bot$ \textbar $left$ \textbar $right$\\
\textbf{Relation Attribute operator(Segments to Value):}\\ \indent $avg\_dist$ \textbar $min\_dist$ \textbar $max\_dist$ \textbar $angle$\\
\textbf{Boolean Operator(Segments to Segment)}:\\ \indent
$union$ \textbar $inter$\\
\hline
\end{tabular}
\caption{DSL Library}
\label{table:dsl}
\vspace{-6mm}
\end{table}

\subsection{Large Language Model for Primary Constraints}
Given user's edit request prompt $T$ and our DSL library $\mathcal{F}$, A pretrained LLM such as GPT-4 or GPT-4V is asked to convert the editing request $T$ using functions from the DSL library $\mathcal{F}$ into a set of the most critical primary constraints $\mathcal{C}^1$ which must be satisfied. To make LLM understand better the context of user's request, we found that providing the image scene information can be crucial for many editing tasks. We represent the image scene as a list of segments, with each segment represented as (1) the semantic label (2) the four vertex positions of the segment's axis-aligned bounding box. (3) Connection relations between segments. This information provides the LLM with the location, size, semantics, and interaction information of the segments in the image scene. We hide low-level geometry details from the LLM, such as each segments boundary path vertices, as we found empirically that it did not help with our task.
We found that including several manually crafted question and answer examples as part of the prompt provided to the LLM can greatly improve the quality of LLM's output and bias the LLM to output answers that are similar to the provided examples. 

\subsection{Relation Detection for Secondary Constraints}
We observed that relying only on the primary constraints from the LLM can frequently result in either an over-constrained or under-constrained problem. We believe this is due to the lack of geometry understanding with current LLMs. To address this issue, we use the LLM to output the most crucial primary constraints (typically under-constraining), and use a geometry constraint detection method to automatically generate additional candidate secondary constraints $\mathcal{C}^2$. These candidate secondary constraints are crucial to preserve the logical and physical coherence of the scene. Given the initial arrangement of the segments, many potential relations can be detected. We choose to detect the most dominant spatial relations, namely: \textit{Inside}, \textit{Contain}, \textit{Overlap(Connect)}. Each has a weak and strong version. The weak version of each relation constraint is constrained by the function $overlap(S_i, S_j)$,  $inside(S_i, S_j)$ and $contain(S_i, S_j)$ respectively, the strong version of each relation additionally includes the $coincide\_on\_point(S_i, p, S_j, p)$ constraint to indicate they are joined at a fixed point. 
Based on the two level hierarchy structure (ObjectName:PartName) of labels $\mathcal{L}$ provided along with the segmentation map, the above mentioned relations can be naturally categorized into intra-object relations $\mathcal{R}_a$ and inter-object relations $\mathcal{R}_i$. The preservation of intra-object relations ensures the integrity of the structure of objects, the preservation of inter-object relations ensures the perception consistency between input scene and edited scene. The intra-object relation $\mathcal{R}_a$ can be further categorized into crucial intra-object relations $\mathcal{R}^{*}_a$ and non-crucial intra-object relations $\mathcal{R}^{'}_a$, for example, consider the basket example shown in fig \ref{fig:optim}, the relations between each handle and the basket body are crucial intra-object relations, the relations between the two handles are non-crucial intra-object relations as they are not real connections but due to the initial spatial arrangements. We use the LLM(GPT-4 Vision) to assign types to these intra-object relations.
Note that not all secondary constraints need to be satisfied, as some initial relation constraints will prevent us from satisfying the primary constraints. We therefore design a search method to find a subset of secondary constraints to satisfy. 
\section{Optimization}
\label{sec:optim}
Given the segments $\mathcal{S}$ and their constraints $\bigcup(C^1, C^2)$ generated. The next step is to find the affine transformation(motion) variables translation vector $t_i \in R^2$, rotation $r_i \in R$ and scaling $s_i \in R^2$ for segments $\mathcal{S}$ with respect to the constraints, thus, with with respect to the editing request. As discussed before, the constraints are categorized as a set of primary constraints $\mathcal{C}^1$ given by LLM and a set of candidate secondary constraints $\mathcal{C}^2$ generated by geometrical relation detection. The primary constraints will always be included in the transformation finding process as they are the most crucial constraints directly associated with the editing request. For the candidate secondary constraints, the challenge is how to identify a maximum subset of them without violating any primary constraint. 

We designed a search method aiming to tackle this challenge problem. Our search method contains two major operations: $Solve()$ and $Flip()$. The $Solve()$ operation takes in the set of segments along with a set of constraints, and optimize the motion parameters of the segments to satisfy the constraints. In order to run $Solve()$, the state of segments' transformation(motion) and the state of constraints need to be determined. This is handled by the $Flip()$ operation, a $Flip()$ operation toggles a segment's motion state from : Static(N), Translate(T), Translate+Rotate(TR), Translate+Scale(TS), Translate+Rotate+Scale(TRS), and it toggles a relation constraint state from: Strong(ST), Weak(WK), Not exist(N). Once the states are specified, the $Solve()$ jointly optimize all the specified motion variables for all segments with respect to (1) all primary constraints and (2) the secondary constraints set to the specified level. Motion variables are optimized using a gradient based optimizer with the Loss as the average of all specified constraints. The output of $Solve()$ are a set of optimized motion parameters to edit the segments and a score indicating how much the constraints are violated. 

\begin{figure}[t!]
    \centering
    \includegraphics[width=0.9\linewidth]{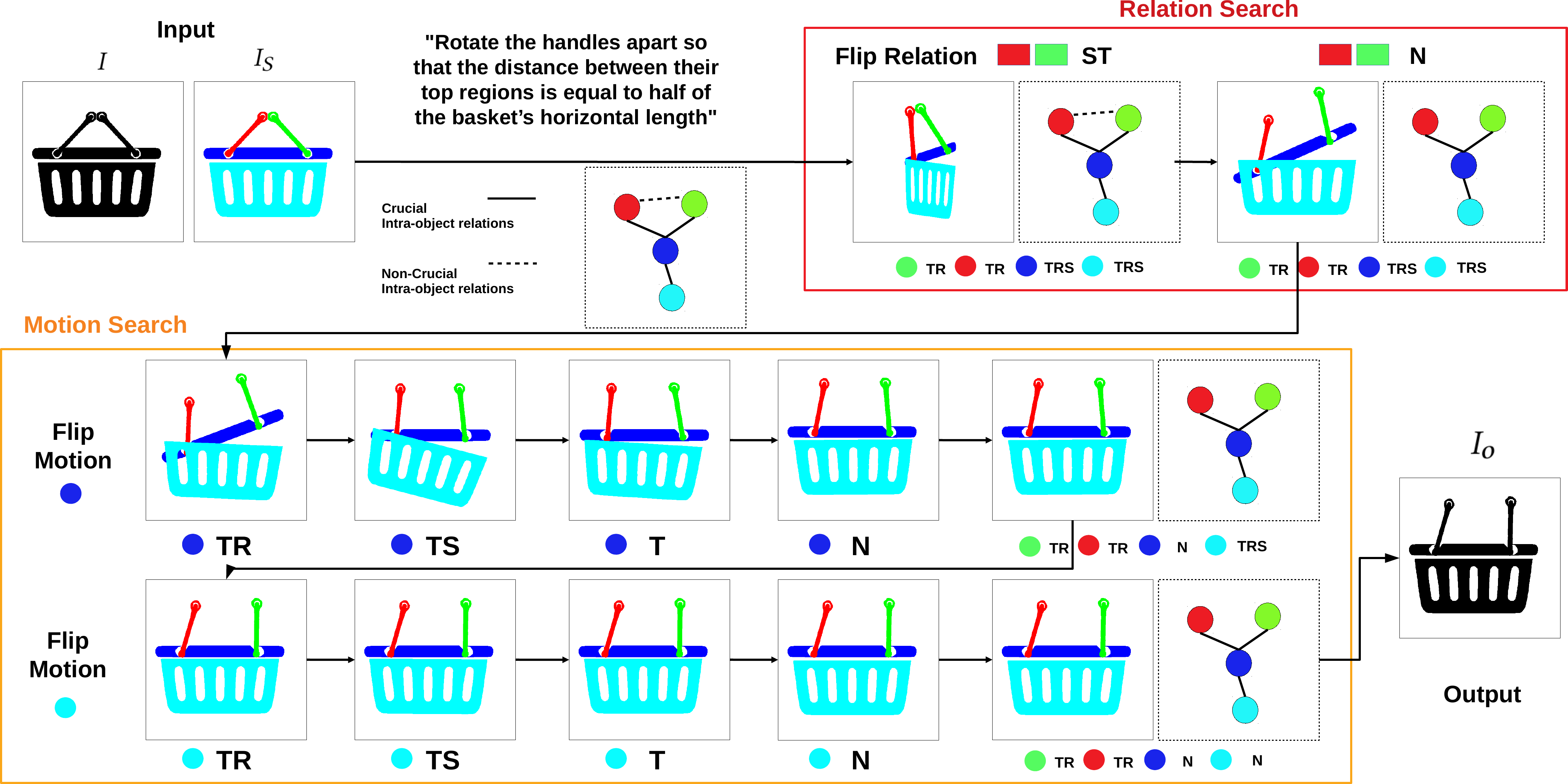}
    \caption{The overview of the search process, the relations between segments are first flipped and tested, once a relation configuration satisfies the constraints is found, the motion state of segments are flipped and tested.}
    \label{fig:optim}
\vspace{-2mm}
\end{figure}

Our method then enforces a search order to apply $Flip()$ and $Solve()$ across all necessary segments. It is helpful to construct a graph structure $G=(\mathcal{V}=\mathcal{S}, \mathcal{E}=(\mathcal{R}^{*}_{a}, \mathcal{R}^{'}_{a}, \mathcal{R}_i))$, with each node $v_i \in \mathcal{V}$ of this graph represents a segment, and each edge $e_i \in \mathcal{E}$ represents a relation constraint between two adjacent segments. The general strategy is to gradually propagate $Flip()$ from an initial set of nodes and edges to the entire graph, and for each flip, run $Solve()$ to evaluate the flipped state. We found that it would greatly reduce the search computation complexity if we perform the search in two sequential passes. In the first pass, we fix the node states and just flip and evaluate the edge states. In the second pass, we fix the edge states from the first stage and just flip and evaluate the node states. In the edge flip pass, we choose the edges that are incident to nodes involved in primary constraints as front edge set, for each subset of the front edge set, the $Flip()$ follows a particular order: firstly, the inter-object edges $\mathcal{R}_i$ are flipped in the order ${ST, WK, N}$, then the non-crucial intra-object edges $\mathcal{R}^{'}_{a}$ are flipped in the order ${ST, N}$, finally the crucial intra-object edges $\mathcal{R}^{*}_{a}$ are flipped in the order ${ST, WK}$. The intuition is that if the strongest interaction relation does not prevent us from satisfying the primary constraints, then it should be kept to preserve the interaction relation between segments as much as possible. Once all combinations of the front edge sets are flipped and evaluated, the neighbor edges are set to be the new front edge set to continue the search. In the node flipping pass, we again choose the nodes that are directly involved in the primary constraints as the initial front nodes, within the front node set, the motion state of each node is flipped in the following order: ${TRS, TR, TS, T, N}$, the intuition is that if one type of motion is not necessary for satisfying the constraints, then it is not needed, so that the original characteristics of objects are preserved as much as possible. Once all nodes in the front node set are flipped and evaluated, the neighbor nodes are set to be the new front nodes to continue the search. Please see Fig. \ref{fig:optim}, Algo. \ref{alg:optim}, and the supplementary material for more details.

After the two search passes, the final optimized motion parameters will be applied to the segments to produce the final edited segments and then re-rendered into the output image, please see the supplementary material for more discussion of re-rendering.

\begin{algorithm}
\caption{Flip and Solve}
\small
\DontPrintSemicolon
\SetKwInOut{Input}{Input}
\SetKwInOut{Output}{Output}
\Input{}
\Output{}
\SetKwProg{Fn}{Function}{:}{}
\SetKwFunction{FOptim}{Optim}
\Fn{\FOptim{$\mathcal{S}, G, \mathcal{C}^1, \mathcal{C}^2$}}{
$\mathbf{M}_X \gets Init()$ \;
$\mathcal{X}_{f} \gets getFront(\mathcal{C}^1, G)$ \;
\While {$Len(\mathcal{X}_{f}) > 0$} {
    $\mathcal{U} \gets Get(\mathcal{X}_{f})$\;
    \For {$U_i \in \mathcal{U}$}{
        $\mathbf{M}^{'}_X \gets Flip(U_i, \mathbf{M}_X)$\;
        $ret \gets Solve(\mathcal{S}, \mathcal{C}^1, \mathcal{C}^2, {M}^{'}_X)$\;
        $\mathbf{M}_X \gets Update(\mathbf{M}^{'}_X, ret, U_i)$\;
    }
    $\mathcal{X}_{f} \gets getNeighbors(\mathcal{X}_{f}, G)$\;
}
\Return $\mathbf{M}_X$ \;
}
\label{alg:optim}
\end{algorithm}
\begin{figure*}[ht!]
    \centering
    \includegraphics[width=0.9\linewidth]{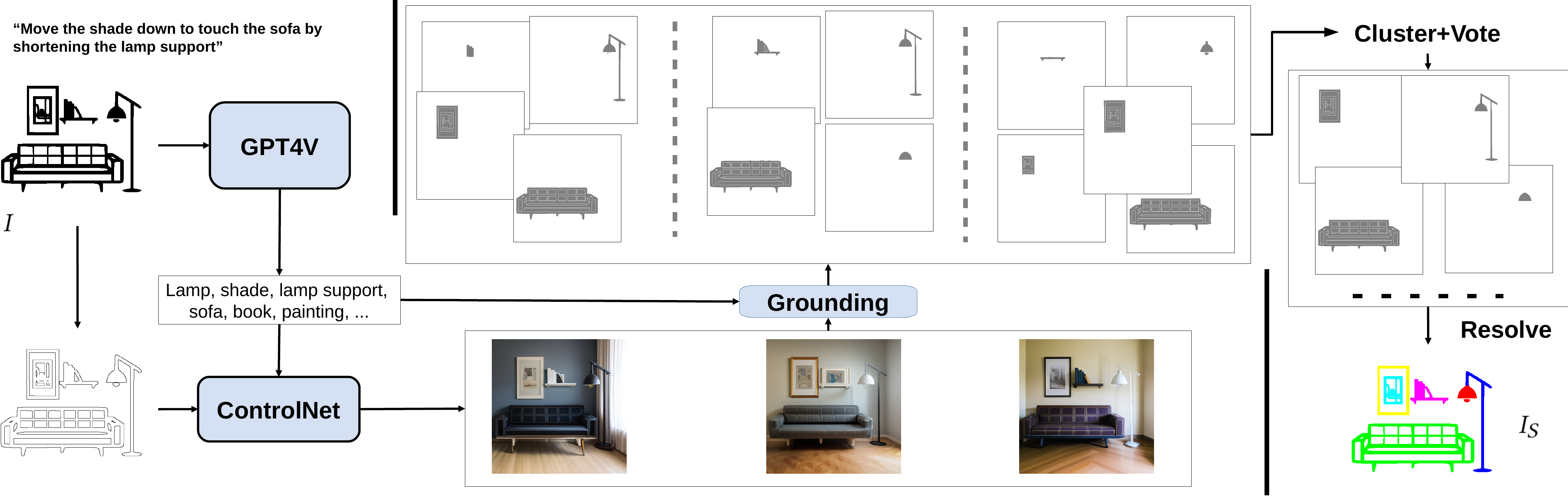}
    \caption{Overview of the segmentation tool pipeline. Given the input icon image $I$, the concept label set is formed with the help of GPT-4V, then the edge map of input image $I$ together with the concept set is feed into a pretrained ControlNet to generate multiple photo-like images. Then a pretrained concept grounding model such as GroundedSam is used to produce predicted segment region masks $\mathcal{M}_i$ for every label $l_i \in \mathcal{L}$. A vote and resolve process is applied to generate the final segmentation map $I_S$.
    }
    \label{fig:pipeline_seg}
\vspace{-3mm}
\end{figure*}

\section{Segmentation Tool}
\label{sec:seg_tool}
In this section we introduce a pipeline that aims to automatically generate the segmentation map $I_S$ given the input icon image $I$ and editing request $T$. We do not claim this tool as part of our main contribution, but rather a tool to speed up the segmentation process and reduce manual effort required by users. Label generation and grounding for icon images is a very hard task as no existing grounding models are trained for icon images, and even for the natural images, current state-of-the-art models are the subject of ongoing research. Our tool is built on top of existing open-vocabulary concept grounding models ~\cite{ren2024grounded, liu2023grounding}, we expect that as the performance of these models advance in the future, the performance of our tool will improve along with them. 

Our segmentation pipeline starts with label generation, given the input icon image $I$, we first use GPT-4V to output the names of objects in the image, we denote this concept set as $\mathcal{L}_{I}$. Then given the editing request $T$, we use GPT-4 to output the name of the entities present in the request, we denote this concept set as $\mathcal{L}_{T}$. The final label/concept set $\mathcal{L} = \bigcup(\mathcal{L}_{I}, \mathcal{L}_{T})$ is a combination these two sets. As the majority of label grounding models are trained on natural images and can perform poorly on icon images, we take the edge map of the input icon image $I$, together with an image description derived from the concept set $\mathcal{L}$, and pass them to a pretrained ControlNet ~\cite{zhang2023adding} to generate realistic natural images. Then for every generated image, a pretrained concept grounding model such as GroundedSam ~\cite{ren2024grounded} is used to produce predicted segment region masks $\mathcal{M}_i$ for every label $l_i \in \mathcal{L}$, and combined to form a set of labelled region masks $\mathcal{M}$. These region masks are then clustered according to their intersection over Union(IOU) ratio $IoU = Inter(M_i, M_j)/Union(M_i, M_j)$. Within each cluster, a representative region mask $M^r$ is picked by choosing the region mask that is closet to the average region shape across all the regions in the cluster. Given the set of all representative region masks $M^{r}_0, M^{r}_1, ... M^{r}_N$, a final resolve process iterate over all pair of representative regions and apply an intersection subtraction operation $M^{r}_i = M^{r}_i - Inter(M^{r}_i, M^{r}_j)$ to ensure no two regions overlap. After the resolve process, a set of labelled segments $(S_0, l_0), (S_1, l_1), ... (S_N, l_N)$ are produced. Please see Fig. \ref{fig:pipeline_seg} and the supplementary material for more discussion and evaluation of our segmentation tool.

\section{Experiments} \label{results}
In order to evaluate our method, we created an evaluation benchmark dataset by selecting 63 images from FIGR-8-SVG \cite{DBLP:journals/corr/abs-1901-02199} dataset, and 65 different editing requests are designed around these images, we manually segment the images and create one possible solution for each editing request as ground truth. 

\subsection{Our method greatly outperforms other LLM based parametric editing baselines}
\begin{figure}
    \centering
    \small
    \setlength{\tabcolsep}{0pt}
    \renewcommand{\arraystretch}{0}
    \newcommand{\resultimg}[1]{\includegraphics[trim={10pt 10pt 10pt 10pt},clip,width=0.14\linewidth]{#1}}
    \begin{tabular}{c|c|c|c|c|c|c}
        \raisebox{0em}{Original} & \raisebox{0em}{gpt4-dm} & \raisebox{0em}{gpt4v-dm} & \raisebox{0em}{gpt4-dc} & \raisebox{0em}{gpt4v-dc} & \raisebox{0em}{Ours} & \raisebox{0em}{Human}\\
        \midrule
        \multicolumn{7}{l}{\parbox{8cm}{\textbf{"Rotate the handles apart so that the distance between their top regions is equal to half of the basket's horizontal length"}}} 
        \\
        \resultimg{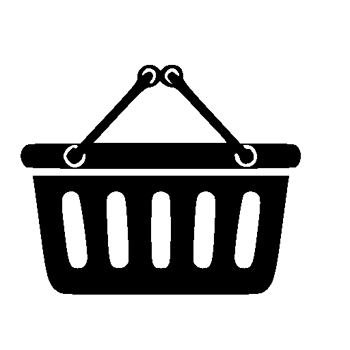} &
        \resultimg{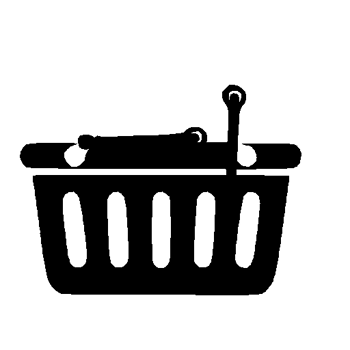} &
        \resultimg{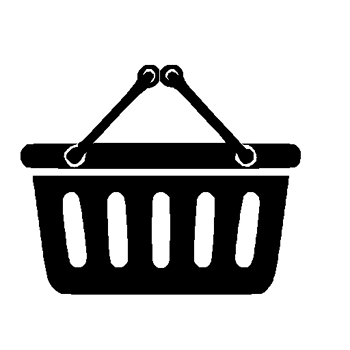} &
        \resultimg{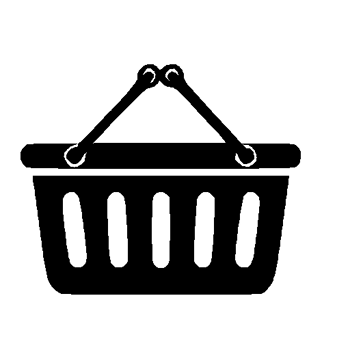} &
        \resultimg{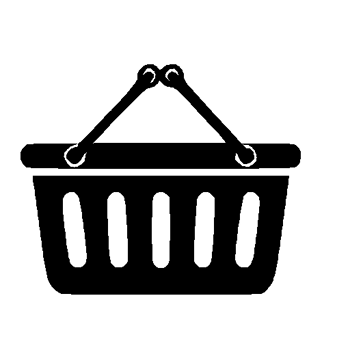}  & 
        \resultimg{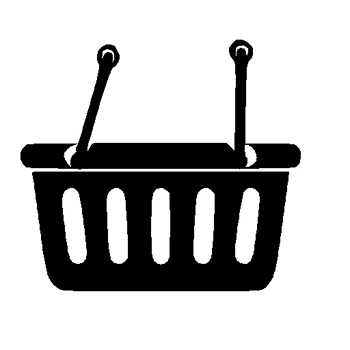}  & 
        \resultimg{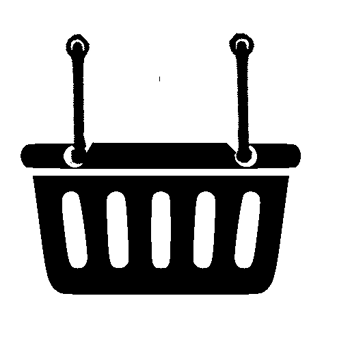}  
        \\
        \midrule
        \multicolumn{7}{l}{\parbox{8cm}{\textbf{"Scale the fish body to reduce its horizontal length by 50\%"}}}
        \\
        \resultimg{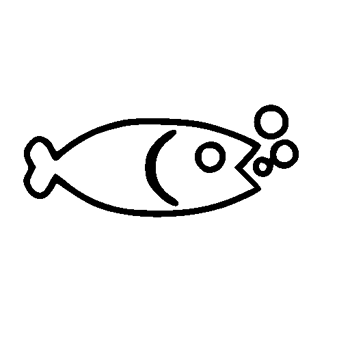} &
        \resultimg{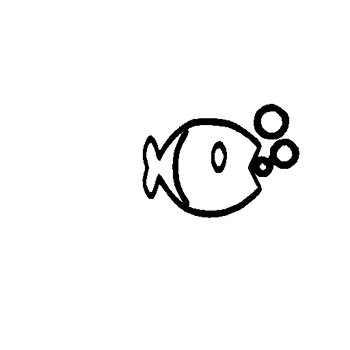} &
        \resultimg{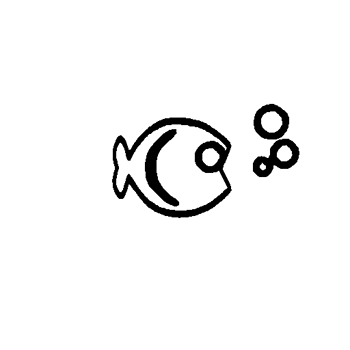} &
        \resultimg{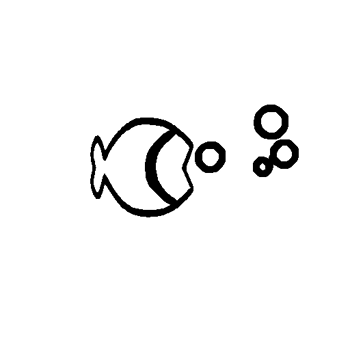} &
        \resultimg{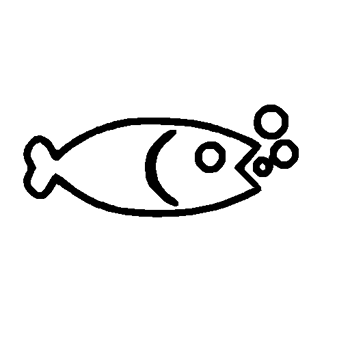}  & 
        \resultimg{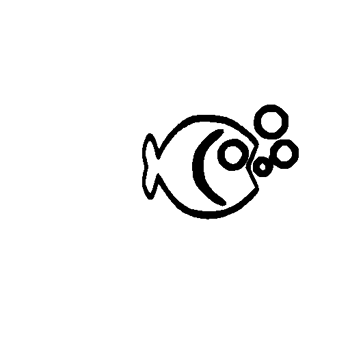}  & 
        \resultimg{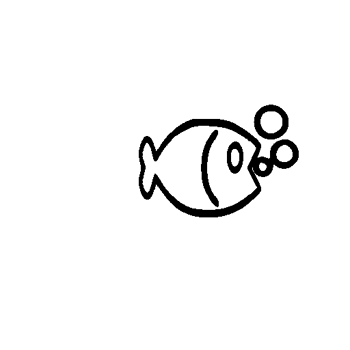}  
        \\
        \midrule
        \multicolumn{7}{l}{\parbox{8cm}{\textbf{"Move the lamp shade to touch the base"}}}
        \\
        \resultimg{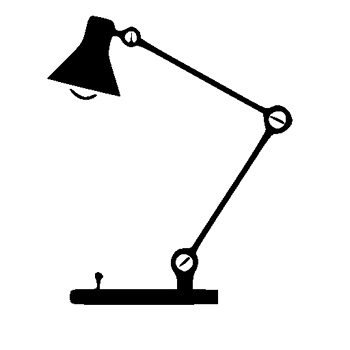} &
        \resultimg{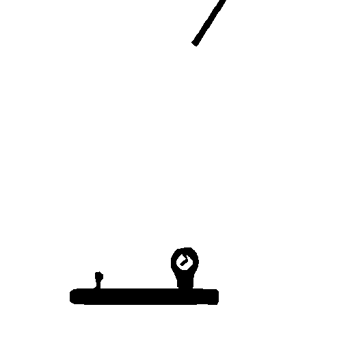} &
        \resultimg{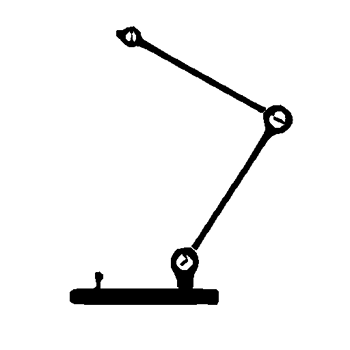} &
        \resultimg{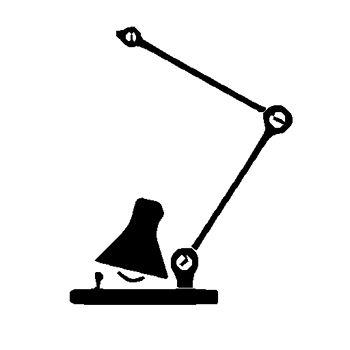} &
        \resultimg{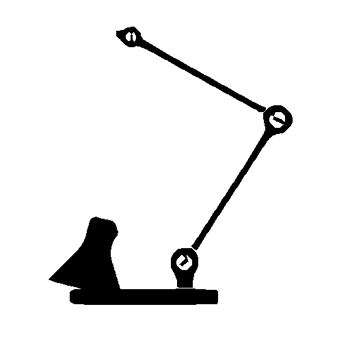}  & 
        \resultimg{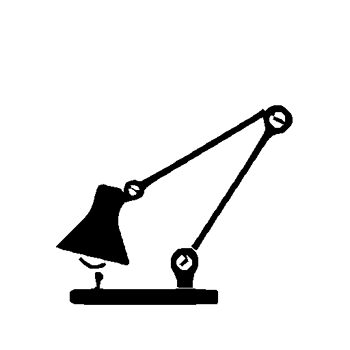}  &
        \resultimg{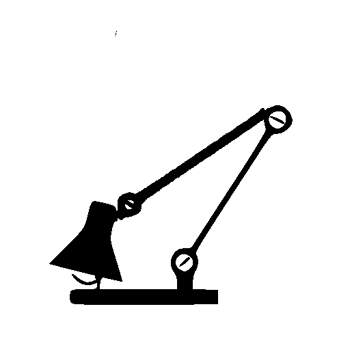}
        \\
        \midrule
        \multicolumn{7}{l}{\parbox{8cm}{\textbf{"Move and rotate the hand so that its bottom region is aligned with the top region of the 2nd cube from the left"}}}
        \\
        \resultimg{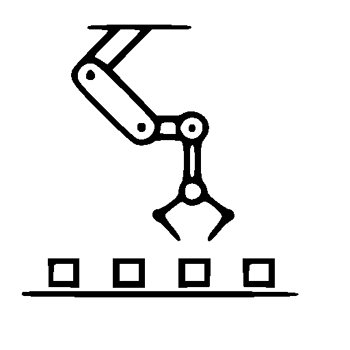} &
        \resultimg{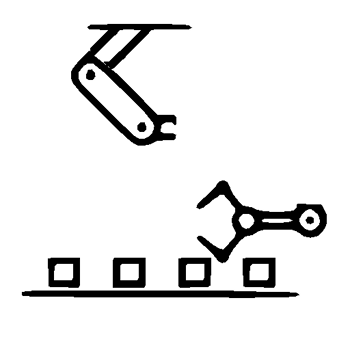} &
        \resultimg{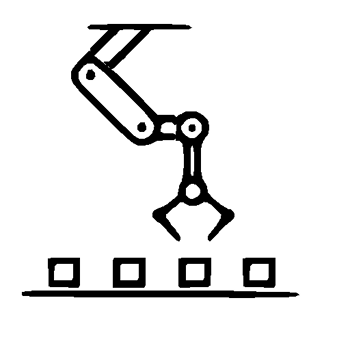} &
        \resultimg{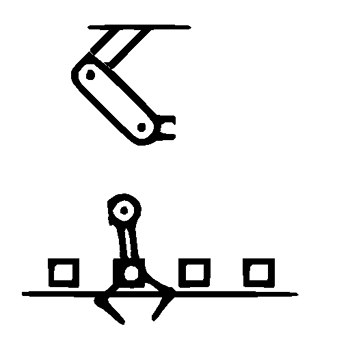} &
        \resultimg{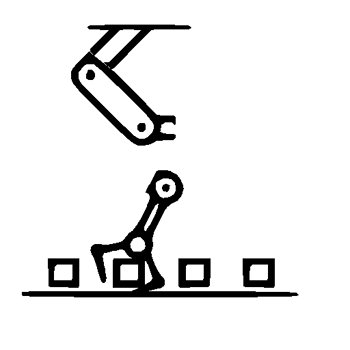}  & 
        \resultimg{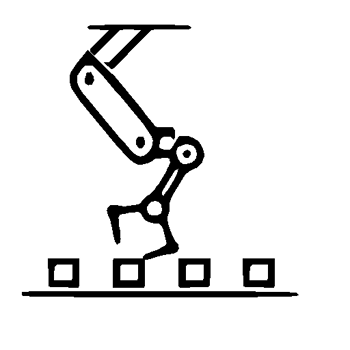} &
        \resultimg{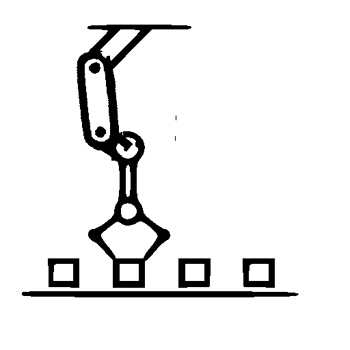}
        \\
        \midrule
        \multicolumn{7}{l}{\parbox{8cm}{\textbf{"Raise the cup to be above the head"}}}
        \\
        \resultimg{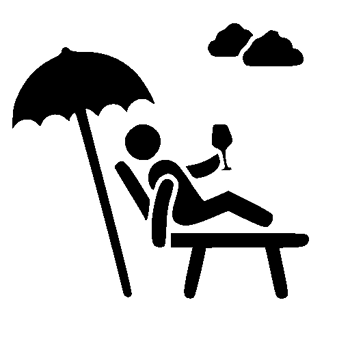} &
        \resultimg{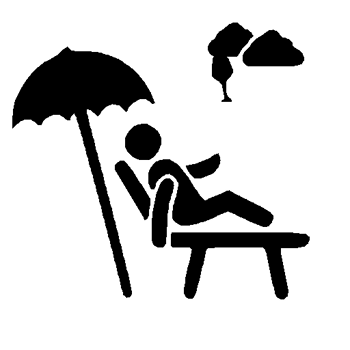} &
        \resultimg{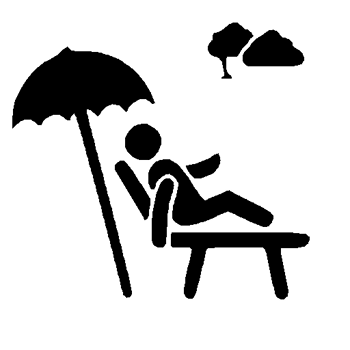} &
        \resultimg{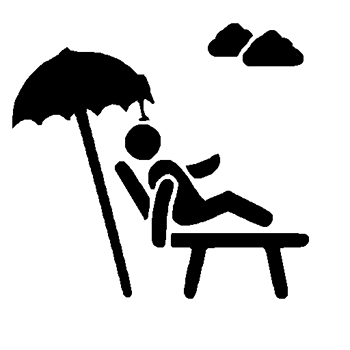} &
        \resultimg{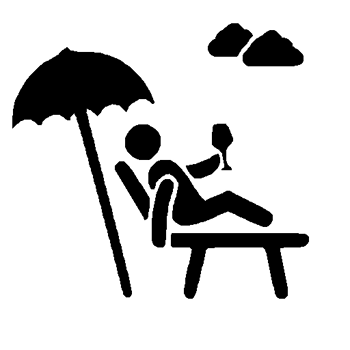}  & 
        \resultimg{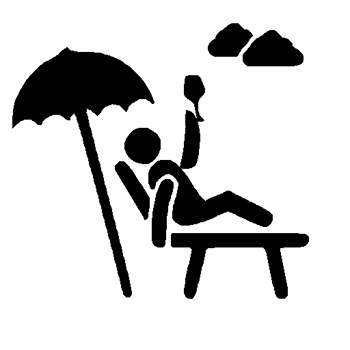}  & 
        \resultimg{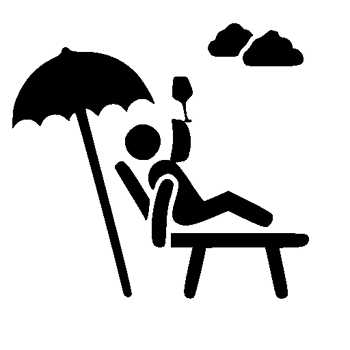}
        \\
        \midrule
        \multicolumn{7}{l}{\parbox{8cm}{\textbf{"Reduce the horizontal length of the table by 40\%"}}}
        \\
        \resultimg{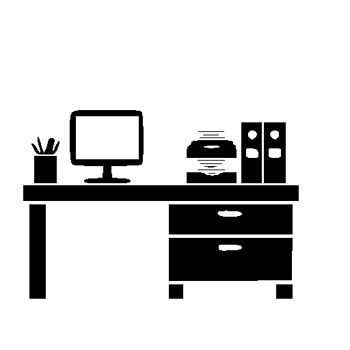} &
        \resultimg{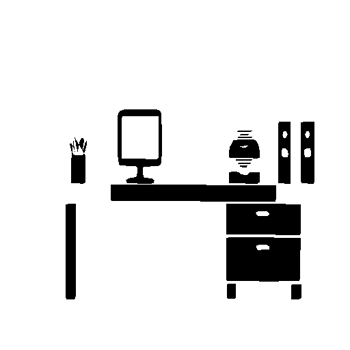} &
        \resultimg{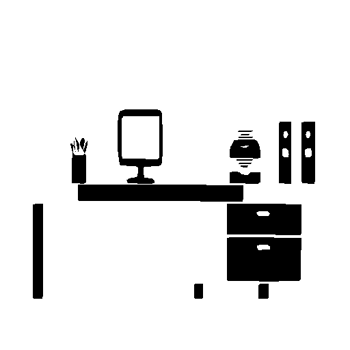} &
        \resultimg{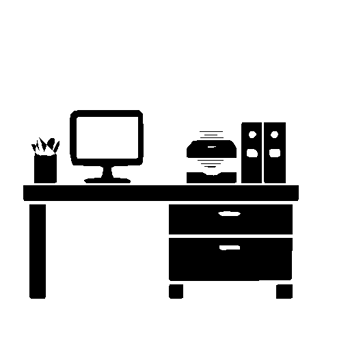} &
        \resultimg{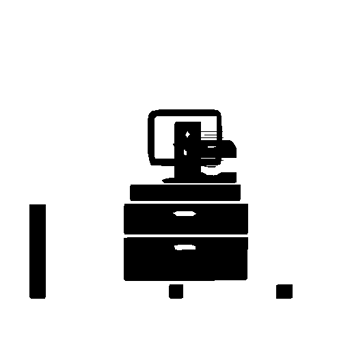}  & 
        \resultimg{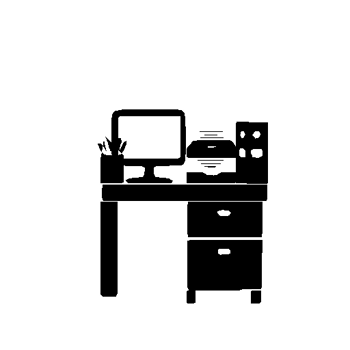} &
        \resultimg{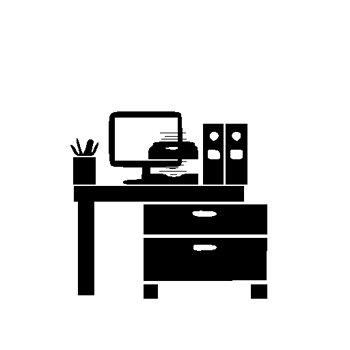}
        \\
    \end{tabular}
    \caption{Qualitative comparison against GPT based editing methods}
    \label{fig:comp_gpt_0}
\end{figure}

\begin{figure}[t!]
    \centering
    \small
    \setlength{\tabcolsep}{0pt}
    \renewcommand{\arraystretch}{0}
    \newcommand{\resultimg}[1]{\includegraphics[trim={0pt 0pt 0pt 0pt},clip,width=0.12\linewidth]{#1}}
    \begin{tabular}{c|c|c|c|c|c|c}
        \raisebox{0em}{Original} & \raisebox{0em}{gpt4-dm} & \raisebox{0em}{gpt4v-dm} & \raisebox{0em}{gpt4-dc} & \raisebox{0em}{gpt4v-dc} & \raisebox{0em}{Ours} & \raisebox{0em}{Human}\\
        \midrule
        \multicolumn{7}{l}{\parbox{7cm}{\textbf{"Move the leftmost candle(with flame) and rightmost candle(with flame) to touch the 2nd layer of the cake"}}}
        \\
        \resultimg{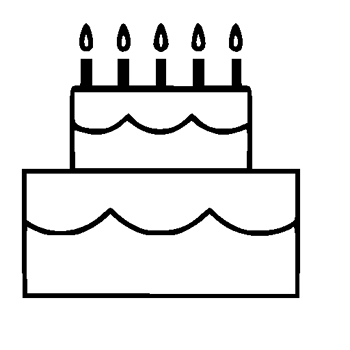} &
        \resultimg{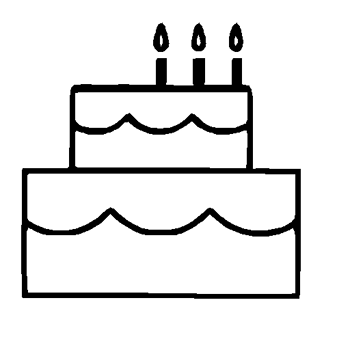} &
        \resultimg{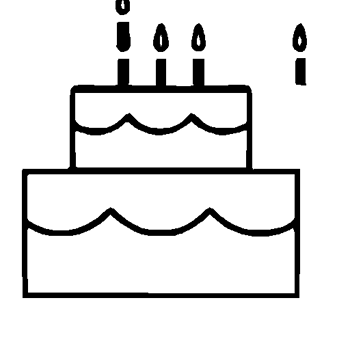} &
        \resultimg{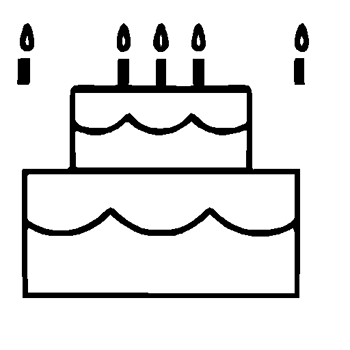} &
        \resultimg{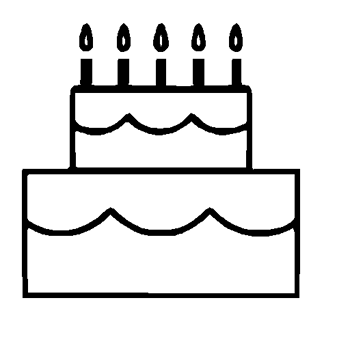}  & 
        \resultimg{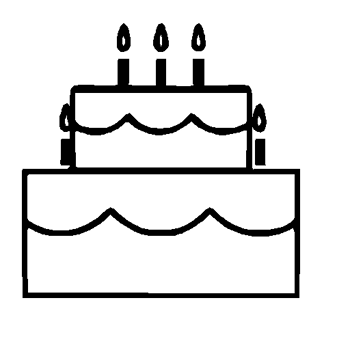} &
        \resultimg{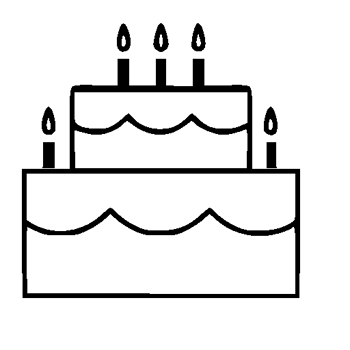}
        \\
        \midrule
        \multicolumn{7}{l}{\parbox{7cm}{\textbf{"Move the shade down to touch the sofa by shortening the lamp support"}}}
        \\
        \resultimg{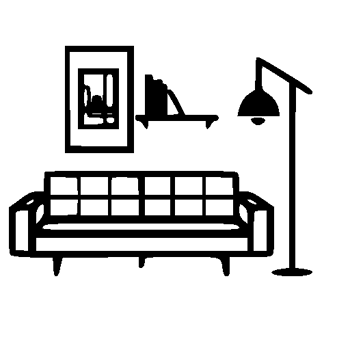} &
        \resultimg{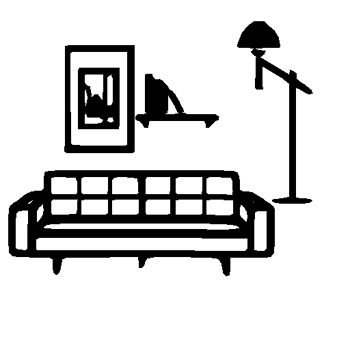} &
        \resultimg{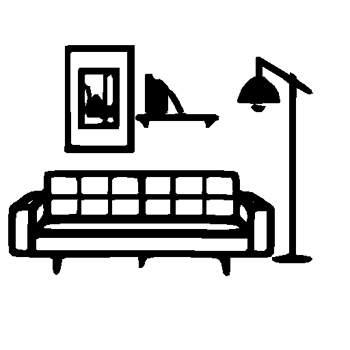} &
        \resultimg{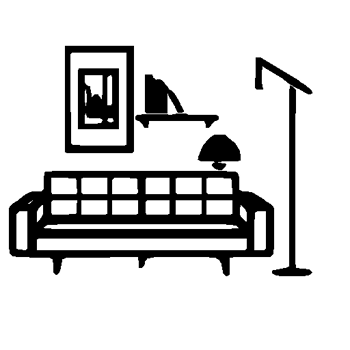} &
        \resultimg{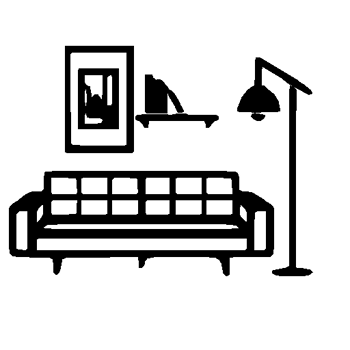}  & 
        \resultimg{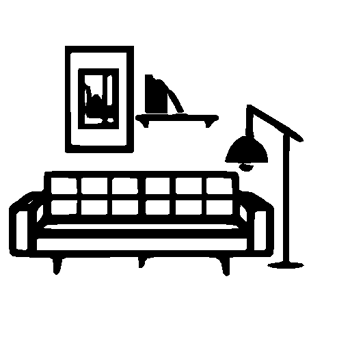} &  
        \resultimg{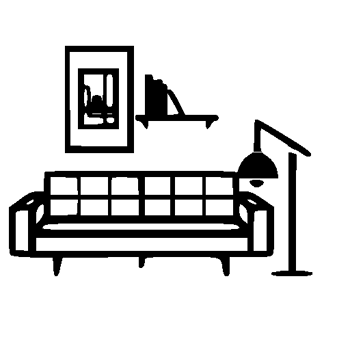}
        \\
        \midrule
        \multicolumn{7}{l}{\parbox{7cm}{\textbf{"Halve the horizontal length of the umbrella head, halve the vertical length of the umbrella support"}}}
        \\
        \resultimg{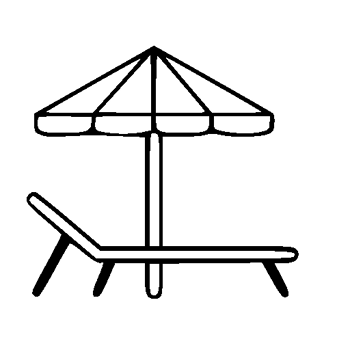} &
        \resultimg{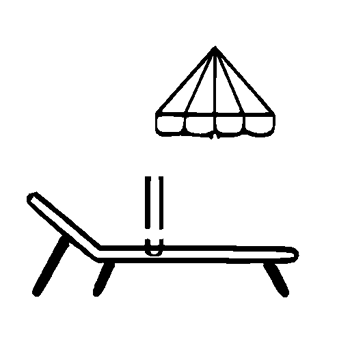} &
        \resultimg{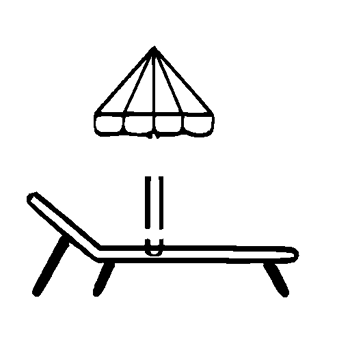} &
        \resultimg{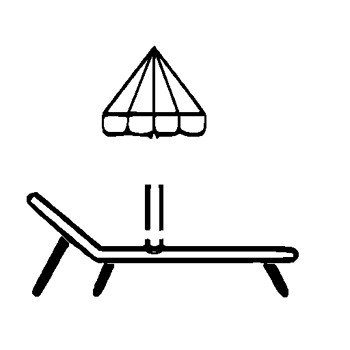} &
        \resultimg{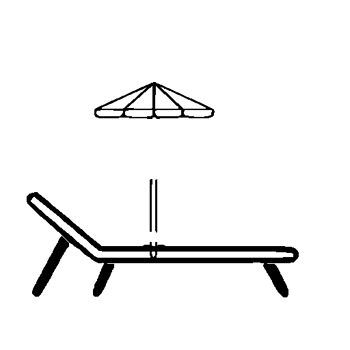}  & 
        \resultimg{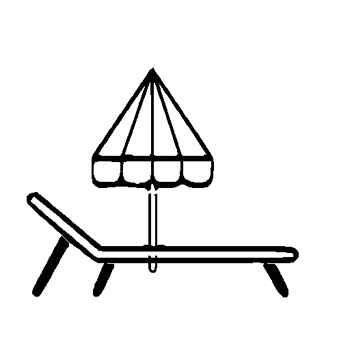} & 
        \resultimg{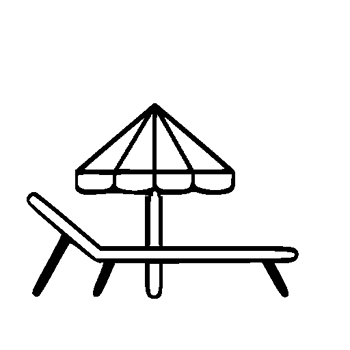}
        \\
        \midrule
        \multicolumn{7}{l}{\parbox{7cm}{\textbf{"Rotate the direction of the claws by 90\%"}}}
        \\
        \resultimg{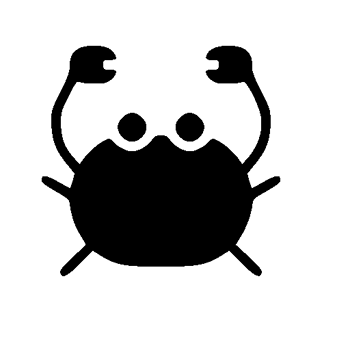} &
        \resultimg{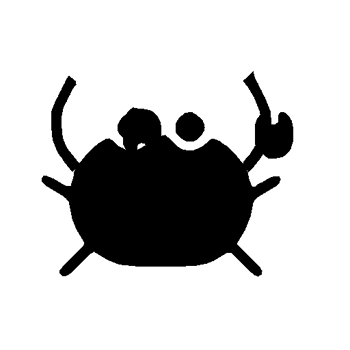} &
        \resultimg{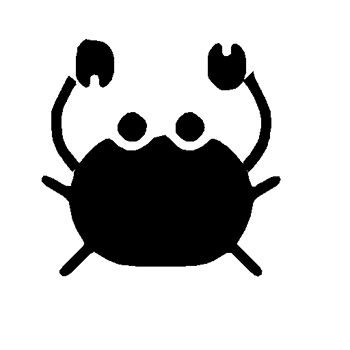} &
        \resultimg{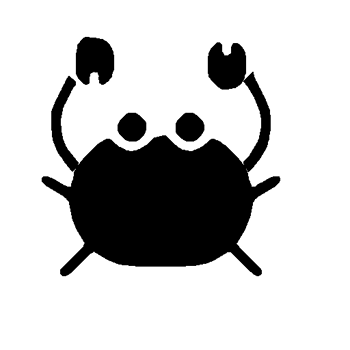} &
        \resultimg{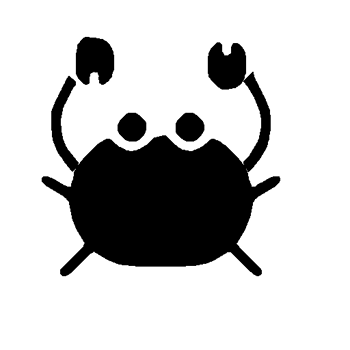}  & 
        \resultimg{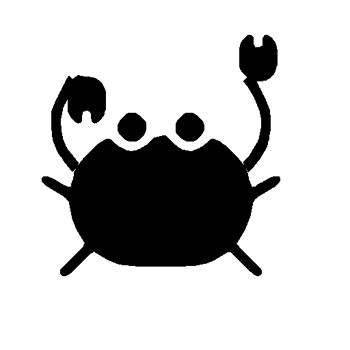} &
        \resultimg{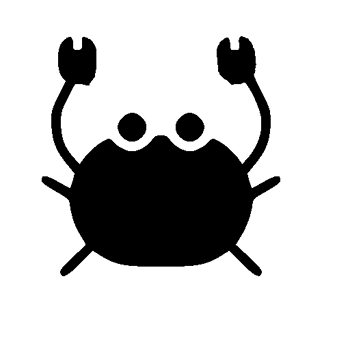}
        \\
        \midrule
        \multicolumn{7}{l}{\parbox{7cm}{\textbf{"Move the right arrow to the left side of the bow"}}}
        \\
        \resultimg{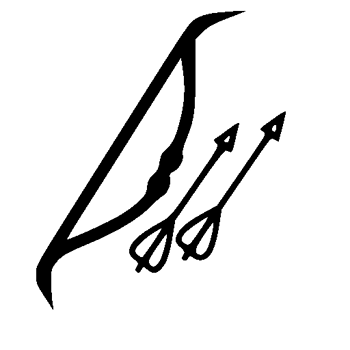} &
        \resultimg{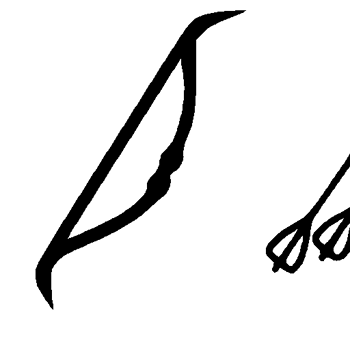} &
        \resultimg{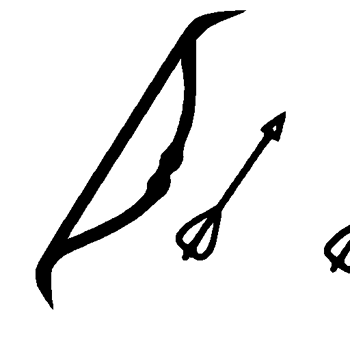} &
        \resultimg{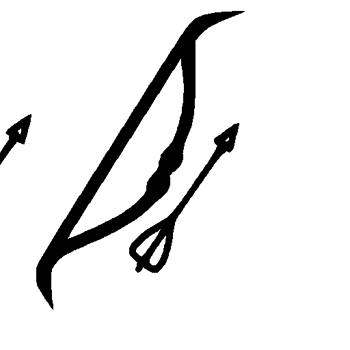} &
        \resultimg{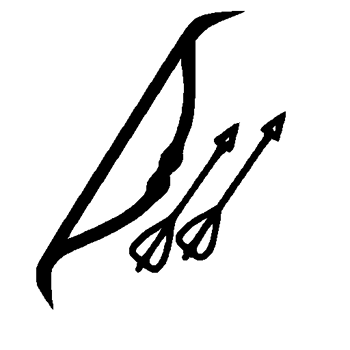}  & 
        \resultimg{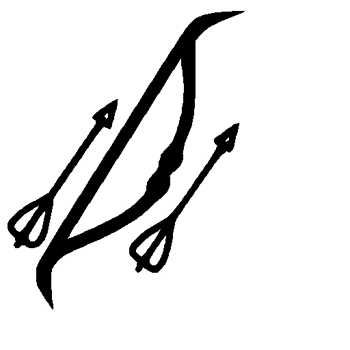} &  
        \resultimg{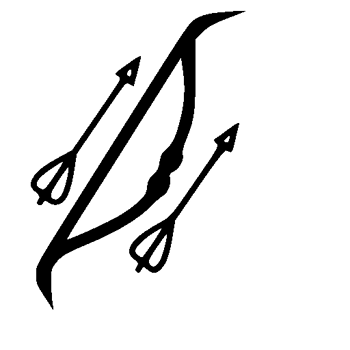}
        \\
    \end{tabular}
    \caption{ Additional qualitative comparison against GPT based editing methods}
    \label{fig:comp_gpt_1}
\end{figure}

In this experiment, we compare our editing method against several LLM based editing methods that we implemented. \textbf{gpt4/gpt4v-dm}: Direct Motion parameters prediction , this method takes in the geometry information(boundary vertex positions) of segments, and ask GPT-4 or GPT-4V to output the affine transformation parameters for every segment. \textbf{gpt4/gpt4v-dc}: Direct Constraints prediction, this method takes in the geometry information(boundary vertex positions) of segments, and ask GPT-4 or GPT4Vision to output the all potential constraints expressed in our DSL. The GPT-4V version of these baselines additionally take in the colored segmented image as input to have more visual information. For all the experiments, the GT segmentation map is given. The relative 2D Chamfer Distance is computed as the Chamfer Distance between the boundary points of predicted edited segment and corresponding GT edited segment then divided by the size of the GT edited segment. Please see Table \ref{tab:comp_gpt}, Fig. \ref{fig:comp_gpt_0} and Fig. \ref{fig:comp_gpt_1} for the comparisons. The results demonstrate our method clearly outperforms all baselines by a large margin. 

\begin{table}
    \centering
    \scriptsize
    \begin{tabular}{lccccc}
        \toprule
        & \raisebox{0em}{gpt4-dm} & \raisebox{0em}{gpt4v-dm} &  \raisebox{0em}{gpt4-dc} & \raisebox{0em}{gpt4v-dc} & \raisebox{0em}{Ours}
        \\
        \midrule
        CD $\downarrow$ & 4.03 & 2.37 & 1.58 & 1.40 & \textbf{0.41}\\
        \bottomrule
    \end{tabular}
    \caption{
    Quantitative comparison against parametric editing methods using LLMs. The error is computed as the relative Chamfer Distance(CD) between boundary points of predicted edited segments and GT edited segments. Computed across 57 images(requests).
    }
    \label{tab:comp_gpt}
\vspace{-4mm}
\end{table}

\subsection{Our method outperforms diffusion based image editing methods by a large margin}
\begin{figure}
    \centering
    \small
    \setlength{\tabcolsep}{0pt}
    \renewcommand{\arraystretch}{0}
    \newcommand{\resultimg}[1]{\includegraphics[trim={0pt 0pt 0pt 0pt},clip,width=0.11\linewidth]{#1}}
    \begin{tabular}{c|c|c|c|c|c|c|c|c}
        \raisebox{0em}{Original} & \raisebox{0em}{sdi2i} & \raisebox{0em}{ip2p} & \raisebox{0em}{imagic} & \raisebox{0em}{sdi2iL} & \raisebox{0em}{ip2pL} & \raisebox{0em}{imagicL} & \raisebox{0em}{Ours} & \raisebox{0em}{Human} 
        \\
        \midrule
        \multicolumn{9}{l}{\parbox{8cm}{\textbf{"Reduce the vertical length of the fuselage by 50\%"}}}
        \\
        \resultimg{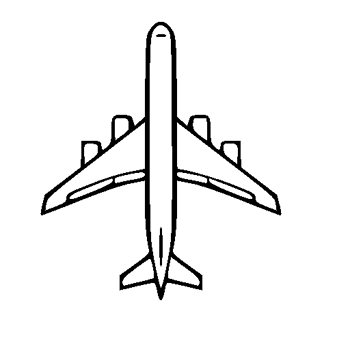} &
        \resultimg{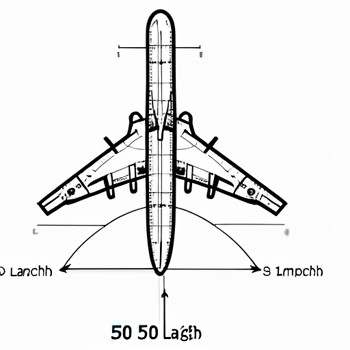} &
        \resultimg{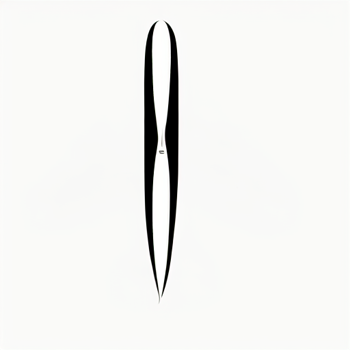} &
        \resultimg{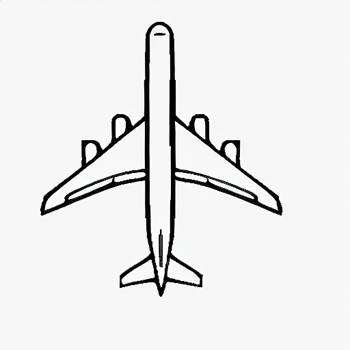} & 
        \resultimg{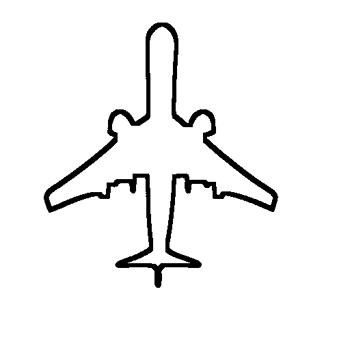} &
        \resultimg{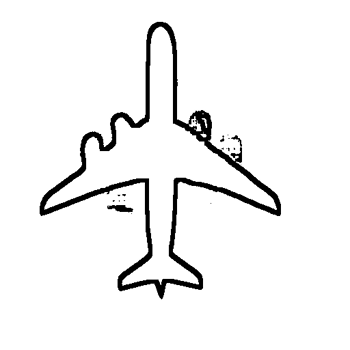} &
        \resultimg{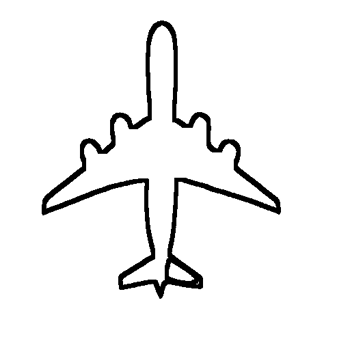} & 
        \resultimg{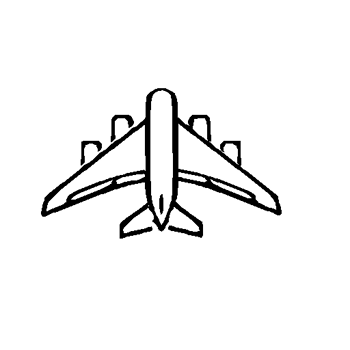} & 
        \resultimg{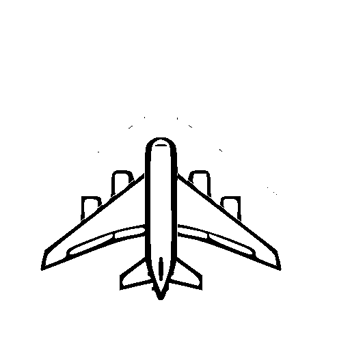}
        \\
        \midrule
        \multicolumn{9}{l}{\parbox{8cm}{\textbf{"Put one bubble on the left of the straw and the other bubble on the right of the straw"}}}
        \\
        \resultimg{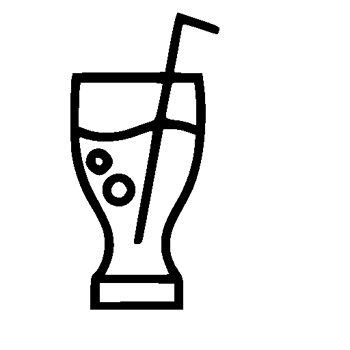} &
        \resultimg{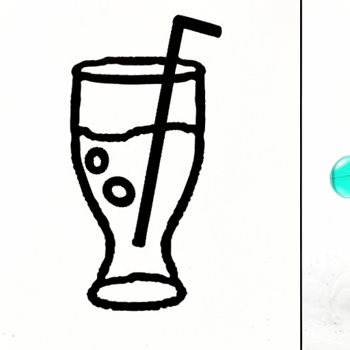} &
        \resultimg{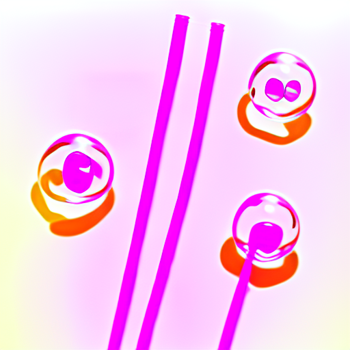} &
        \resultimg{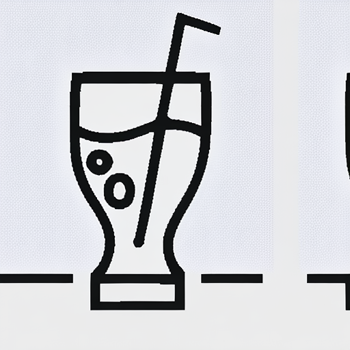} & 
        \resultimg{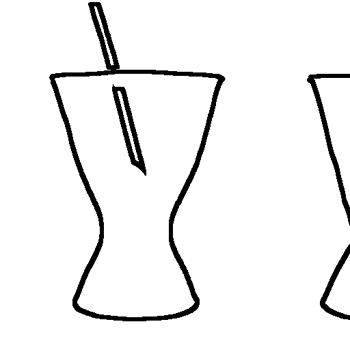} &
        \resultimg{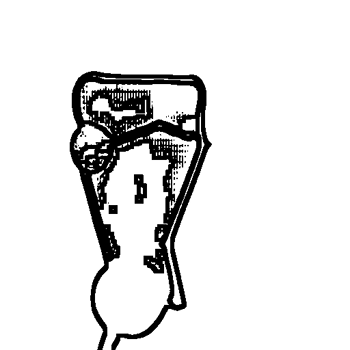} &
        \resultimg{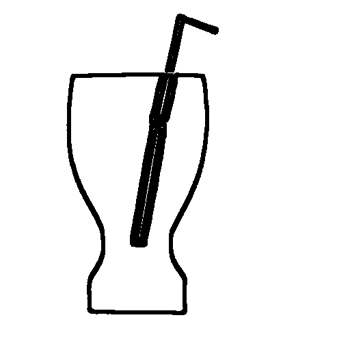} & 
        \resultimg{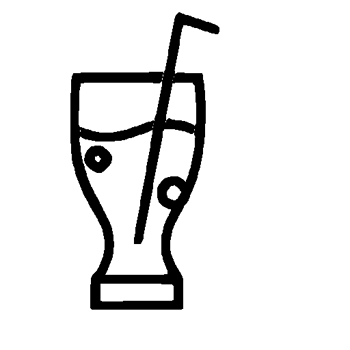} & 
        \resultimg{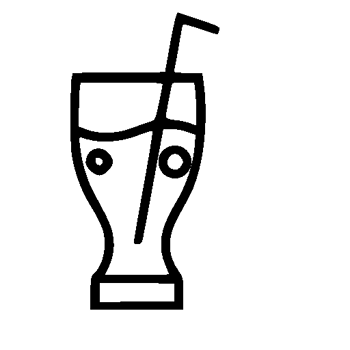}
        \\
        \midrule
        \multicolumn{9}{l}{\parbox{8cm}{\textbf{"Raise the cup to be above the head"}}}
        \\
        \resultimg{figs/results/original_white/11.png} &
        \resultimg{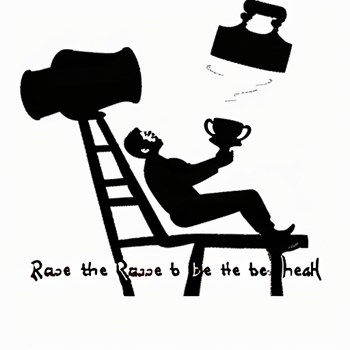} &
        \resultimg{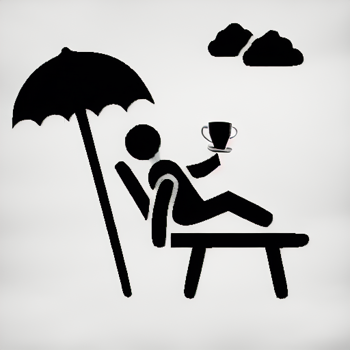} &
        \resultimg{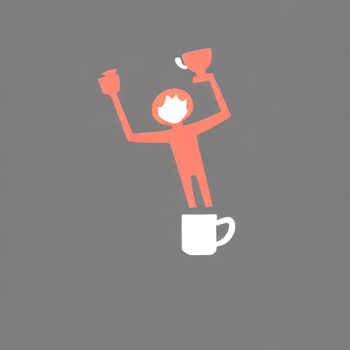} & 
        \resultimg{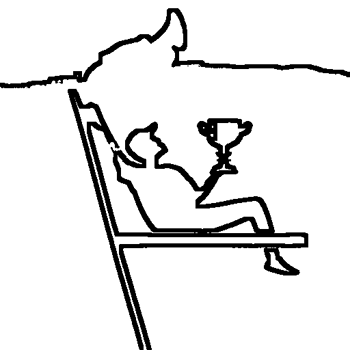} &
        \resultimg{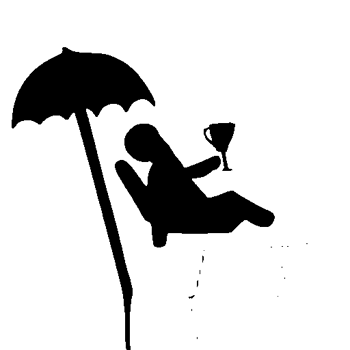} &
        \resultimg{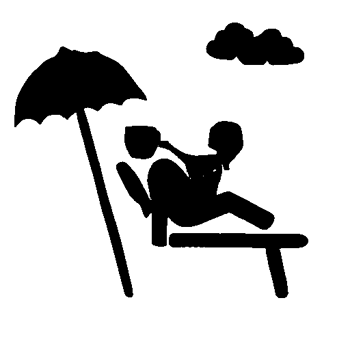} & 
        \resultimg{figs/results/ours_c1m1/11.png} &
        \resultimg{figs/results/gt/11.png}
        \\
        \midrule
        \multicolumn{9}{l}{\parbox{8cm}{\textbf{"Move and rotate the right leaf of the plant to touch the sofa"}}}
        \\
        \resultimg{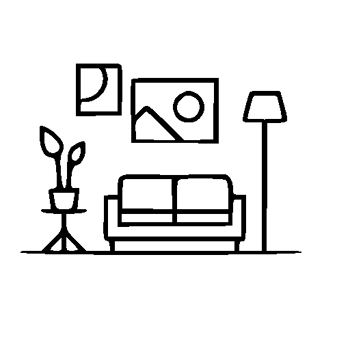} &
        \resultimg{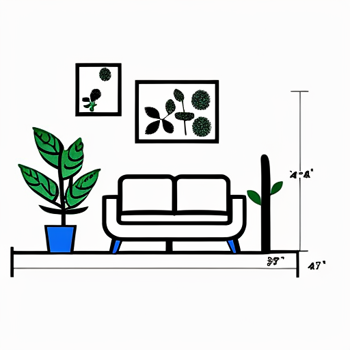} &
        \resultimg{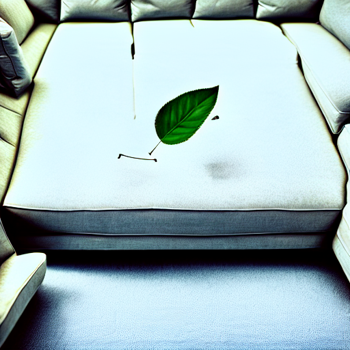} &
        \resultimg{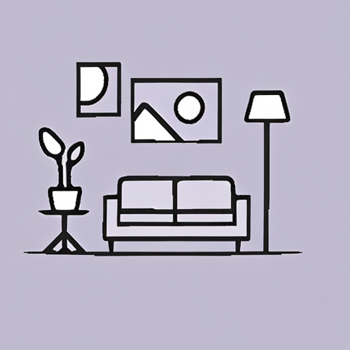} & 
        \resultimg{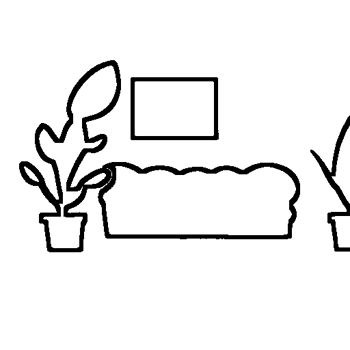} &
        \resultimg{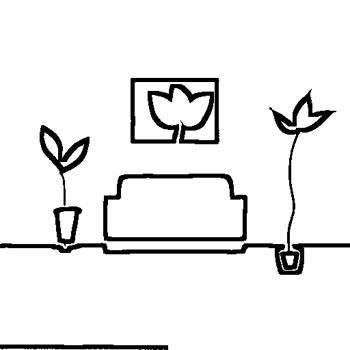} &
        \resultimg{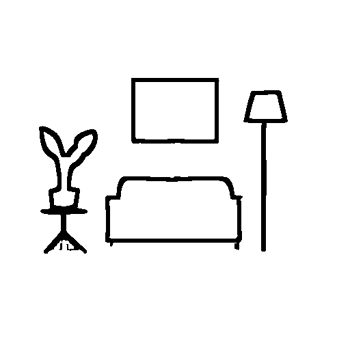} & 
        \resultimg{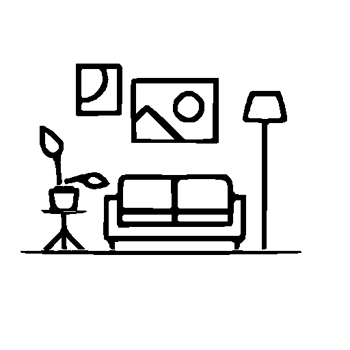} &  
        \resultimg{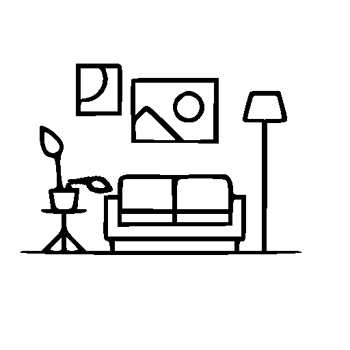}
        \\
        \midrule
        \multicolumn{9}{l}{\parbox{8cm}{\textbf{"Rotate the umbrella by 45 degrees"}}}
        \\
        \resultimg{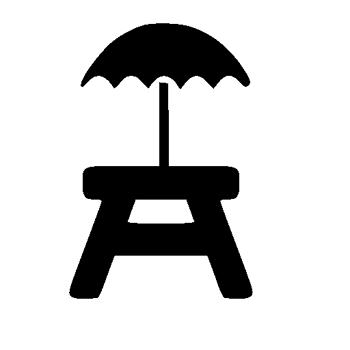} &
        \resultimg{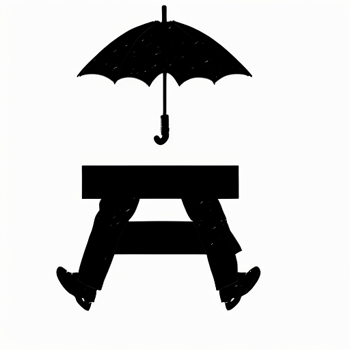} &
        \resultimg{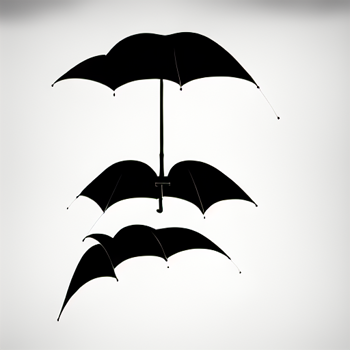} &
        \resultimg{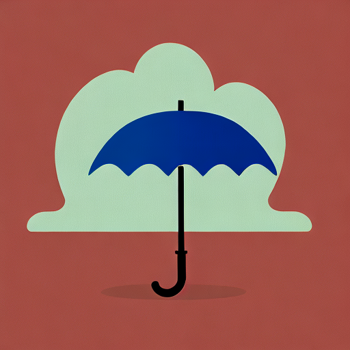} & 
        \resultimg{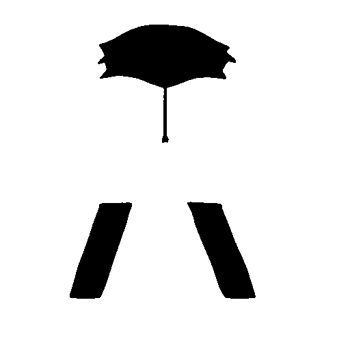} &
        \resultimg{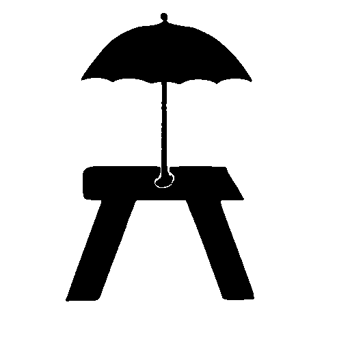} &
        \resultimg{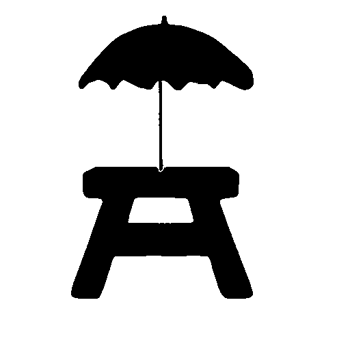} & 
        \resultimg{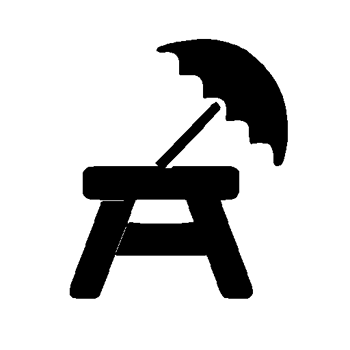} &  
        \resultimg{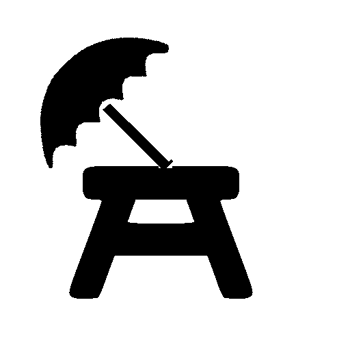}
        \\
        \midrule
        \multicolumn{9}{l}{\parbox{8cm}{\textbf{"Reduce the vertical length of the vase by 50\%, move down the flowers to keep touching the vase"}}}
        \\
        \resultimg{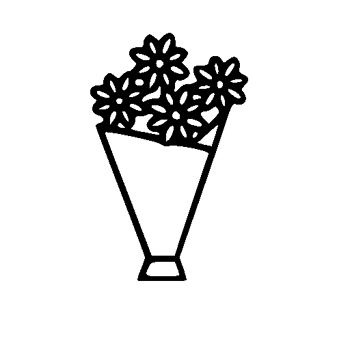} &
        \resultimg{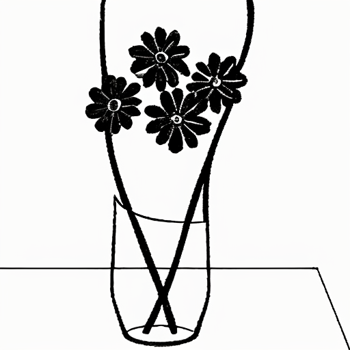} &
        \resultimg{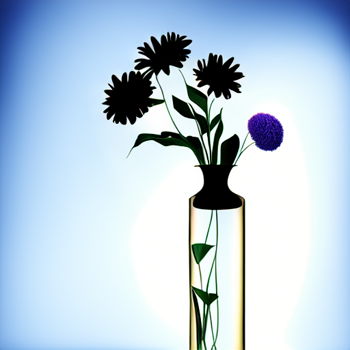} &
        \resultimg{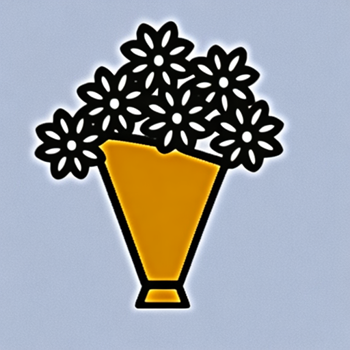} & 
        \resultimg{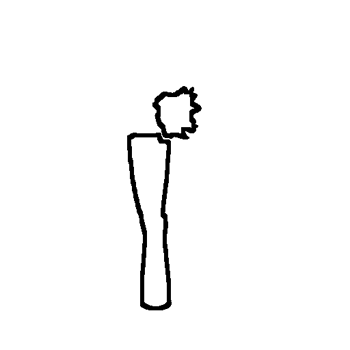} &
        \resultimg{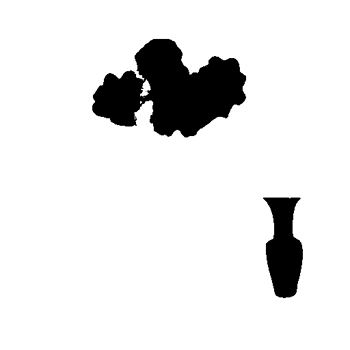} &
        \resultimg{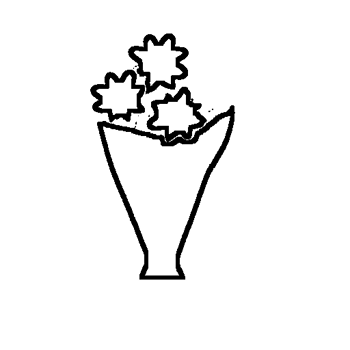} & 
        \resultimg{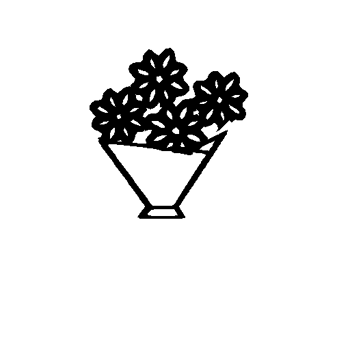} & 
        \resultimg{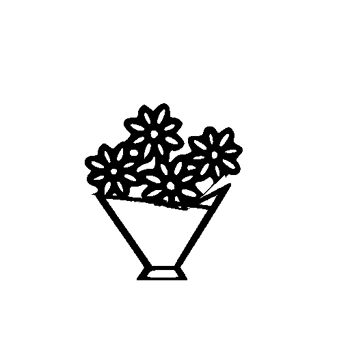}
        \\
    \end{tabular}
    \caption{Qualitative comparison against diffusion based editing methods}
    \label{fig:comp_diffusion_0}
\vspace{-4mm}
\end{figure}

In this experiment, we compare our editing method against several state-of-the-art diffusion based image editing methods: \textbf{sdi2i}: Stable Diffusion Image2Image ~\cite{rombach2021highresolution} , \textbf{ip2p}: InstrauctPix2Pix ~\cite{brooks2022instructpix2pix}, \textbf{imagic}: Imagic ~\cite{kawar2023imagic}, for each of these methods, we first apply the methods (as implemented by Hugging Face) directly on the input icon image to produce results. Additionally, we perform comparisons on variations of these methods. Considered that these methods are mostly trained on natural images and do not take semantic segmentation as input (which our methods have access to), we try to partially address this for a fairer comparison. We designed a lift-and-project variation of these methods, this process uses the labels from the segmentation along with the edge map of the input image to generate a more realistic image using ControlNet~\cite{zhang2023adding}, then we apply these editing methods to edit the generated images, we then use GroundedSam~\cite{ren2024grounded} to detect the shape of the labels and rasterize them back to an icon image for error computation, please see the supplementary material for more details. Method names with an "L" appended (\textbf{sdi2iL}, \textbf{ip2pL} and \textbf{imagicL}) indicates the lift-and-project version is used. We compute the image space mean-squared-error between the edited image and the GT image. The Chamfer Distance used in the previous experiment cannot be used here because we do not have the paired segments between the original image and the edited image. Please see Table \ref{tab:comp_diffusion} and Fig. \ref{fig:comp_diffusion_0} for comparisons. The results show that our method outperforms all baselines by a large margin. 

\begin{table}
    \centering
    \scriptsize
    \begin{tabular}{lccccccc}
        \toprule
        & \raisebox{0em}{sdi2i} & \raisebox{0em}{sdi2iL} & \raisebox{0em}{ip2p} & \raisebox{0em}{ip2pL} & \raisebox{0em}{imagic} & \raisebox{0em}{imagicL} & \raisebox{0em}{Ours}
        \\
        \midrule
        Image MSE $\downarrow$ & 5056 & 7222 & 8808 & 6979 & 16707 & 5540 & \textbf{3148}
        \\
        \bottomrule
    \end{tabular}
    \caption{
    Quantitative comparison against diffusion based image editing methods. The error is computed as the Mean Squared Error(MSE) between edited image and GT image. Computed across 64 images(requests). The pixel value range is [0, 255]. 
    }
    \label{tab:comp_diffusion}
\vspace{-4mm}
\end{table}

\subsection{Ablation Study}
We study the effectiveness of our state search components described in Section~\ref{sec:optim} and report these results in Table~\ref{tab:comp_ablation} and Fig.~\ref{tab:comp_ablation}. We find that without relation constraint search, where all the candidate relation constraints exist, the optimization can become over-constrained and prevent us from achieving the editing goal. Without motion search, some segments are unnecessary modified which leads to less preservation of object features. 

\begin{table}
    \centering
    \scriptsize
    \begin{tabular}{lcccc}
        \toprule
        & \raisebox{0em}{Ours} & \raisebox{0em}{-MS} & \raisebox{0em}{-RS} & \raisebox{0em}{-MS, -RS}
        \\
        \midrule
        CD $\downarrow$ & \textbf{0.475} & 0.509 & 0.997 & 1.040
        \\
        \bottomrule
    \end{tabular}
    \caption{
    Ablation study with various components removed from our system, the error is computed as the relative Chamfer Distance between predicted edited segments and GT edited segments. Computed across 61 images (requests). MS: Motion Search, RS: Relation Search.
    }
    \label{tab:comp_ablation}
\vspace{-4mm}
\end{table}

\subsection{Full Automatic Pipeline}
\begin{figure}
    \centering
    \small
    \setlength{\tabcolsep}{0pt}
    \renewcommand{\arraystretch}{0}
    \newcommand{\resultimg}[1]{\includegraphics[trim={10pt 10pt 10pt 10pt},clip,width=0.14\linewidth]{#1}}
    \begin{tabular}{ccc|ccc}
        \raisebox{0em}{Original} & \raisebox{0em}{Ours Seg} & \raisebox{0em}{Ours Edit} & \raisebox{0em}{Original} & \raisebox{0em}{Ours Seg} & \raisebox{0em}{Ours Edit} 
        \\
        \midrule
        \multicolumn{3}{c}{\parbox{3cm}{\textbf{"Move the lamp shade to touch the base"}}} & 
        \multicolumn{3}{c}{\parbox{3cm}{\textbf{"Reduce the vertical length of the fuselage by 50\%"}}}
        \\
        \resultimg{figs/results/original_white/5.png} &
        \resultimg{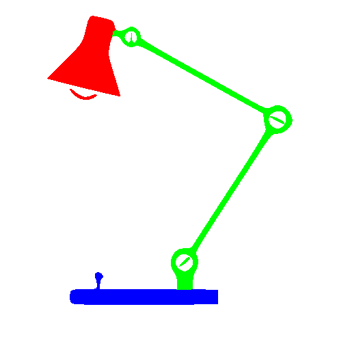} &
        \resultimg{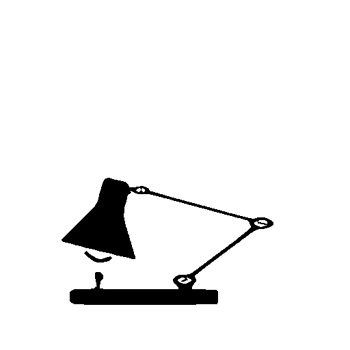} &
        \resultimg{figs/results/original_white/13.png} &
        \resultimg{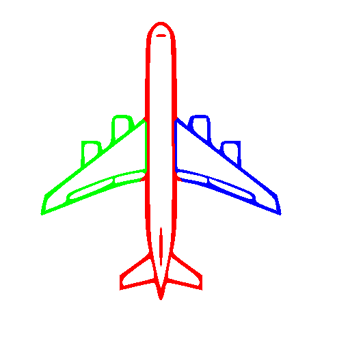} &
        \resultimg{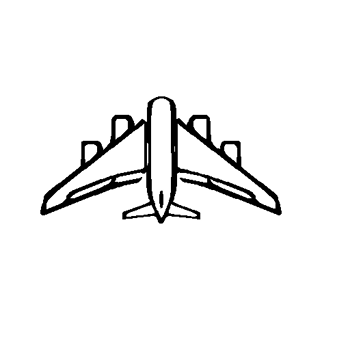} 
        \\
        \midrule
        \multicolumn{3}{c}{\parbox{3cm}{\textbf{"Move the right arrow to the left side of the bow"}}} & 
        \multicolumn{3}{c}{\parbox{3cm}{\textbf{"Move the smallest star to the left of all other stars"}}}
        \\
        \resultimg{figs/results/original_white/16.png} &
        \resultimg{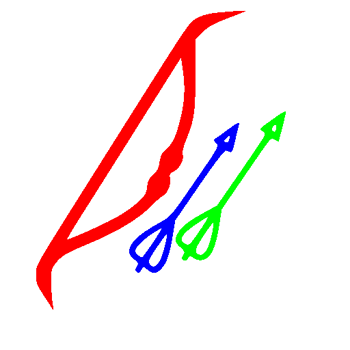} &
        \resultimg{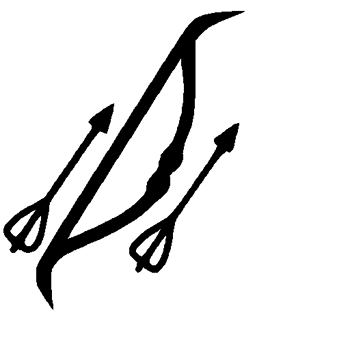} &
        \resultimg{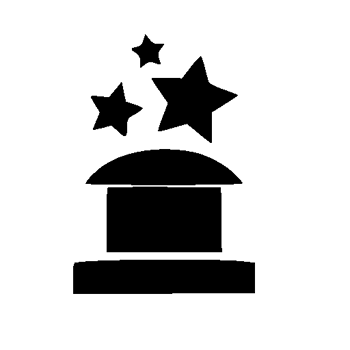} &
        \resultimg{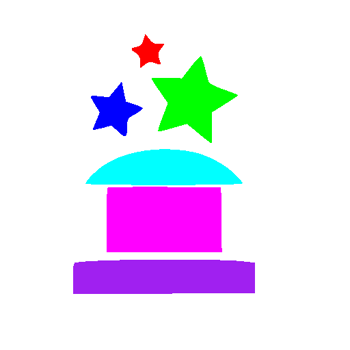} &
        \resultimg{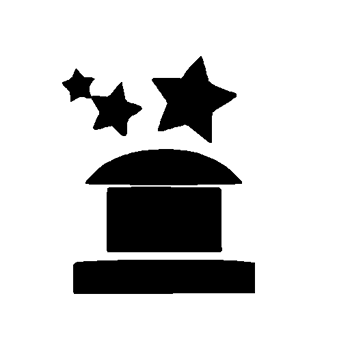} 
        \\
        \midrule
        \multicolumn{3}{c}{\parbox{3cm}{\textbf{"Halve the vertical length of the right building..."}}} & 
        \multicolumn{3}{c}{\parbox{3cm}{\textbf{"Rotate the umbrella by 45 degrees"}}}
        \\
        \resultimg{figs/results/original_white/20.png} &
        \resultimg{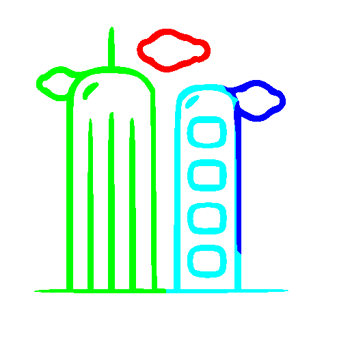} &
        \resultimg{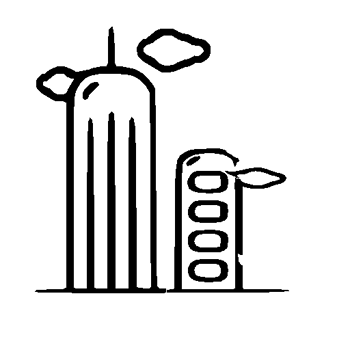} &
        \resultimg{figs/results/original_white/27.png} &
        \resultimg{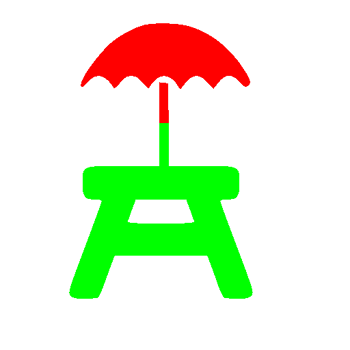} &
        \resultimg{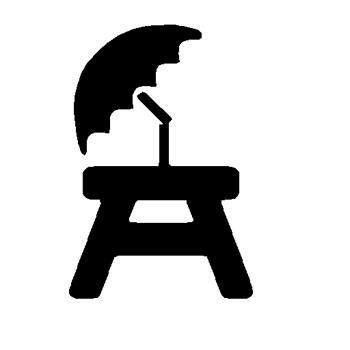} 
        \\
        \midrule
        \multicolumn{3}{c}{\parbox{3cm}{\textbf{"Put the book to touch and on top of the cup"}}} & 
        \multicolumn{3}{c}{\parbox{3cm}{\textbf{"... grass vertical length is equal to the vertical length of the box support column"}}}
        \\
        \resultimg{figs/results/original_white/28.png} &
        \resultimg{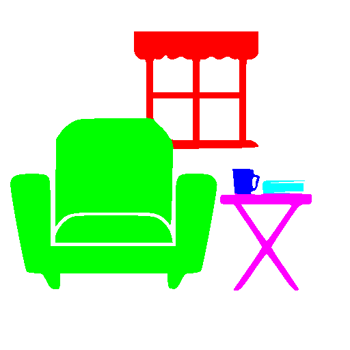} &
        \resultimg{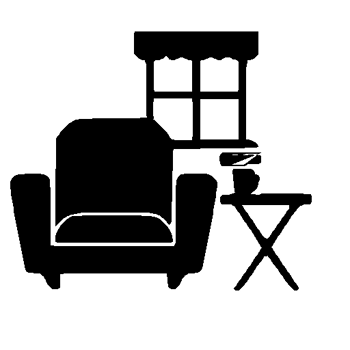} &
        \resultimg{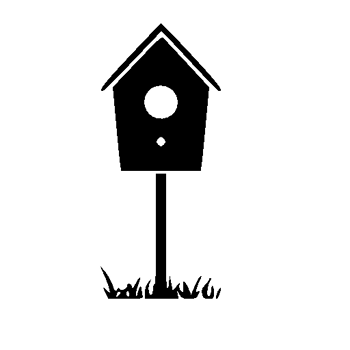} &
        \resultimg{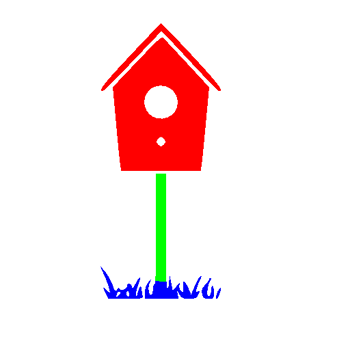} &
        \resultimg{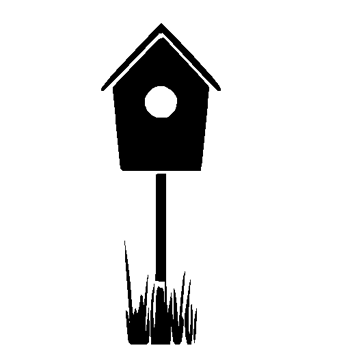} 
        \\
        \midrule
        \multicolumn{3}{c}{\parbox{3cm}{\textbf{"Move and scale the moon to touch the bed"}}} & 
        \multicolumn{3}{c}{\parbox{3cm}{\textbf{"Reduce the vertical length of bread by 50\%"}}}
        \\
        \resultimg{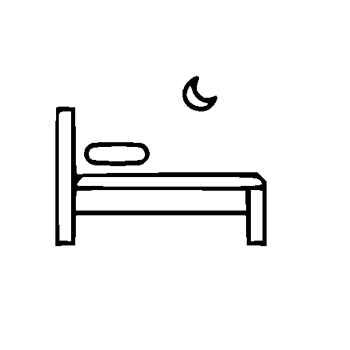} &
        \resultimg{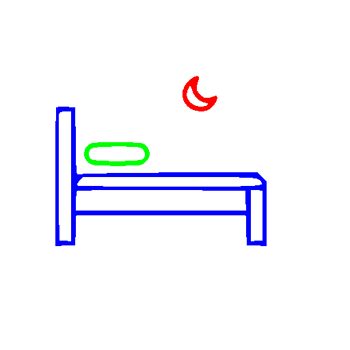} &
        \resultimg{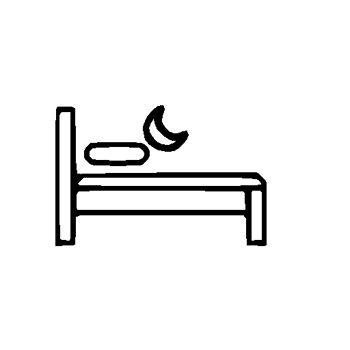} &
        \resultimg{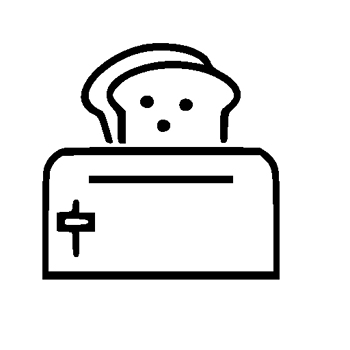} &
        \resultimg{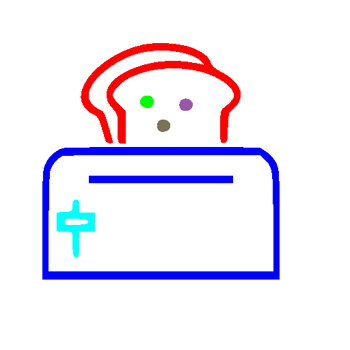} &
        \resultimg{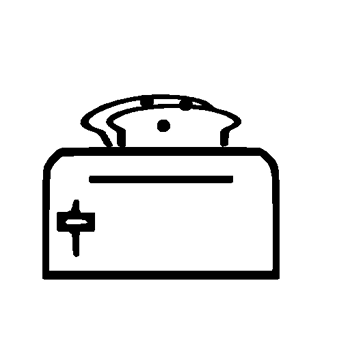} 
        \\
    \end{tabular}
    \caption{Editing results when the full automatic pipeline is enabled}
    \label{fig:full}
\end{figure}

Please see the Fig.~\ref{fig:full} for the editing results of our full automatic pipeline. In this experiment, the only input is the icon image and the editing request, without any GT segmentation or label provided. The segmentation is first automatically predicted by our segmentation tool and then used with the editing pipeline. As you can see, our full automatic pipeline can produce reasonable editing results for some images and its performance can be further improved when segmentation tool pipeline it rely on gets improved.
Please see the supplementary material for the discussion of how to evaluate our segmentation tool and how much it can reduce the manual effort required from users.

\subsection{Limitations and Failure Cases}
The main failure modes are associated with some fundamental challenges of the 2D icon editing task: One failure mode occurs when the editing request is too abstract resulting in the LLM failing to output the correct constraints. 
Another failure mode occurs when the icon image depicts a perspective rendered 3D scene where the spatial relations become too complex. Refer to the supplementary material for extended discussion of limitations.

\subsection{Additional Results}
\begin{figure}
    \centering
    \small
    \renewcommand{\arraystretch}{0}
    \newcommand{\resultimg}[1]{\includegraphics[trim={10pt 10pt 10pt 10pt},clip,width=0.22\linewidth]{#1}}
    \begin{tabular}{ccc}
        \raisebox{0em}{Original} & \raisebox{0em}{Interpretation 1} & \raisebox{0em}{Interpretation 2} 
        \\
        \midrule
        \multicolumn{3}{l}{\parbox{7cm}{\textbf{"Raise the cup to be above the head"}}} 
        \\
        \resultimg{figs/results/original_white/11.png} &
        \resultimg{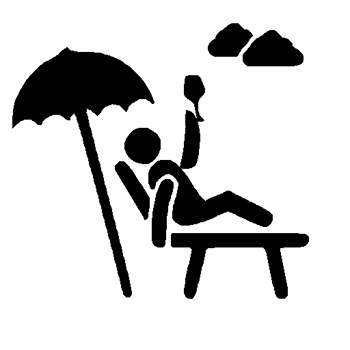} &
        \resultimg{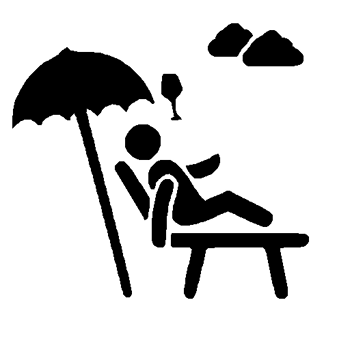} 
        \\
         \midrule
        \multicolumn{3}{l}{\parbox{7cm}{\textbf{"Halve the vertical length of the right building while keep it connected to the ground"}}}
        \\
        \resultimg{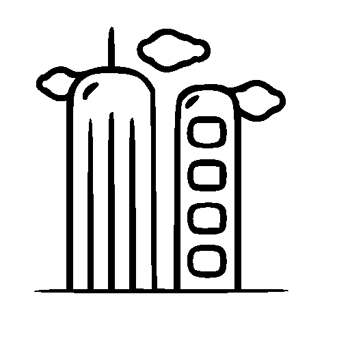} &
        \resultimg{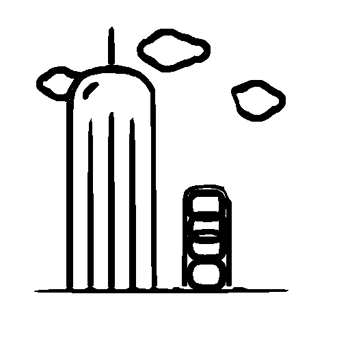} &
        \resultimg{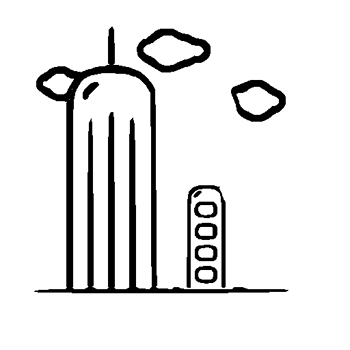} 
        \\
         \midrule
        \multicolumn{3}{l}{\parbox{7cm}{\textbf{"Reduce the horizontal length of the table by 40\%"}}}
        \\
        \resultimg{figs/results/original_white/15.png} &
        \resultimg{figs/results/ours_c1m1/15.png} &
        \resultimg{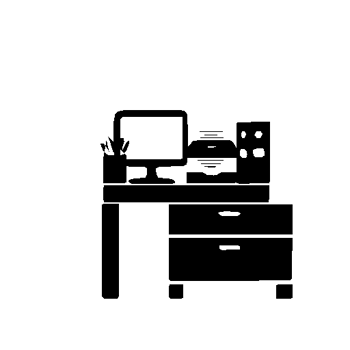} 
        \\
         \midrule
        \multicolumn{3}{l}{\parbox{7cm}{\textbf{"Scale the fish body to make its horizontal length shorter"}}}
        \\
        \resultimg{figs/results/original_white/1.png} &
        \resultimg{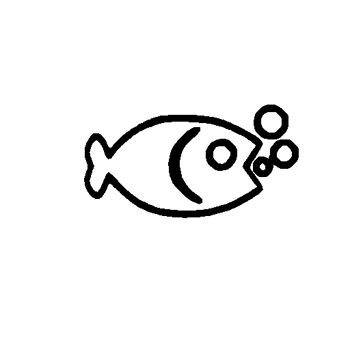} &
        \resultimg{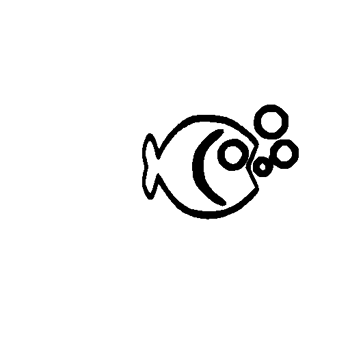} 
        \\
    \end{tabular}
    \caption{Editing results for when our system is asked to give multiple interpretation for the same editing request}
    \label{fig:multiinterp}
\end{figure}

Please see Fig,~\ref{fig:multiinterp} for multiple editing strategies(given by LLM) of the same editing request and Fig.~\ref{fig:comp_complete} for results with segment completion enabled. Please see the supplementary material for more results, such as re-use of the same segmentation for multiple requests, examples of LLM constraint formation, and more qualitative comparison results.
\section{Conclusion}
We presented a language-driven editing system to create spatial variations of icon images, our method greatly outperforms several alternative editing baselines and proven to be the only method (among all demonstrated methods) that is capable of natural editing of icon images. For future work we are interested to generalize our editing method to support more editing operations (such as local deformation, split and merge of objects) that are beyond affine transformations.

\begin{figure}[htb]
    \centering
    \small
    \setlength{\tabcolsep}{0pt}
    \renewcommand{\arraystretch}{0}
    \newcommand{\resultimg}[1]{\includegraphics[trim={0pt 0pt 0pt 0pt},clip,width=0.10\linewidth]{#1}}
    \begin{tabular}{c|c|c|c|c|c|c|c|c}
        \raisebox{0em}{Original} & \raisebox{0em}{sdi2i} & \raisebox{0em}{ip2p} & \raisebox{0em}{imagic} & \raisebox{0em}{sdi2iL} & \raisebox{0em}{ip2pL} & \raisebox{0em}{imagicL} & \raisebox{0em}{Ours}  & \raisebox{0em}{Human}
        \\
        \midrule
        \multicolumn{9}{l}{\parbox{7cm}{\textbf{"Move the person to the left inner region of the bus"}}}
        \\
        \resultimg{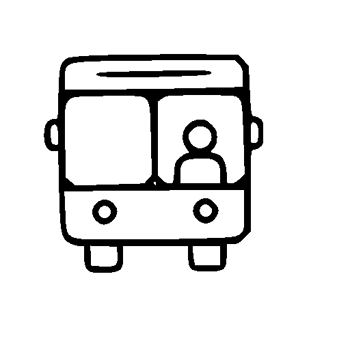} &
        \resultimg{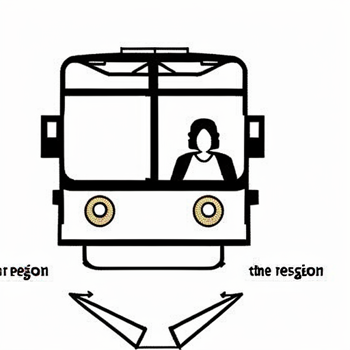} &
        \resultimg{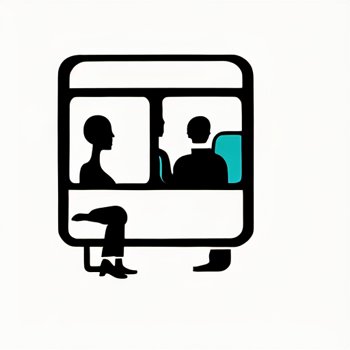} &
        \resultimg{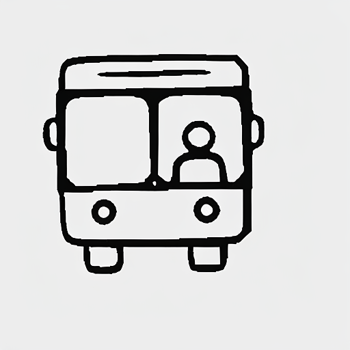} & 
        \resultimg{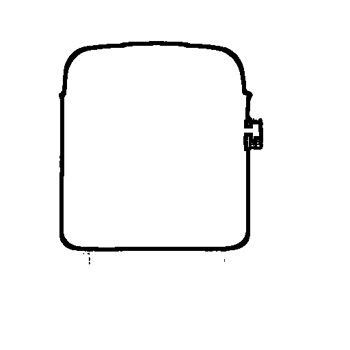} &
        \resultimg{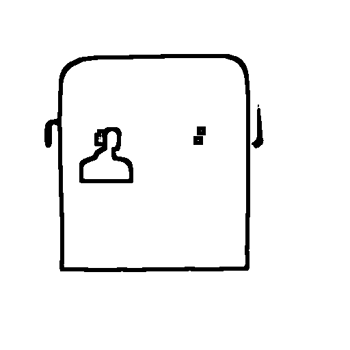} &
        \resultimg{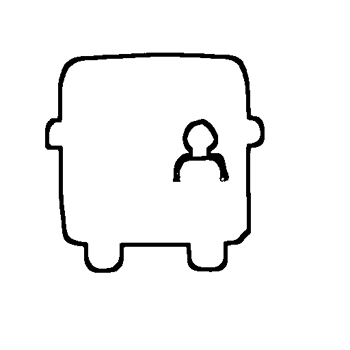} & 
        \resultimg{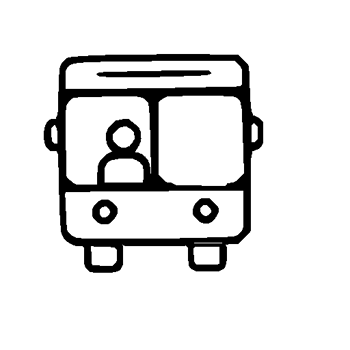} &
        \resultimg{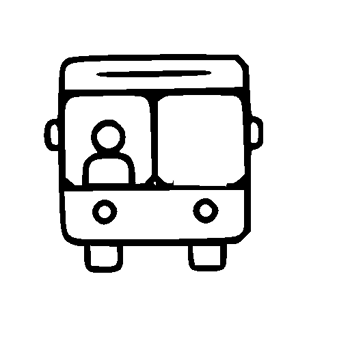}        
        \\
        \midrule
        \multicolumn{9}{l}{\parbox{7cm}{\textbf{"Make the two ice cubes at the same vertical position"}}}
        \\
        \resultimg{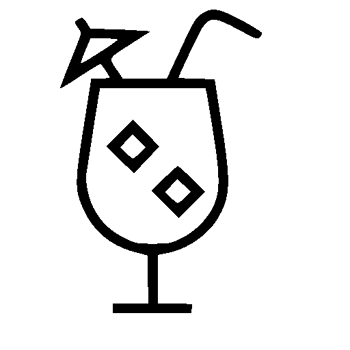} &
        \resultimg{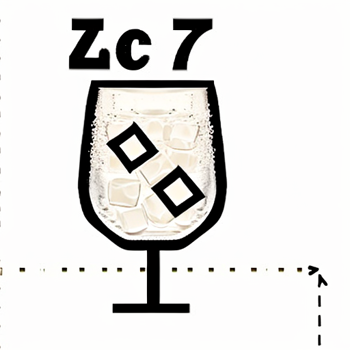} &
        \resultimg{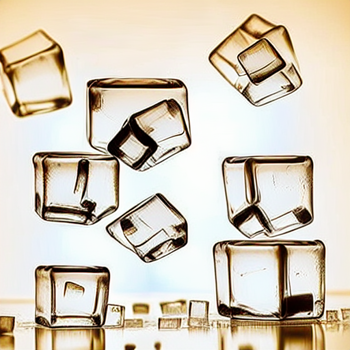} &
        \resultimg{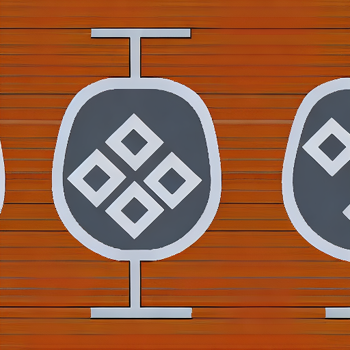} & 
        \resultimg{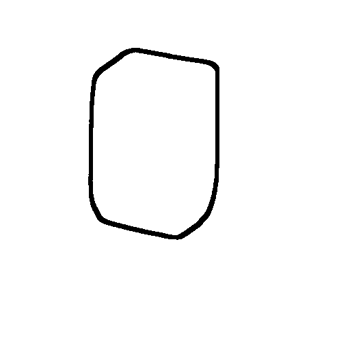} &
        \resultimg{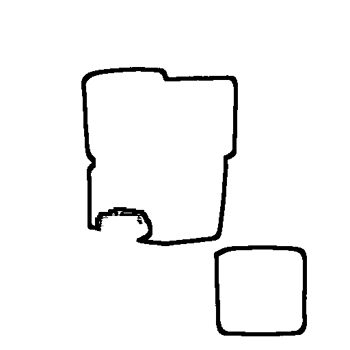} &
        \resultimg{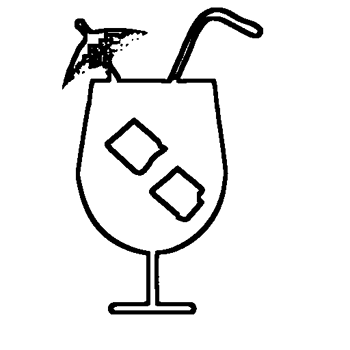} & 
        \resultimg{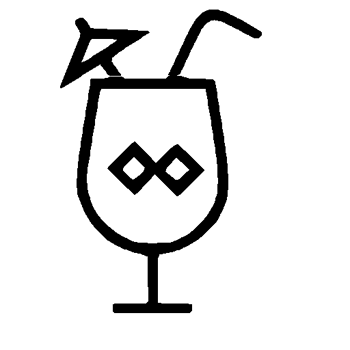} & 
        \resultimg{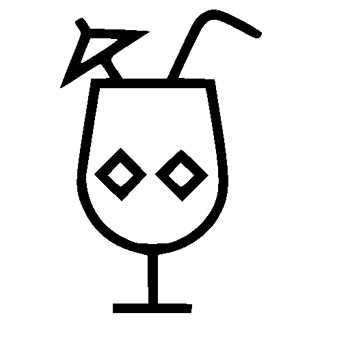}        
        \\
        \midrule
        \multicolumn{9}{l}{\parbox{7cm}{\textbf{"Swap the position of the lollipop and the upper candy"}}}
        \\
        \resultimg{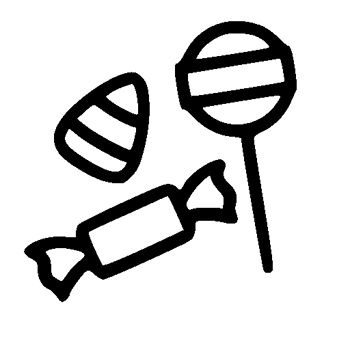} &
        \resultimg{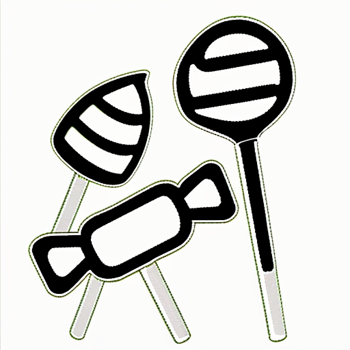} &
        \resultimg{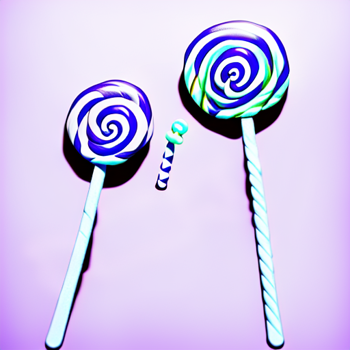} &
        \resultimg{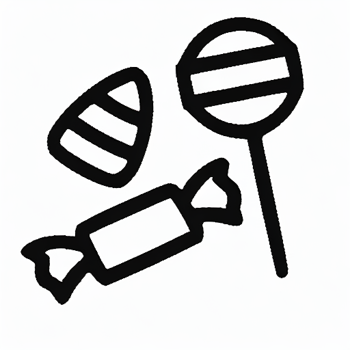} & 
        \resultimg{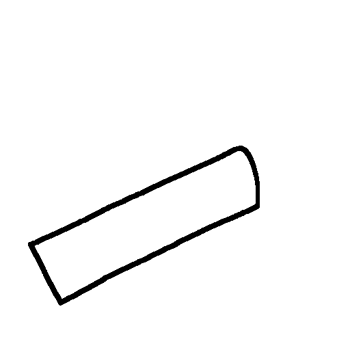} &
        \resultimg{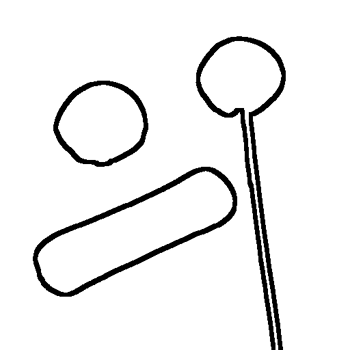} &
        \resultimg{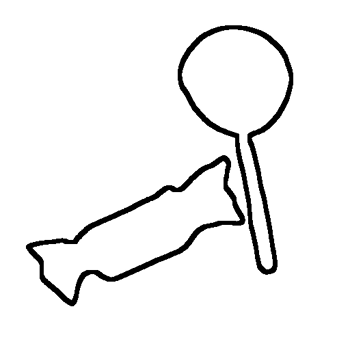} & 
        \resultimg{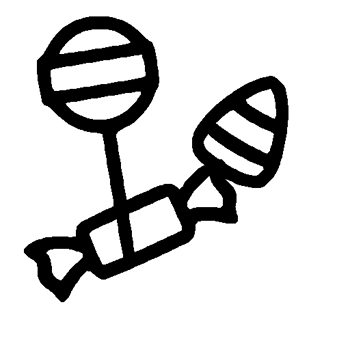} &  
        \resultimg{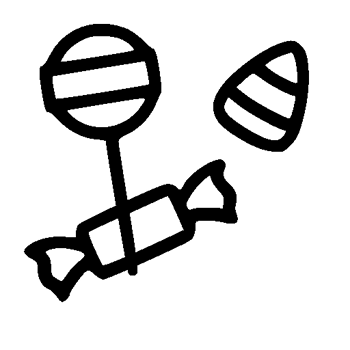}        
        \\
        \midrule
        \multicolumn{9}{l}{\parbox{7cm}{\textbf{"Move the lamp to touch and to be on top of the sofa"}}}
        \\
        \resultimg{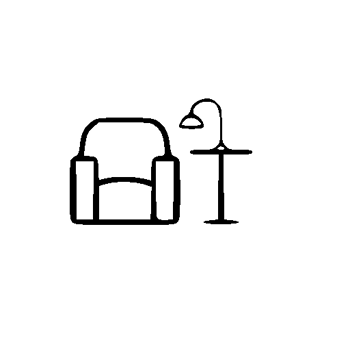} &
        \resultimg{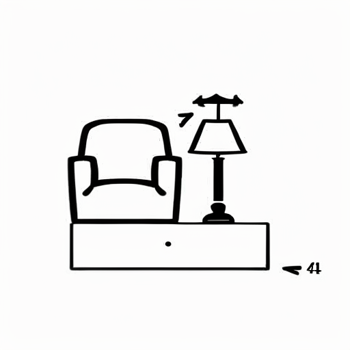} &
        \resultimg{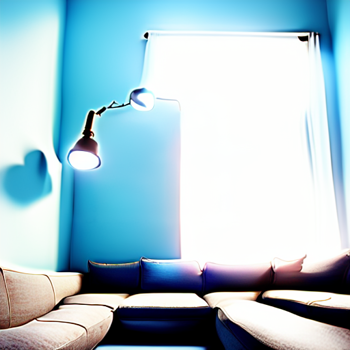} &
        \resultimg{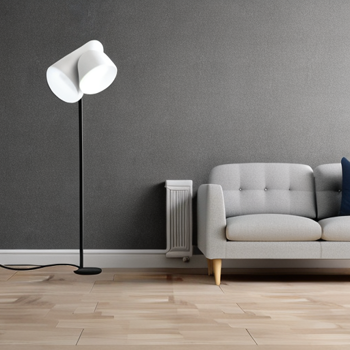} & 
        \resultimg{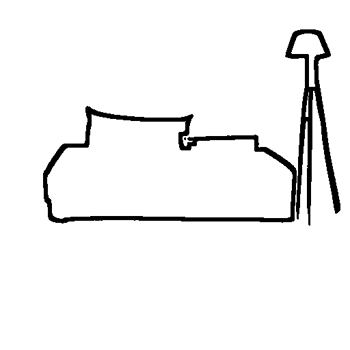} &
        \resultimg{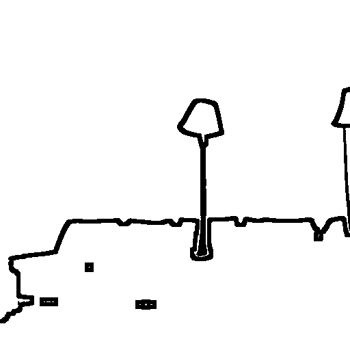} &
        \resultimg{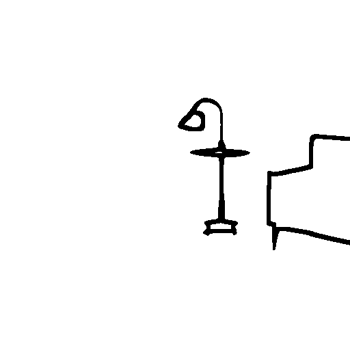} & 
        \resultimg{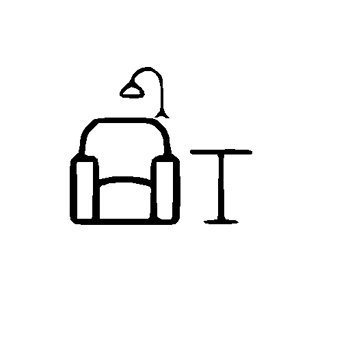} &  
        \resultimg{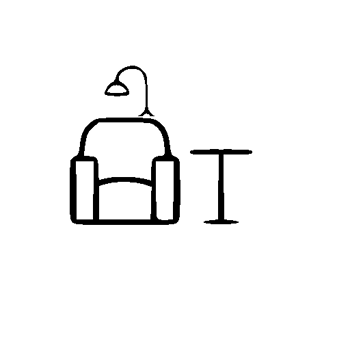}        
        \\
        \midrule
        \multicolumn{9}{l}{\parbox{7cm}{\textbf{"Make all trees the same vertical length"}}}
        \\
        \resultimg{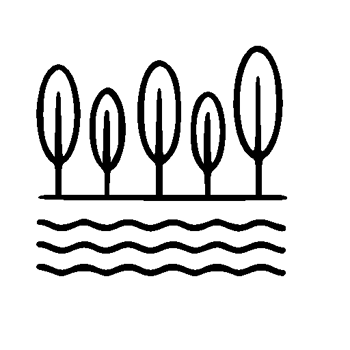} &
        \resultimg{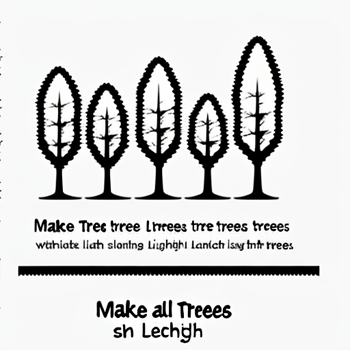} &
        \resultimg{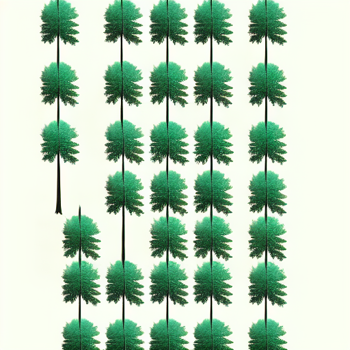} &
        \resultimg{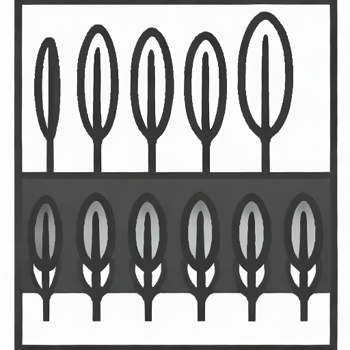} & 
        \resultimg{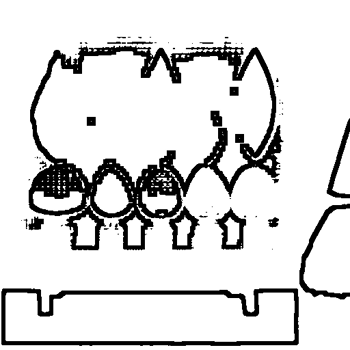} &
        \resultimg{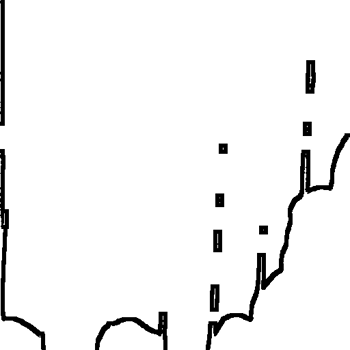} &
        \resultimg{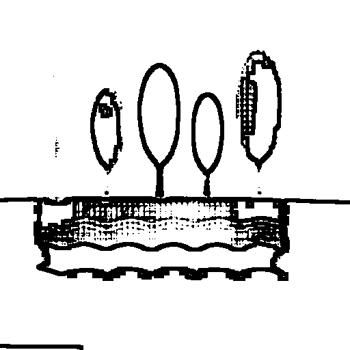} & 
        \resultimg{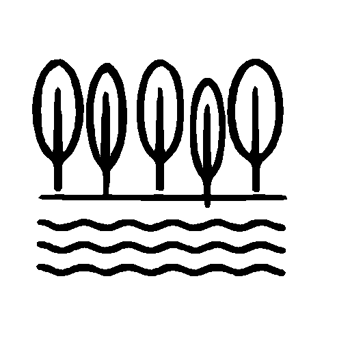}  &
        \resultimg{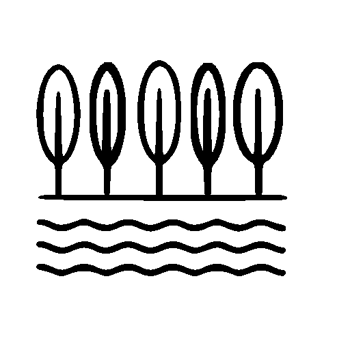}        
        \\
        \midrule
        \multicolumn{9}{l}{\parbox{7cm}{\textbf{"Rotate the fan blade by 60 degrees"}}}
        \\
        \resultimg{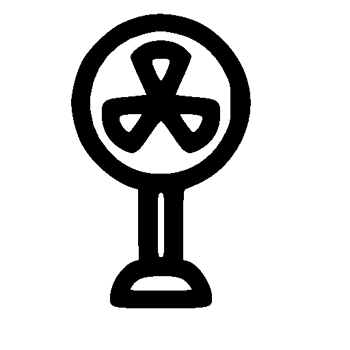} &
        \resultimg{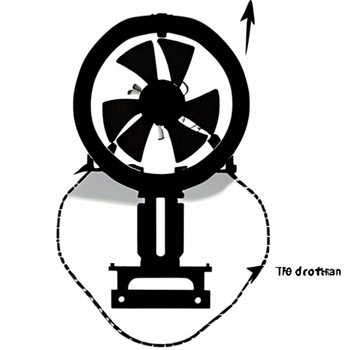} &
        \resultimg{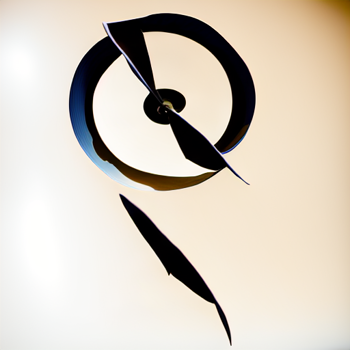} &
        \resultimg{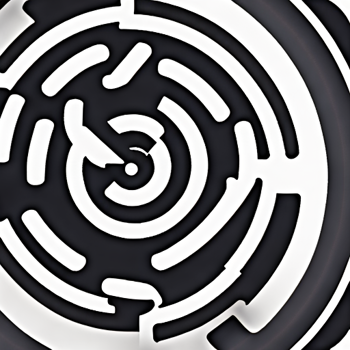} & 
        \resultimg{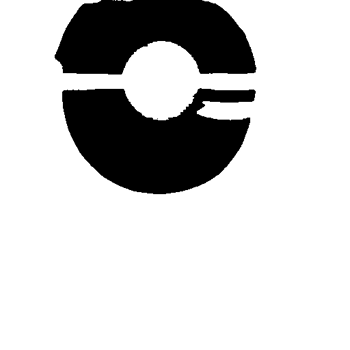} &
        \resultimg{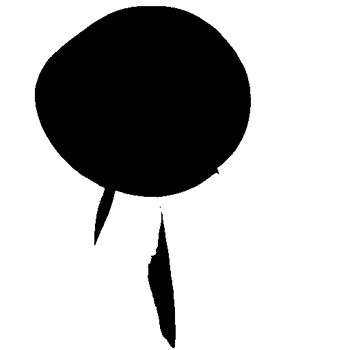} &
        \resultimg{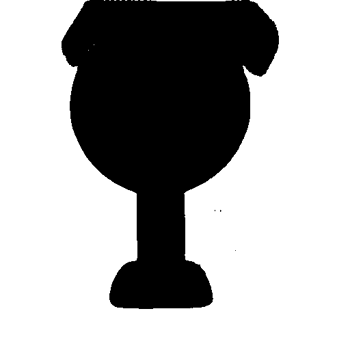} & 
        \resultimg{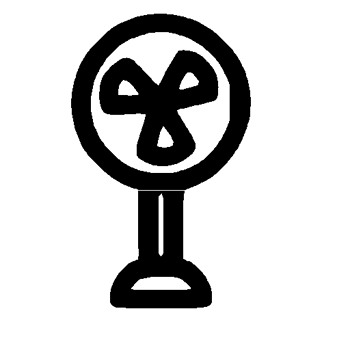} &
        \resultimg{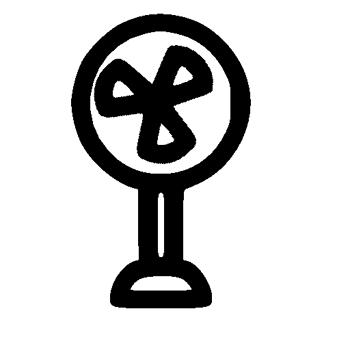}        
        \\
    \end{tabular}
    \caption{Additional qualitative comparison against diffusion based editing methods }
    \label{fig:comp_diffusion_1}
\end{figure}

\begin{figure}
    \centering
    \small
    \renewcommand{\arraystretch}{0}
    \newcommand{\resultimg}[1]{\includegraphics[trim={10pt 10pt 10pt 10pt},clip,width=0.14\linewidth]{#1}}
    \begin{tabular}{ccccc}
        \raisebox{0em}{Original} & \raisebox{0em}{-MS, -RS} & \raisebox{0em}{-RS} & \raisebox{0em}{-MS} & \raisebox{0em}{Ours} 
        \\
        \midrule
        \multicolumn{5}{l}{\parbox{7cm}{\textbf{"Rotate the handles apart so that the distance between their top regions is equal to half of the basket's horizontal length"}}}
        \\
        \resultimg{figs/results/original_white/0.png} &
        \resultimg{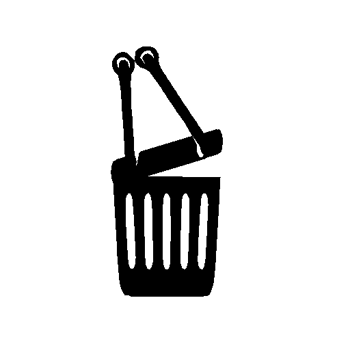} &
        \resultimg{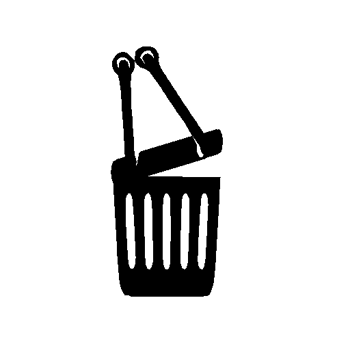} &
        \resultimg{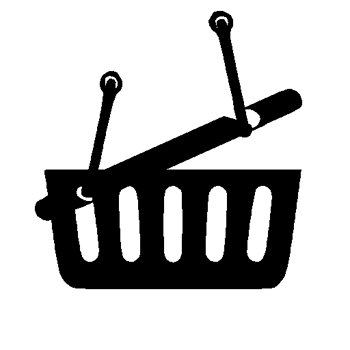} & 
        \resultimg{figs/results/ours_c1m1/0.png}  
        \\
        \midrule
        \multicolumn{5}{l}{\parbox{7cm}{\textbf{"Swap the positions of the door and the leftmost window"}}}
        \\
        \resultimg{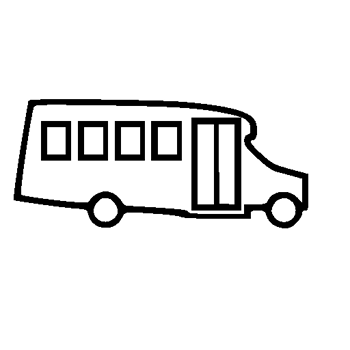} &
        \resultimg{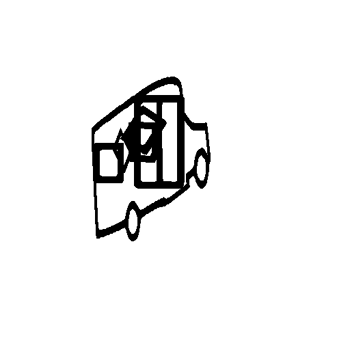} &
        \resultimg{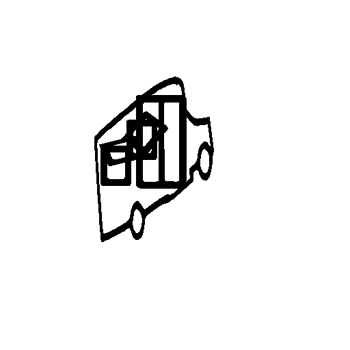} &
        \resultimg{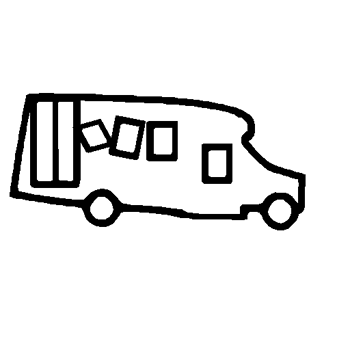} & 
        \resultimg{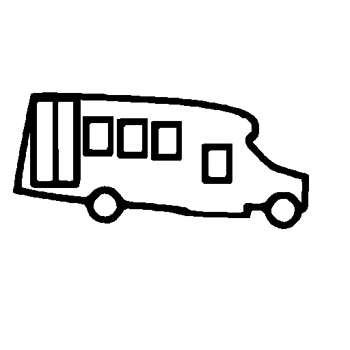}  
        \\
        \midrule
        \multicolumn{5}{l}{\parbox{7cm}{\textbf{"Put one bubble on the left of the straw and the other bubble on the right of the straw"}}}
        \\
        \resultimg{figs/results/original_white/26.png} &
        \resultimg{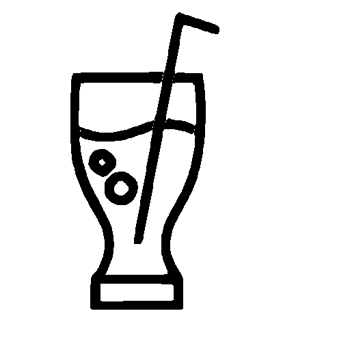} &
        \resultimg{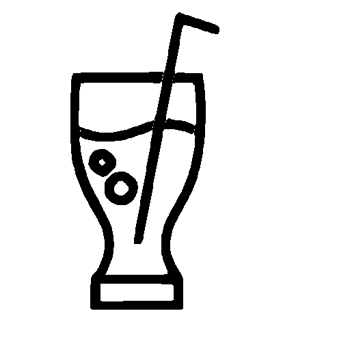} &
        \resultimg{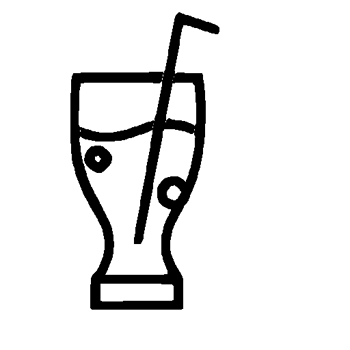} & 
        \resultimg{figs/results/ours_c1m1/26.png}  
        \\
        \midrule
        \multicolumn{5}{l}{\parbox{7cm}{\textbf{"Reduce the vertical length of the fuselage by 50\%"}}}
        \\
        \resultimg{figs/results/original_white/13.png} &
        \resultimg{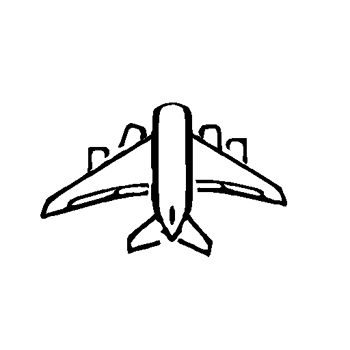} &
        \resultimg{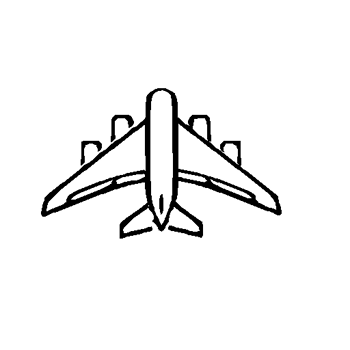} &
        \resultimg{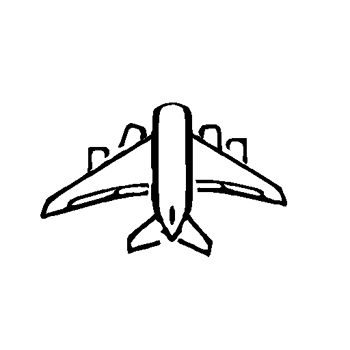} & 
        \resultimg{figs/results/ours_c1m1/13.png}  
        \\
        \midrule
        \multicolumn{5}{l}{\parbox{7cm}{\textbf{"Reduce the horizontal length of the table by 40\%"}}}
        \\
        \resultimg{figs/results/original_white/15.png} &
        \resultimg{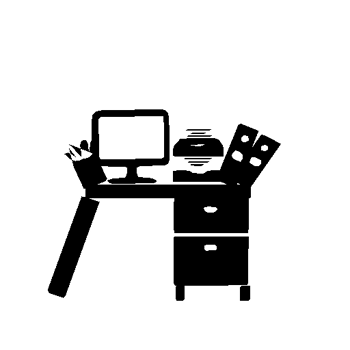} &
        \resultimg{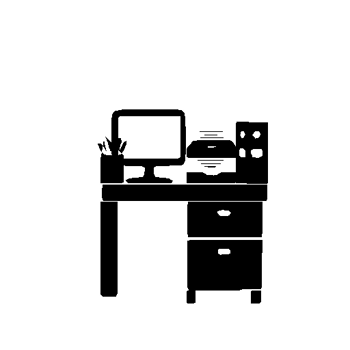} &
        \resultimg{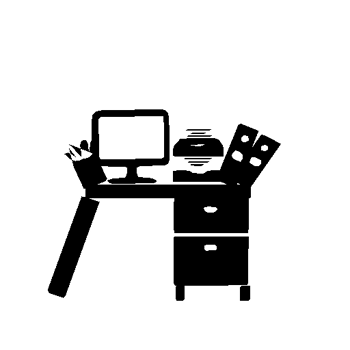} & 
        \resultimg{figs/results/ours_c1m1/15.png}  
        \\
        \midrule
        \multicolumn{5}{l}{\parbox{7cm}{\textbf{"Scale the fish body to reduce its horizontal length by 50\%"}}}
        \\
        \resultimg{figs/results/original_white/1.png} &
        \resultimg{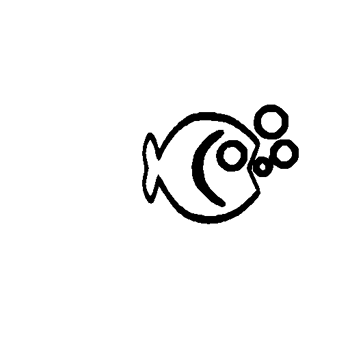} &
        \resultimg{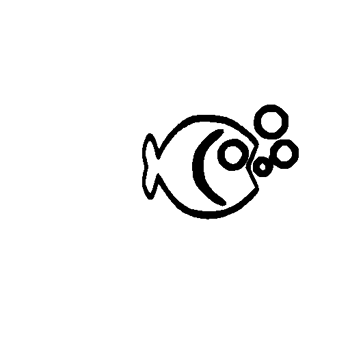} &
        \resultimg{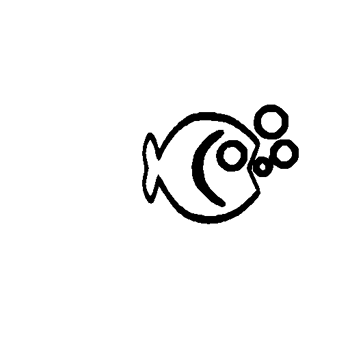} & 
        \resultimg{figs/results/ours_c1m1/1.png}  
        \\
        \midrule
        \multicolumn{5}{l}{\parbox{7cm}{\textbf{"Move the lamp shade to touch the base"}}}
        \\
        \resultimg{figs/results/original_white/5.png} &
        \resultimg{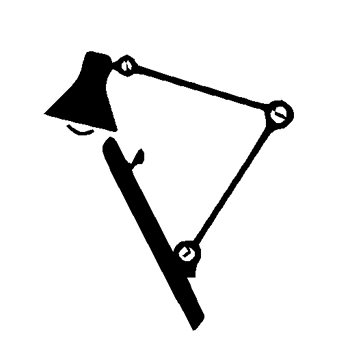} &
        \resultimg{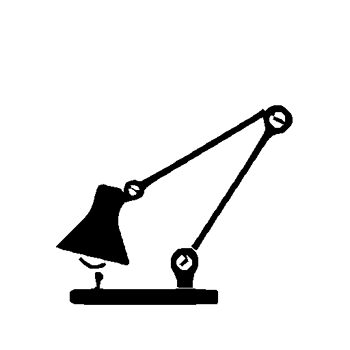} &
        \resultimg{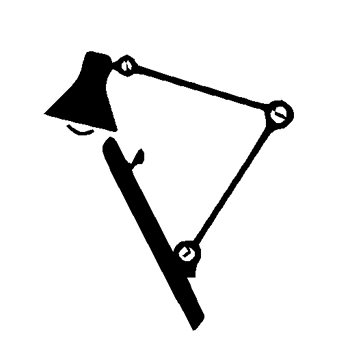} & 
        \resultimg{figs/results/ours_c1m1/5.png}  
        \\
        \midrule
        \multicolumn{5}{l}{\parbox{7cm}{\textbf{"Put the book to touch and on top of the cup"}}}
        \\
        \resultimg{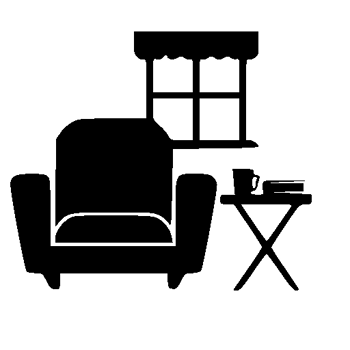} &
        \resultimg{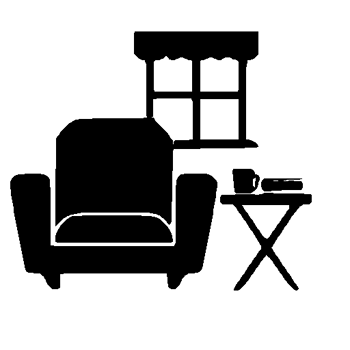} &
        \resultimg{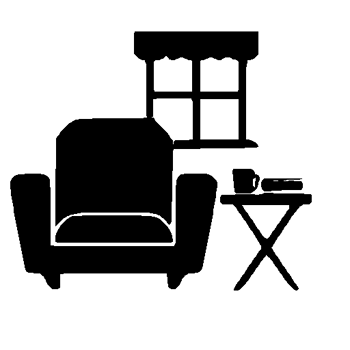} &
        \resultimg{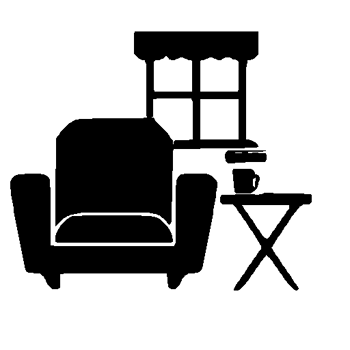} & 
        \resultimg{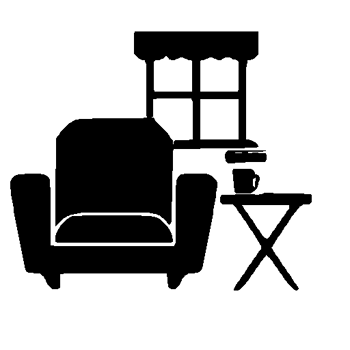}  
        \\
        \midrule
        \multicolumn{5}{l}{\parbox{7cm}{\textbf{"Raise the cup to be above the head"}}}
        \\
        \resultimg{figs/results/original_white/11.png} &
        \resultimg{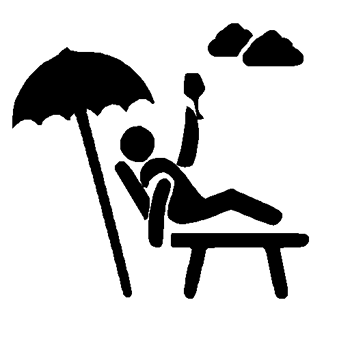} &
        \resultimg{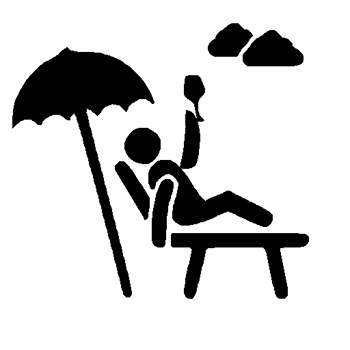} &
        \resultimg{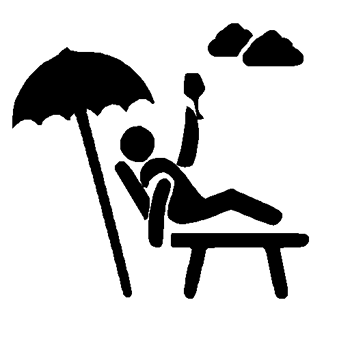} & 
        \resultimg{figs/results/ours_c1m1/11.png}  
        \\
        \midrule
        \multicolumn{5}{l}{\parbox{7cm}{\textbf{"Move and rotate the right leaf of the plant to touch the sofa"}}}
        \\
        \resultimg{figs/results/original_white/12.png} &
        \resultimg{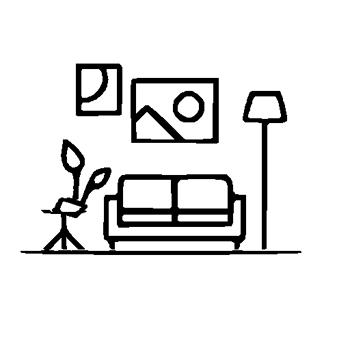} &
        \resultimg{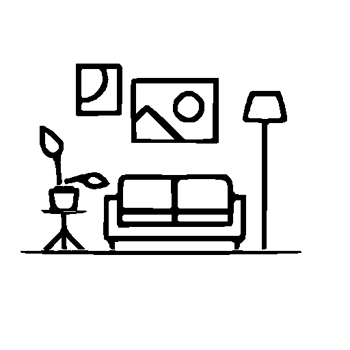} &
        \resultimg{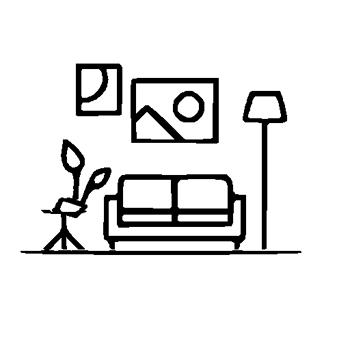} & 
        \resultimg{figs/results/ours_c1m1/12.png}  
        \\
    \end{tabular}
    \caption{Ablation study. MS: Motion Search, RS: Relation Search.}
    \label{fig:comp_ablation_1}
\end{figure}

\begin{figure}
    \centering
    \small
    \renewcommand{\arraystretch}{0}
    \newcommand{\resultimg}[1]{\includegraphics[trim={10pt 10pt 10pt 10pt},clip,width=0.22\linewidth]{#1}}
    \begin{tabular}{ccc}
        \raisebox{0em}{Original} & \raisebox{0em}{Without Occlusion Guess} & \raisebox{0em}{With Occlusion Guess} 
        \\
        \midrule
        \multicolumn{3}{l}{\parbox{7cm}{\textbf{"Move the leftmost candle(with flame) and rightmost candle(with flame) to touch the 2nd layer of the cake"}}}
        \\
        \resultimg{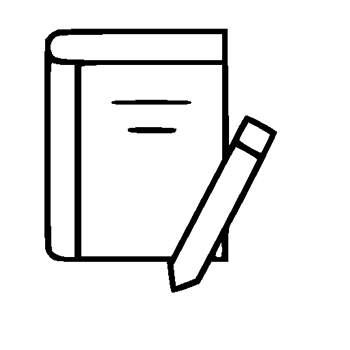} &
        \resultimg{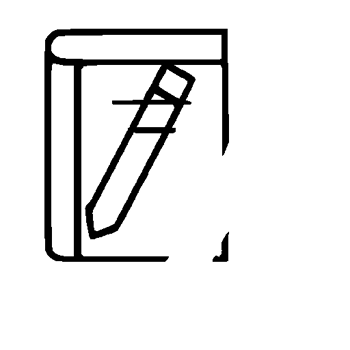} &
        \resultimg{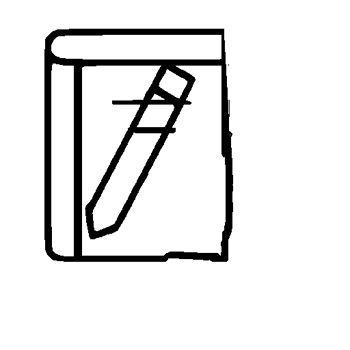} 
        \\
        \midrule
        \multicolumn{3}{l}{\parbox{7cm}{\textbf{"Move and rotate the paddle to the right region of the boat"}}}
        \\
        \resultimg{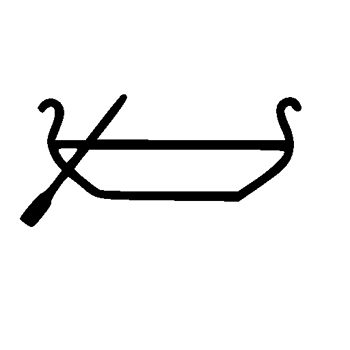} &
        \resultimg{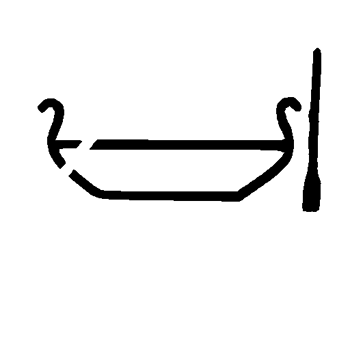} &
        \resultimg{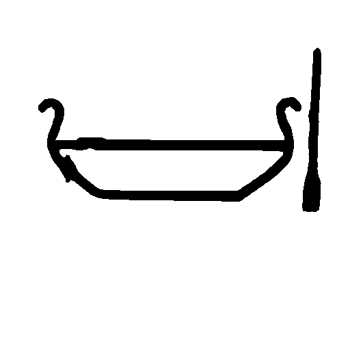} 
        \\
        \midrule
        \multicolumn{3}{l}{\parbox{7cm}{\textbf{"Pull the carrot out of the bowl"}}}
        \\
        \resultimg{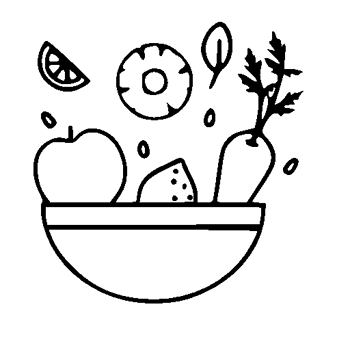} &
        \resultimg{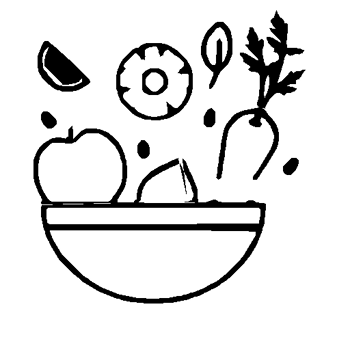} &
        \resultimg{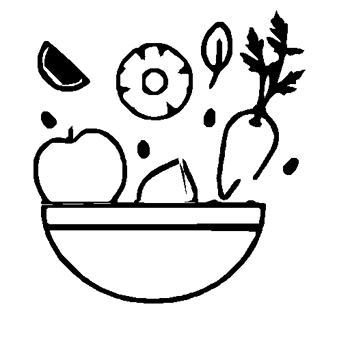} 
        \\
    \end{tabular}
    \caption{Editing results with the occlusion completion component enabled}
    \label{fig:comp_complete}
\end{figure}

\begin{figure}
    \centering
    \small
    \setlength{\tabcolsep}{0pt}
    \renewcommand{\arraystretch}{0}
    \newcommand{\resultimg}[1]{\includegraphics[trim={10pt 10pt 10pt 10pt},clip,width=0.13\linewidth]{#1}}
    \begin{tabular}{ccc|ccc}
        \raisebox{0em}{Original} & \raisebox{0em}{Seg-Tool} & \raisebox{0em}{GT} & \raisebox{0em}{Original} & \raisebox{0em}{Seg-Tool} & \raisebox{0em}{GT} 
        \\
        \resultimg{figs/results/original_white/0.png} &
        \resultimg{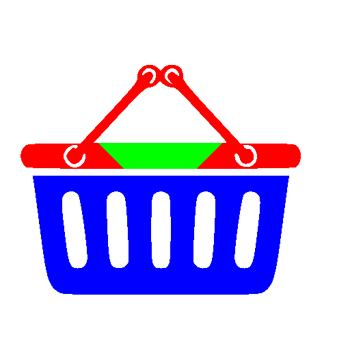} &
        \resultimg{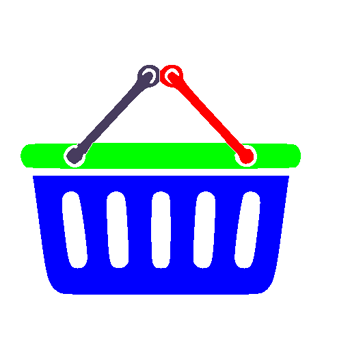} &
        \resultimg{figs/results/original_white/1.png} &
        \resultimg{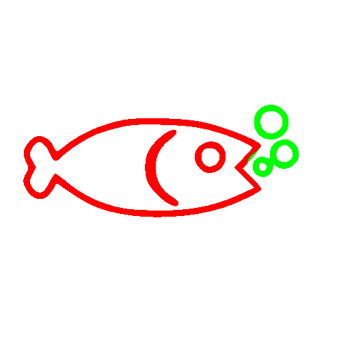} &
        \resultimg{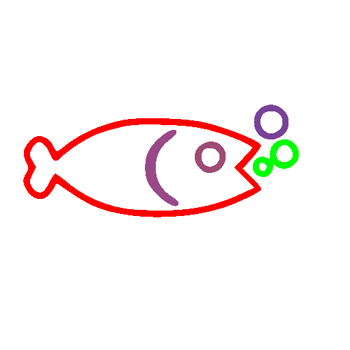} 
        \\
        \resultimg{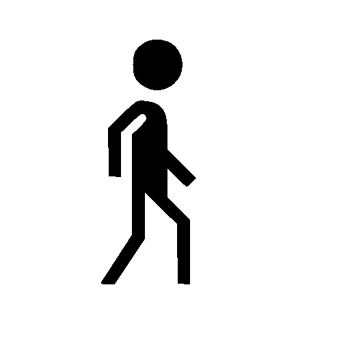} &
        \resultimg{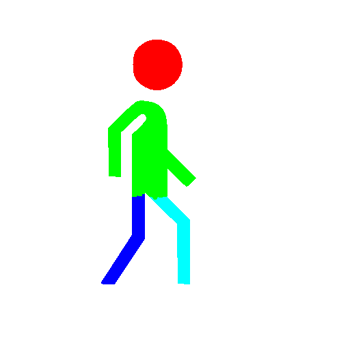} &
        \resultimg{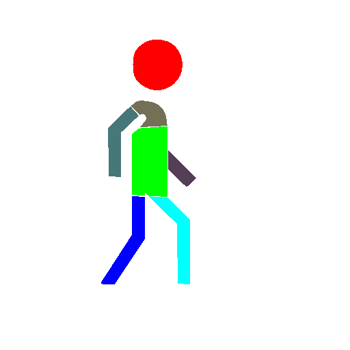} &
        \resultimg{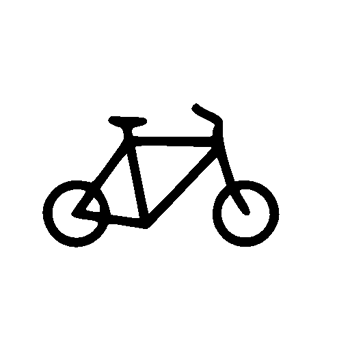} &
        \resultimg{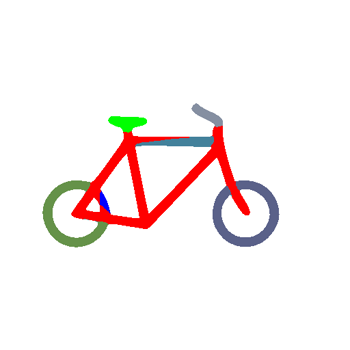} &
        \resultimg{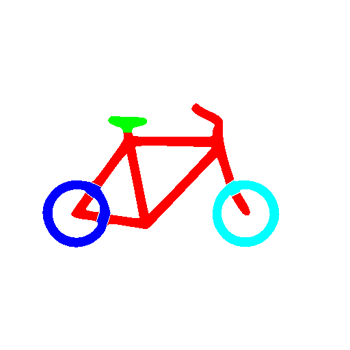} 
        \\
        \resultimg{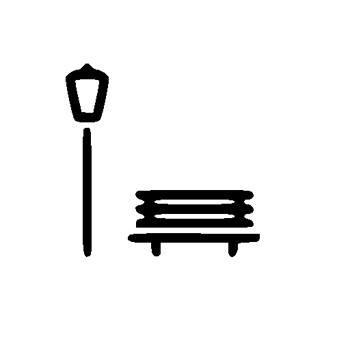} &
        \resultimg{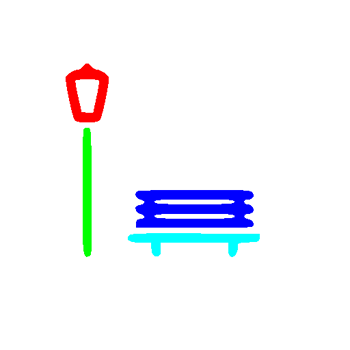} &
        \resultimg{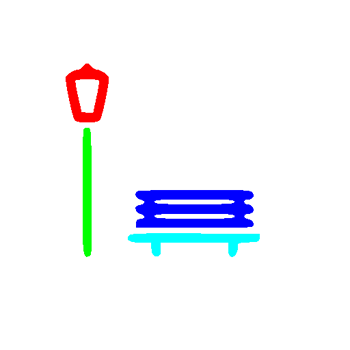} &
        \resultimg{figs/results/original_white/11.png} &
        \resultimg{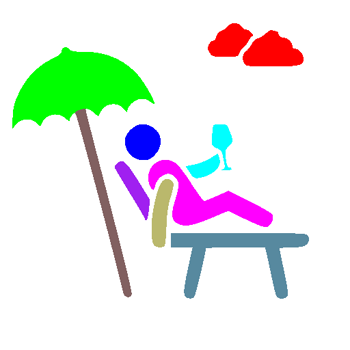} &
        \resultimg{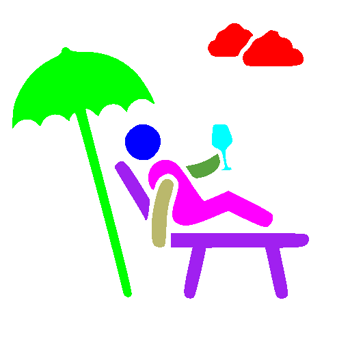} 
        \\
        \resultimg{figs/results/original_white/12.png} &
        \resultimg{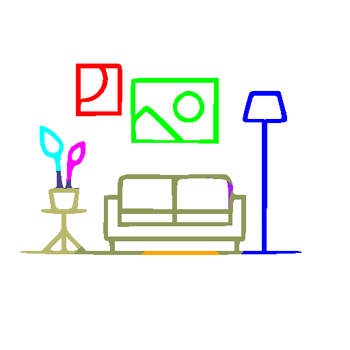} &
        \resultimg{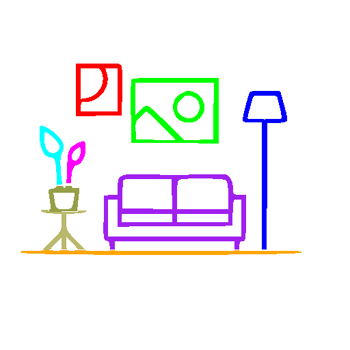} &
        \resultimg{figs/results/original_white/13.png} &
        \resultimg{figs/results/seg_pred/13.png} &
        \resultimg{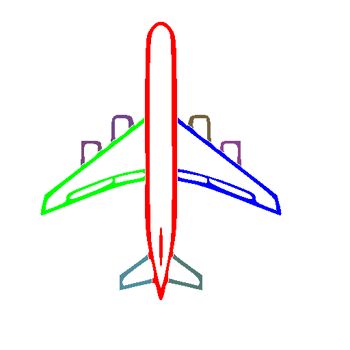} 
        \\
        \resultimg{figs/results/original_white/16.png} &
        \resultimg{figs/results/seg_pred/16.png} &
        \resultimg{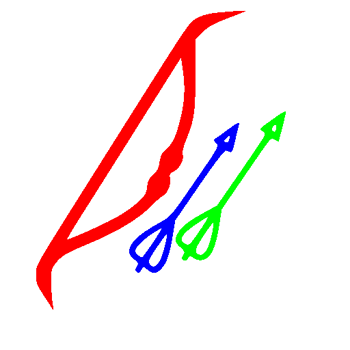} &
        \resultimg{figs/results/original_white/18.png} &
        \resultimg{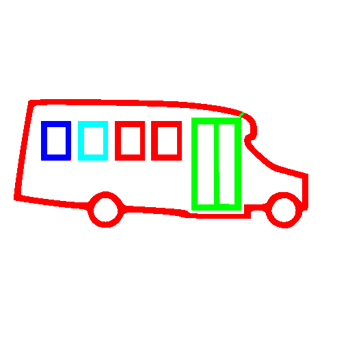} &
        \resultimg{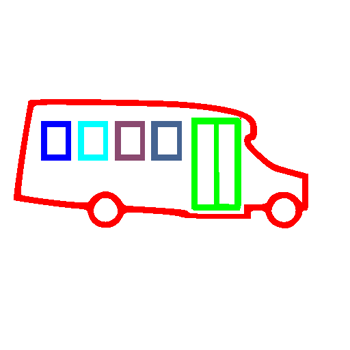} 
        \\
        \resultimg{figs/results/original_white/19.png} &
        \resultimg{figs/results/seg_pred/19.png} &
        \resultimg{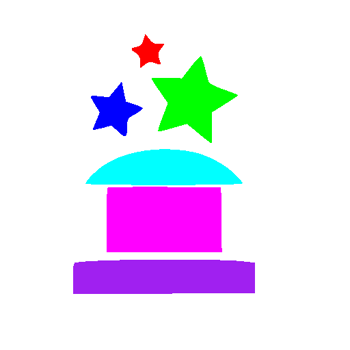} &
        \resultimg{figs/results/original_white/25.png} &
        \resultimg{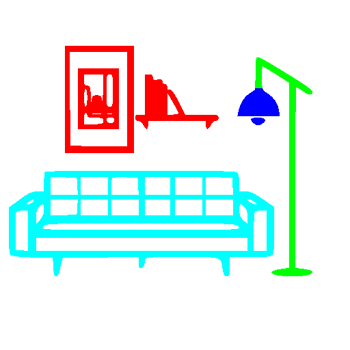} &
        \resultimg{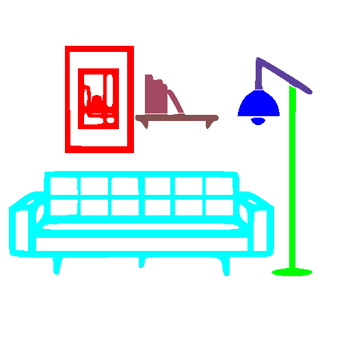} 
        \\
    \end{tabular}
    \caption{Predicted Segmentation(Seg-Tool) vs. GT Segmentation}
    \label{fig:comp_seg_0}
\end{figure}

\clearpage

\section{Supplementary Materials for Creating language-driven spatial variations of icon images}

\subsection{Occlusion Completion}
\begin{figure}[htb!]
    \centering
    \includegraphics[width=0.9\linewidth]{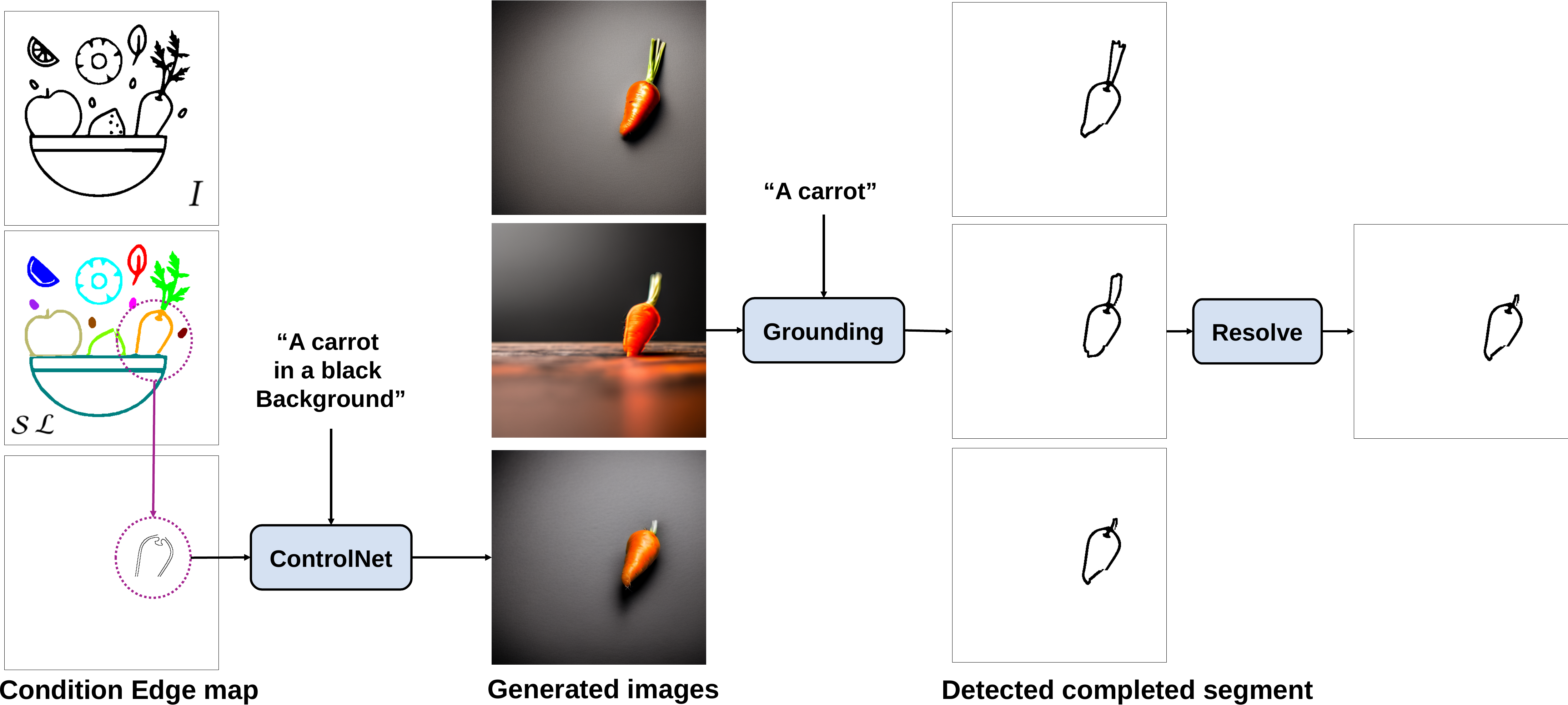}
    \caption{The overview of segment completion pipeline, the edge map of the occluded segment $S_i$ is feed it into ControlNet to generate guessed shape of segment $S_i$. A pretrained grounding model is then used to extract the shape.}
    \label{fig:complete}
\end{figure}

Given the initial arrange of the segments in the scene, some segments $S_i$ may be occluded by other segment $S_j$, as the consequence of this occlusion, the geometry information of the occluded region in segment $S_i$ is lost. During the editing, if the front segment $S_j$ is moved to a different location, the occluded region in segment $S_i$ will become visible. In such case, user may want the missing geometry to be filled for the naturalness of the scene. In order to do so, we designed a segment completion method and allow user to choose if some guessed geometry should be generated and appended to the occluded segments. The method start by extracting the boundary edge map of the occluded segment $S_i$ and feed it into a pretrained ControlNet ~\cite{zhang2023adding} together with its semantic label $l_i$ to generate multiple images of the guessed complete version of segment $S_i$. Then a pre-trained label detection model such as GroundedSam or others ~\cite{ren2024grounded, kirillov2023segany, liu2023grounding, li2021grounded, li2023semantic} is used to extract the shape of the complete version of $S_i$, a voting procedure is then applied to choose the shape that is closet to the mean geometry of all guessed shapes as the final complete shape of $S_i$. Please see fig \ref{fig:complete} for the illustration of the method. 

\subsection{More details for optimization}
Please see algo \ref{alg:optim} for more details of the optimization process. Two state matrices $\mathbf{M}_M$ for motion states and $\mathbf{M}_R$ for relation states are initialized according to the LLM's primary constraint specification. In the relation search stage, the front edges are firstly identified, then the combinations of the front edges are iterated, for each combination, the FlipRelation() flip the relations states as $\mathbf{M}^{(1)}_R$ : The inter-object relations are set from Strong(ST) to Weak(WK). $\mathbf{M}^{(2)}_R$ : The inter-object relations are set from Weak to Not Exist. $\mathbf{M}^{(3)}_R$ : The non-crucial intra-object relations are set from Strong(ST) to Not Exist(N). $\mathbf{M}^{(4)}_R$ : The crucial intra-object relations are set from Strong(ST) to Weak(WK). The Solve() is then applied to evaluate the flipped states in order, the score outputs from $Solve()$ is the violation value of all primary constraints and secondary constraints set to the flipped state. Each flipped state inherit the value from the previous state. Optional Early stop can be performed if the score is low enough. If not stopped, the front edges are updated to non-visited neighbor edges and repeat the process. In the motion search stage, the front nodes are firstly identified, then for each segment in the front nodes, the FlipMotion() flips the motion states as : $\mathbf{M}^{(1)}_M$ : the motion type is set to Translate+Rotate+Scale(TRS), $\mathbf{M}^{(2)}_M$ : the motion type is set to Translate+Rotate(TR), $\mathbf{M}^{(3)}_M$ : the motion type is set to Translate+Scale(TS), $\mathbf{M}^{(4)}_M$ : the motion type is set to Translate(T), $\mathbf{M}^{(5)}_M$ : the motion type is set to Not Exist(N). If not stopped, the front nodes are updated to non-visited neighbor nodes and repeat the process. Once both search stages end, the output optimized segments are the final edited segments.

Under the hood of $Solve()$, the constraints input to the solver will be firstly converted into a constraint tree, with leaf nodes as segments or constant values, with the non-leaf nodes as operations that act on one or multiple segments or values (its children nodes). The root node always corresponds to one of the constraint specifier, this is to ensure the final output value of every constraint indicates how much this constraint is violated. During the optimization, the solver will perform a post-order traversal of the tree to compute the final constraint violation value as a loss value. We use Adam optimizer with a gradually decreasing(starting from lr=10) learning rate for optimizing the affine transformation variables. Each optimization run takes maximum 150 iterations. Additionally, we can choose to create a proxy segment $S^{'}_i$ for each segment $S_i$ to run the optimization, the proxy segment is constructed by a ray-shooting process, for each pixel, we shoot K 2D rays(uniformly spaced around the circle centered on the pixel) outward from the pixel, if a majority of these rays(ratio>=0.9) hit the original segment, the pixel is marked as part of the proxy segment. With this process, we obtained a set of pixels used to create the proxy segment. One benefit of doing so is to make the inside check of segments be consistent with our DSL (so that anything inside the inner holes of an segment is still considered as inside of the element).  

\begin{algorithm}
\small
\DontPrintSemicolon
\SetKwInOut{Input}{Input}
\SetKwInOut{Output}{Output}
\SetKwProg{Fn}{Function}{:}{}
\SetKwFunction{FOptim}{Optim}
\Fn{\FOptim{$\mathcal{S}, G, \mathcal{C}^1, \mathcal{C}^2$}}{
$\mathbf{M}_M \gets Init()$\;
$\mathbf{M}_R \gets Init()$\;
$score^{opt} \gets Init()$\;
$\mathcal{S}^{opt} \gets \mathcal{S}$\;
\;
$\mathcal{R}_{f} \gets getFrontEdges(\mathcal{C}^1, G)$\;
\While {$Len(\mathcal{R}_{f}) > 0$} {
    $\mathcal{U} \gets getCombs(\mathcal{R}_{f})$\;
    \For {$U_i \in \mathcal{U}$}{
        $\mathbf{M}^{(1)}_{R}, \mathbf{M}^{(2)}_{R}, \mathbf{M}^{(3)}_{R}, \mathbf{M}^{(4)}_{R} \gets FlipRelation(U_i, \mathbf{M}_R)$\;
        \For {$\mathbf{M}^{'}_R \in \bigcup{(\mathbf{M}^{(1)}_{R}, \mathbf{M}^{(2)}_{R}, \mathbf{M}^{(3)}_{R}, \mathbf{M}^{(4)}_{R})}$}{
            $\mathcal{S}^{'}, score \gets Solve(\mathcal{S}, \mathcal{C}^1, \mathcal{C}^2, \mathbf{M}_M, \mathbf{M}^{'}_{R})$\;
            \If {$score < score^{opt}$}{
                $\mathbf{M}_R \gets \mathbf{M}^{'}_R$\;
                $score^{opt} \gets score$\;
                $\mathcal{S}^{opt} \gets \mathcal{S}^{'}$\;
            }
        }
    }
    $CheckEarlyStop(score^{opt})$\;
    $\mathcal{R}_{f} \gets getNeighbors(\mathcal{R}_{f}, G)$\; 
}
$\mathbf{M}^{opt}_{R} \gets \mathbf{M}_{R}$\;
\;

$\mathcal{M}_{f} \gets getFrontNodes(\mathcal{C}^1, G)$\;
\While {$Len(\mathcal{M}_{f}) > 0$} {
    $\mathcal{U} \gets getItems(\mathcal{M}_{f})$\;
    \For {$S_i \in \mathcal{U}$}{
        $\mathbf{M}^{(1)}_{M}, \mathbf{M}^{(2)}_{M}, \mathbf{M}^{(3)}_{M}, \mathbf{M}^{(4)}_{M}, \mathbf{M}^{(5)}_{M} \gets FlipMotion(S_i, \mathbf{M}_M)$\;
        \For {$M^{'}_M \in \bigcup{(\mathbf{M}^{(1)}_M, \mathbf{M}^{(2)}_M, \mathbf{M}^{(3)}_M, \mathbf{M}^{(4)}_M)}$}{
            $\mathcal{S}^{'}, score \gets Solve(\mathcal{S}, \mathcal{C}^1, \mathcal{C}^2, \mathbf{M}^{'}_M, \mathbf{M}^{opt}_{R})$\;
            \If {$score < score^{opt}$}{
                $\mathbf{M}_M \gets \mathbf{M}^{'}_M$\;
                $score^{opt} \gets score$\;
                $\mathcal{S}^{opt} \gets \mathcal{S}^{'}$\;
            }
        }
    }
    $\mathcal{M}_{f} \gets getNeighbors(\mathcal{M}_{f}, G)$\; 
}
\;
\Return $\mathcal{S}^{opt}$\;
}
\caption{Optimization}
\label{alg:optim}
\end{algorithm}

\subsection{Re-render}
\begin{figure}[htb!]
    \centering
    \includegraphics[width=0.9\linewidth]{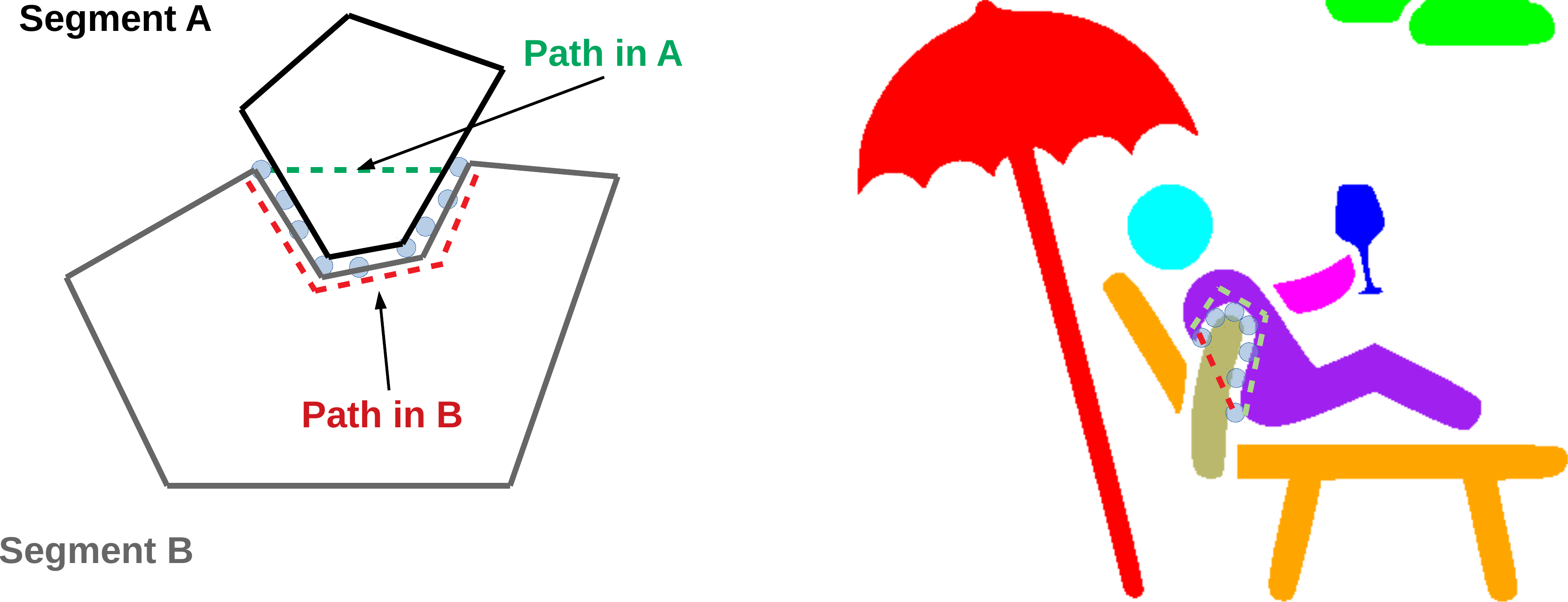}
    \caption{The illustration of our depth order determination method. The segment with a longer inner path is usually perceived as being behind of the other segment}
    \label{fig:depth}
\end{figure}

As described in the main paper, once the transformation for segments are updated. The final step is to re-render them back to an image. The depth ordering of the segments will be crucial for the re-rendering of the edited segments. We developed an algorithm to decide the ordering relation between two segments given their initial spatial configurations. Given two segments $S_i$ and $S_j$, Our method starts by extracting the boundary region $b$ around the intersection region between $S_i$ and $S_j$, then for each segment, we compute the farthest path length $L(b)$ between any two points in the boundary region $b$. The path must be formed inside the segment, for each point pair $(p_s, p_e) \in b$, we run a breath-first-search(BFS) from $p_s$ to $p_e$, the number of search steps will be saved as the path length for $(p_a, p_b)$. the longest path length for segment $S_i$ is indicated as $Length(b_i)$ and the longest path length for segment $S_j$ is indicated as $Length(b_j)$. The shorter path indicates that this object is convex and extruding into the other object, that we usually perceive as this object occluded the other object. This rule is summarized as the following, If $Length(b_i) > Length(b_j)$ : segment i is occluded by segment j, If $Length(b_i) < Length(b_j)$ : segment j is occluded by segment i, If $Length(b_i) \sim Length(b_j)$ : The depth order is not determined. 
Once the depth ordering of segments are determined, we can raster the segments into a color image buffer with the help of a depth image buffer. Please see fig \ref{fig:depth} for illustration.

\subsection{Evaluation for Segmentation Tool}
\begin{figure}[htb!]
    \centering
    \includegraphics[width=0.9\linewidth]{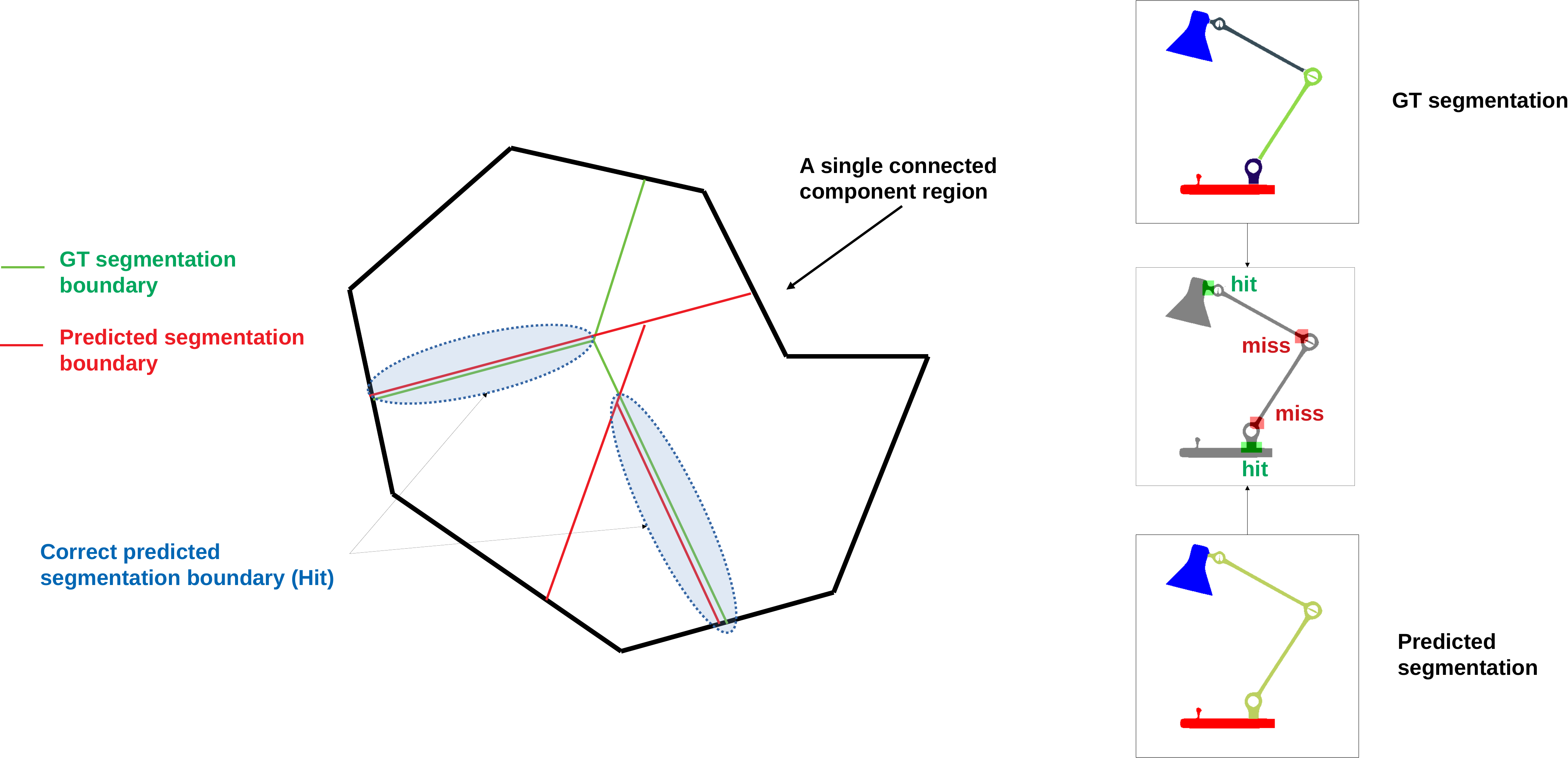}
    \caption{Illustration of segmentation fixing path error computation. if a predicted segmentation boundary is approximately partially aligned with a GT segmentation boundary, the aligned part of the GT boundary is considered as being covered(hit), the path in GT boundary that is not aligned with predicted path is considered as missed(miss) and can be manually fixed later.}
    \label{fig:seg_fix}
\end{figure}

The output labelled segments of the segmentation tool pipeline described in the main paper may not always be well aligned with user's understanding of the image scene. This is caused by multiple sources: (1) The label generated by GPT4Vision may not be accurate, (2) The ControlNet generated image may not be an accurate reflection of the input image scene, (3) The performance of label grounding models such as GroundedSam or GLIP is still far from ideal. As a result, some regions are over-split while others are under-split, and the assigned labels may also need to be adjusted. Based on these observations, we designed some metrics to measure how much our segmentation tool can close the gap between raw input and GT segmentation, in other words, how much our tool can help the user to spend less manual effort on labeling and segmentation. We designed two metrics for evaluating the amount of effort that user may need to manually spent based on the auto-generated labelled segments from our tool. We assume that a simple segmentation and label fixing tool would allow users to do two operations: (1) Drawing a path in the image indicating the boundary path of a segment, (2) Changing an incorrect label assigned to one segment into a different label. As you can see, in the case of under-split, user can draw additional paths on a segment to reach the desired split. In the case of over-split, user can changing the labels of adjacent segments so that they are merged as one segment. In the case of incorrect label, only changing the label is enough. As the consequences of these two fixing operations, two evaluation metrics are naturally emerged: (1) The length of the additional boundary paths drawn on top of the predicted boundary paths. (2) The number of label correcting operation. Please see fig \ref{fig:seg_fix} for the illustration of these metrics. For metric (1), we compute it by measuring how many GT segment boundary pixels are not covered by predicted segment boundary pixels, for each GT segment boundary pixel, we search if there is a predicted boundary pixel within a radius(r = 20) of this GT pixel. For metric (2), we count the number of mismatched labels assigned to predicted segments and GT segments. We also include the numbers of these two metrics computed on the original input (indicates the total manual effort a user need to spent given the original unsegmented image) for comparison. Please see table \ref{tab:seg_fix} for this comparison. As you can see, the path need to be drawn(fix) become much shorter after using our segmentation tool, the label correcting numbers are slightly higher but it is a relative cheap operation. In summary, our segmentation tool can greatly reduce the user's manual effort for creating the labelled segmentation that is required by the editing process.

\begin{figure}[htb!]
    \centering
    \includegraphics[width=0.9\linewidth]{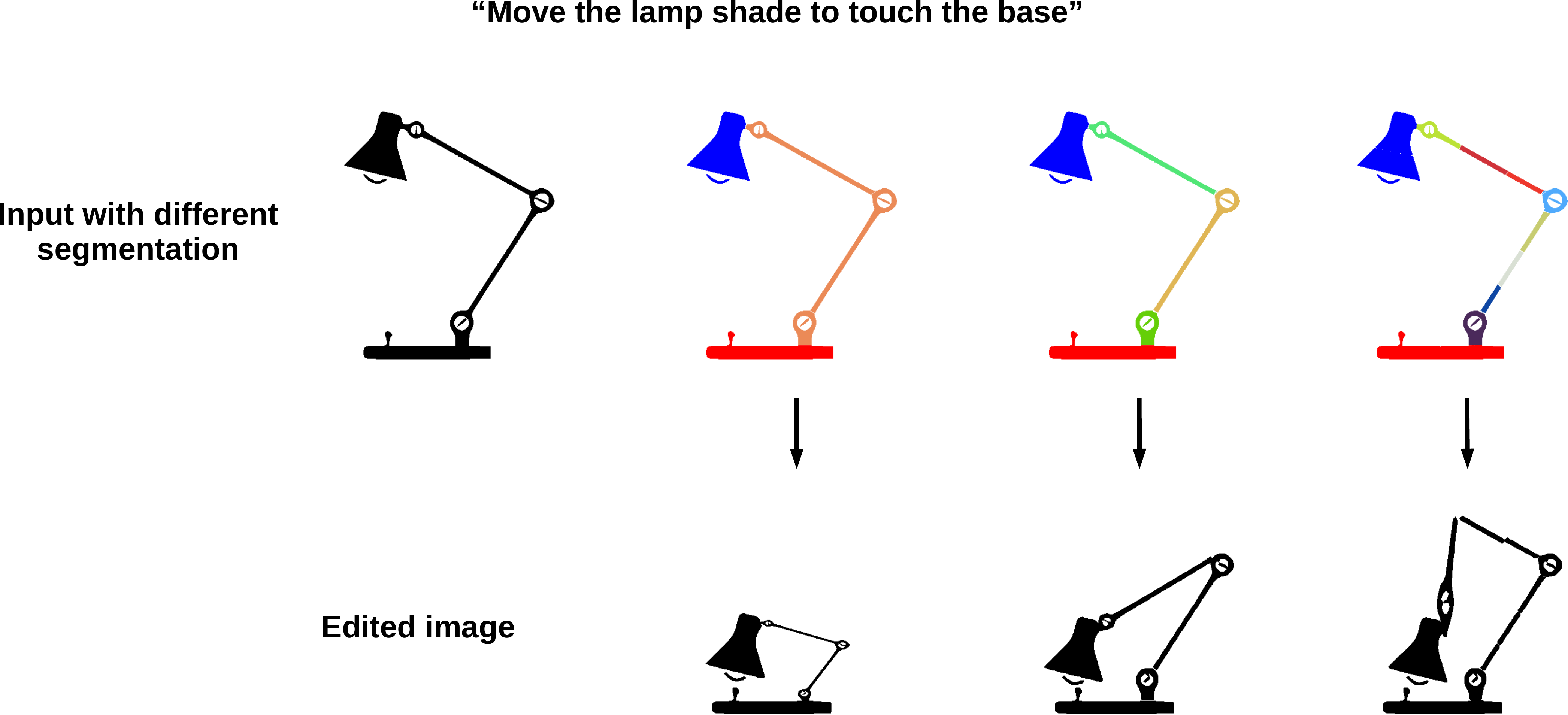}
    \caption{The illustration of how different segmentation resolution can affect the editing results.
    }
    \label{fig:multiseg}
\end{figure}

Another thing to notice is that an edit can be achieved with different segmentation results, please see fig \ref{fig:multiseg} for illustration of the editing results generated by our editing system with different segmentation configuration.

\begin{table}[htb!]
    \centering
    \scriptsize
    \begin{tabular}{lcc}
        \toprule
        & \raisebox{0em}{Manual} & \raisebox{0em}{Ours}
        \\
        \midrule
        Missing path length & 612 & 366
        \\
        Mismatched label count & 6.6 & 8.1
        \\
        \bottomrule
    \end{tabular}
    \caption{
    Segmentation Tool Performance, Manual: The total effort needed for creating the GT labelled segmentation. Ours: The remaining effort needed for creating the GT labelled segmentation.
    }
    \label{tab:seg_fix}
\end{table}

\begin{figure}[htb!]
    \centering
    \includegraphics[width=0.9\linewidth]{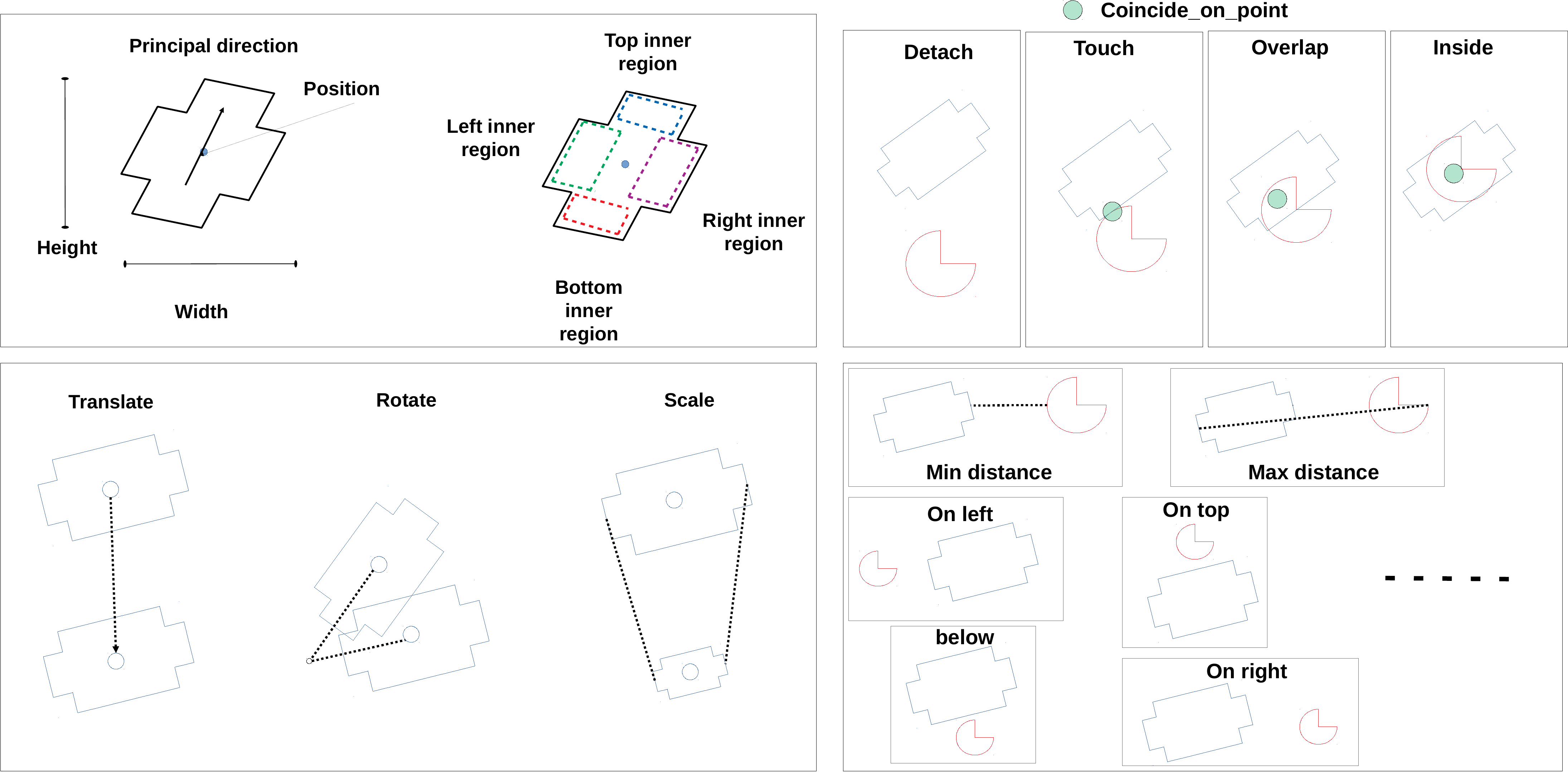}
    \caption{The visual illustration of some operators included in our DSL operator library.}
    \label{fig:dsl}
\end{figure}

\begin{figure}[htb!]
    \centering
    \includegraphics[width=0.9\linewidth]{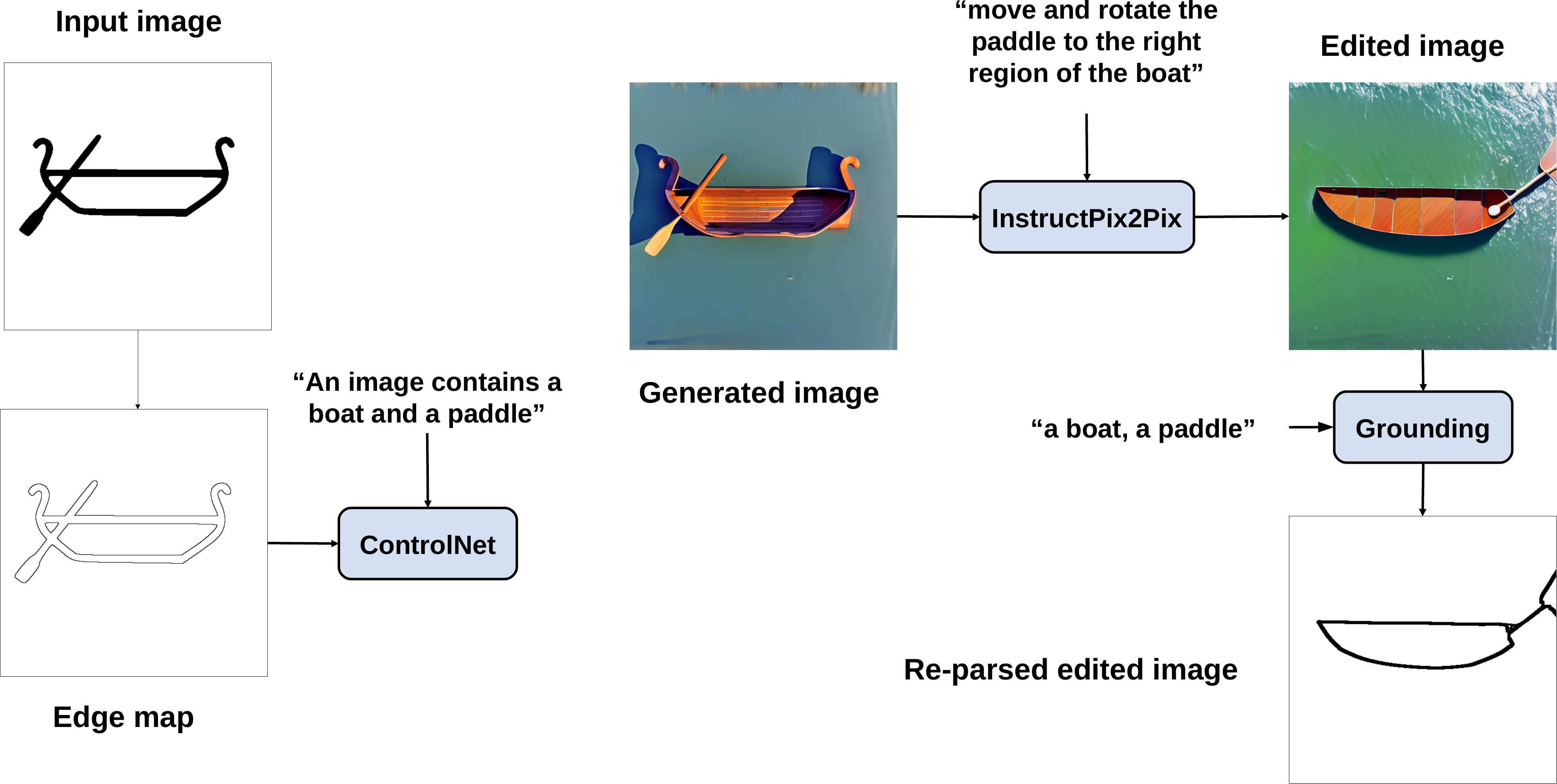}
    \caption{The overview of our lift-and-project process, the edge map together with the labels of the input icon image is feed into a pretrained ControlNet to generate a more realistic image, then the diffusion based editing methods such as StableDiffusion Image2Image, InstructPix2Pix and Imagic will be applied to edit this generated image given the editing request. The edited image is then projected back to the icon image domain by using a label grounding model to detect relevant labels.}
    \label{fig:lift}
\end{figure}

\subsection{Additional Results}

\begin{figure}
    \centering
    \small
    \setlength{\tabcolsep}{0pt}
    \renewcommand{\arraystretch}{0}
    \newcommand{\resultimg}[1]{\includegraphics[trim={0pt 0pt 0pt 0pt},clip,width=0.16\linewidth]{#1}}
    \begin{tabular}{c|c|c|c|c|c}
        \raisebox{0em}{Original} & \raisebox{0em}{gpt4-dm} & \raisebox{0em}{gpt4v-dm} & \raisebox{0em}{gpt4-dc} & \raisebox{0em}{gpt4v-dc} & \raisebox{0em}{Ours}\\
        \midrule
        \multicolumn{6}{l}{\parbox{7cm}{\textbf{"Rotate the left arm to be on the same altitude as the shoulder"}}} 
        \\
        \resultimg{figs/results/original_white/7.png} &
        \resultimg{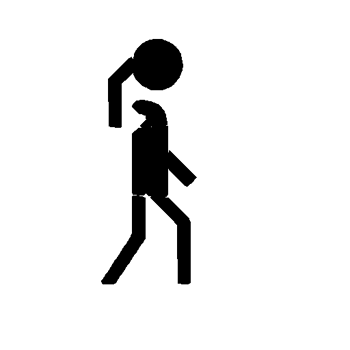} &
        \resultimg{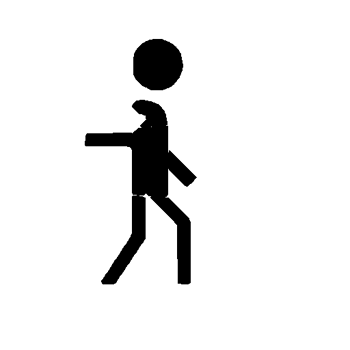} &
        \resultimg{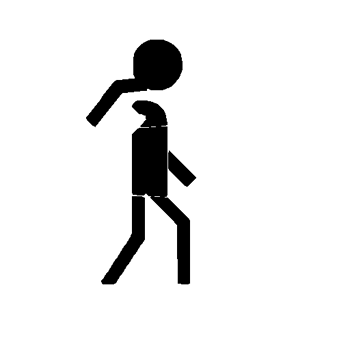} &
        \resultimg{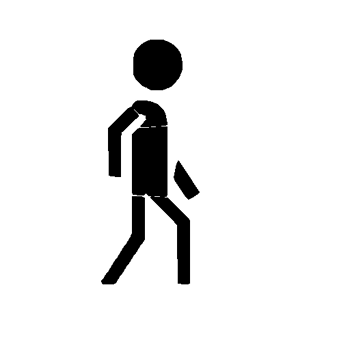}  & 
        \resultimg{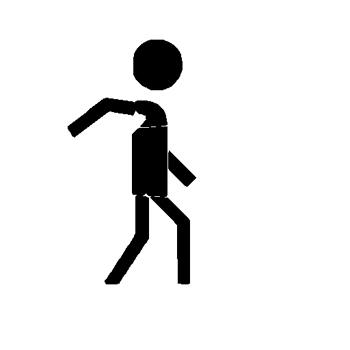}  
        \\
        \midrule
        \multicolumn{6}{l}{\parbox{7cm}{\textbf{"Reduce the vertical length of the fuselage by 50\%"}}}
        \\
        \resultimg{figs/results/original_white/13.png} &
        \resultimg{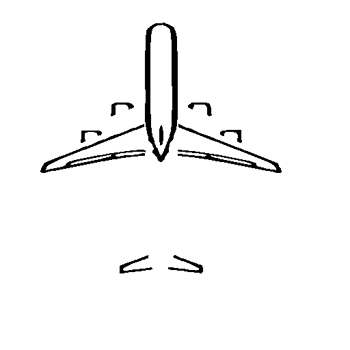} &
        \resultimg{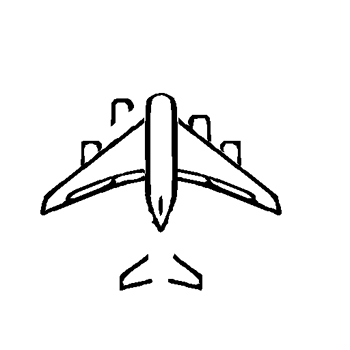} &
        \resultimg{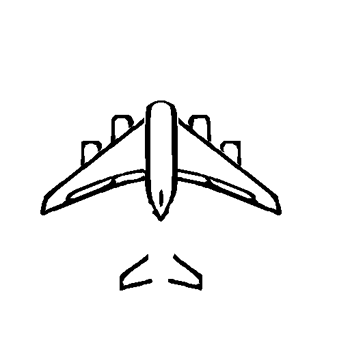} &
        \resultimg{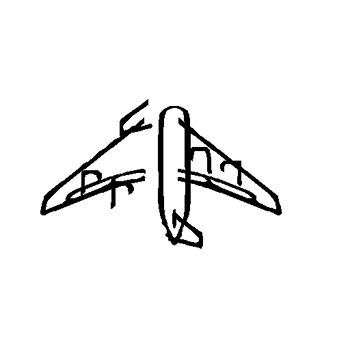}  & 
        \resultimg{figs/results/ours_c1m1/13.png}  
        \\
        \midrule
        \multicolumn{6}{l}{\parbox{7cm}{\textbf{"Reduce the vertical length of the vase by 50\%, move down the flowers to keep touching the vase"}}}
        \\
        \resultimg{figs/results/original_white/68.png} &
        \resultimg{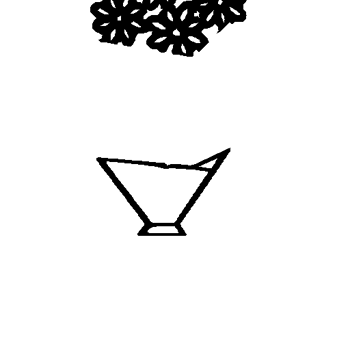} &
        \resultimg{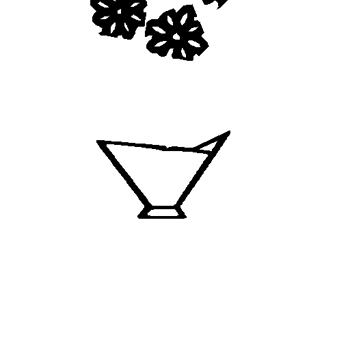} &
        \resultimg{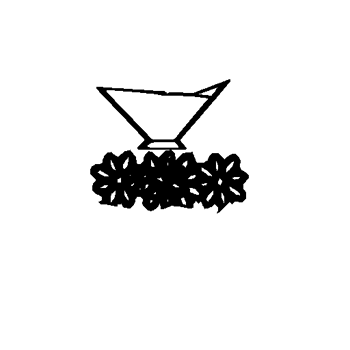} &
        \resultimg{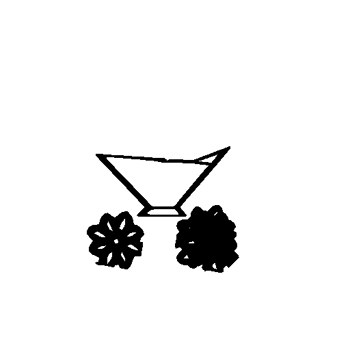}  & 
        \resultimg{figs/results/ours_c1m1/68.png}
        \\
        \midrule
        \multicolumn{6}{l}{\parbox{7cm}{\textbf{"Swap the position of the lollipop and the upper candy"}}}
        \\
        \resultimg{figs/results/original_white/14.png} &
        \resultimg{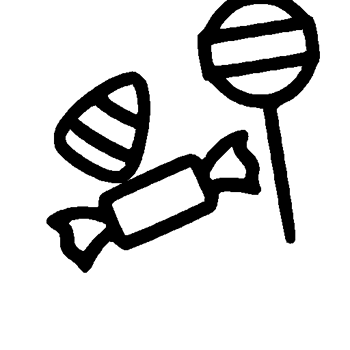} &
        \resultimg{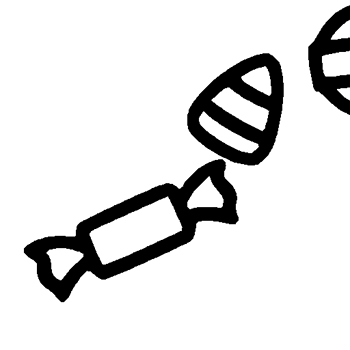} &
        \resultimg{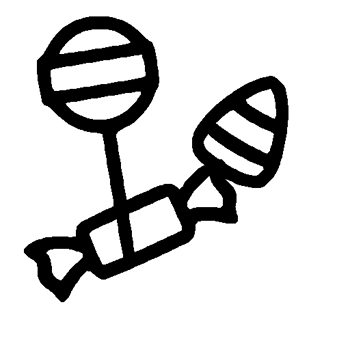} &
        \resultimg{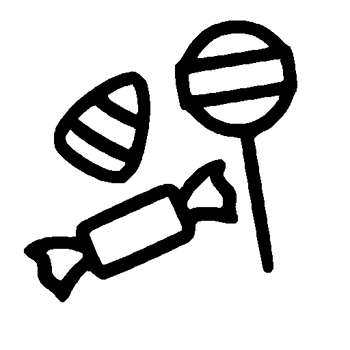}  & 
        \resultimg{figs/results/ours_c1m1/14.png}  
        \\
        \midrule
        \multicolumn{6}{l}{\parbox{7cm}{\textbf{"Move the shade down to touch the sofa by shortening the lamp support"}}}
        \\
        \resultimg{figs/results/original_white/25.png} &
        \resultimg{figs/results/direct_motion_gpt4/25.png} &
        \resultimg{figs/results/direct_motion_gpt4v/25.png} &
        \resultimg{figs/results/direct_constraint_gpt4/25.png} &
        \resultimg{figs/results/direct_constraint_gpt4v/25.png}  & 
        \resultimg{figs/results/ours_c1m1/25.png}  
        \\
        \midrule
        \multicolumn{6}{l}{\parbox{7cm}{\textbf{"Reduce the horizontal length of the bus by 50\%"}}}
        \\
        \resultimg{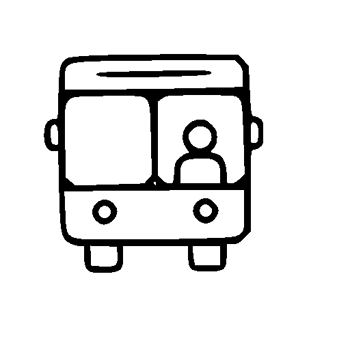} &
        \resultimg{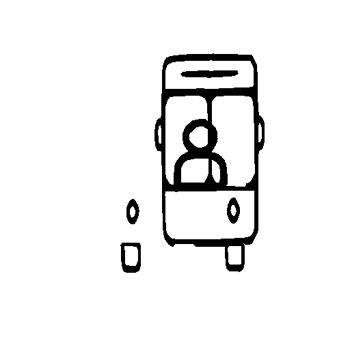} &
        \resultimg{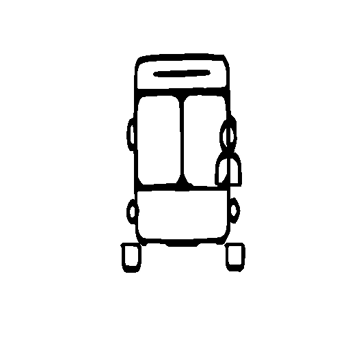} &
        \resultimg{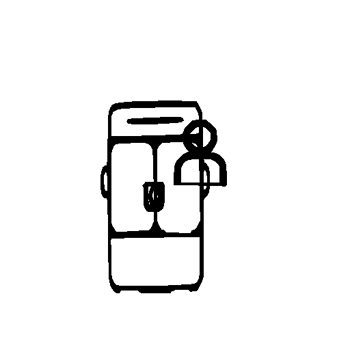} &
        \resultimg{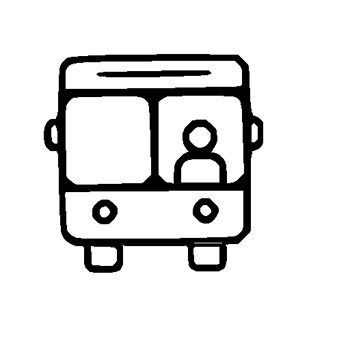}  & 
        \resultimg{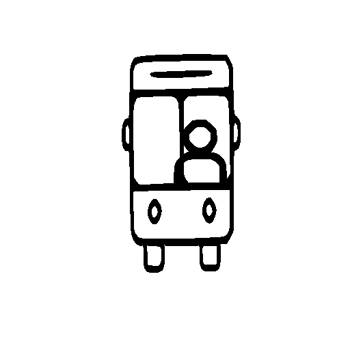}  
        \\
        \midrule
        \multicolumn{6}{l}{\parbox{7cm}{\textbf{"Increase the horizontal length of the stool table by 50\%"}}}
        \\
        \resultimg{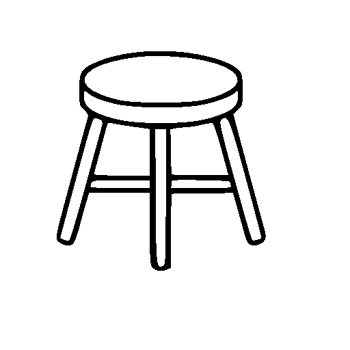} &
        \resultimg{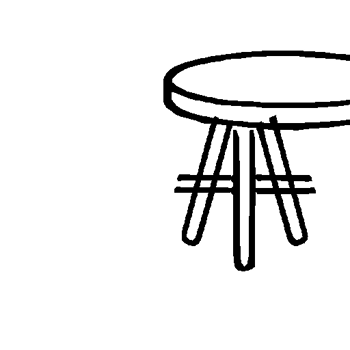} &
        \resultimg{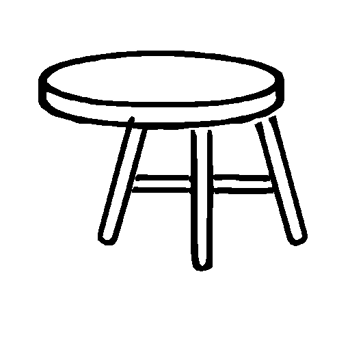} &
        \resultimg{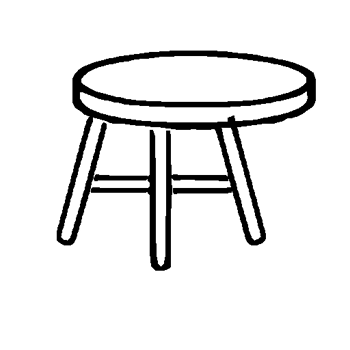} &
        \resultimg{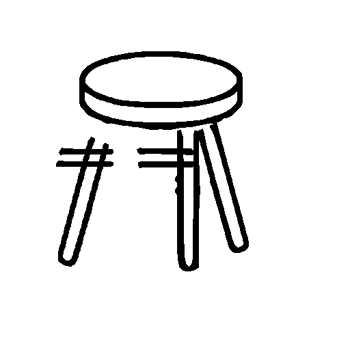}  & 
        \resultimg{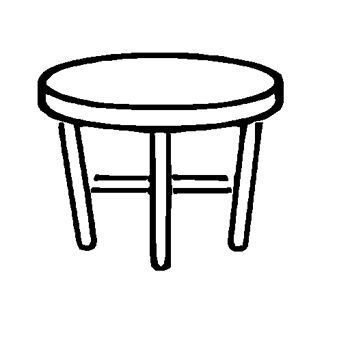}  
        \\
        \midrule
        \multicolumn{6}{l}{\parbox{7cm}{\textbf{"Halve the horizontal length of the umbrella head, halve the vertical length of the umbrella support"}}}
        \\
        \resultimg{figs/results/original_white/36.png} &
        \resultimg{figs/results/direct_motion_gpt4/36.png} &
        \resultimg{figs/results/direct_motion_gpt4v/36.png} &
        \resultimg{figs/results/direct_constraint_gpt4/36.png} &
        \resultimg{figs/results/direct_constraint_gpt4v/36.png}  & 
        \resultimg{figs/results/ours_c1m1/36.png}  
        \\
        \midrule
        \multicolumn{6}{l}{\parbox{7cm}{\textbf{"Swap the position of the door and the window"}}}
        \\
        \resultimg{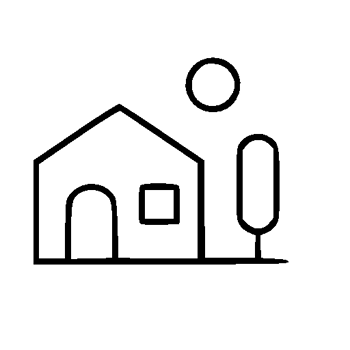} &
        \resultimg{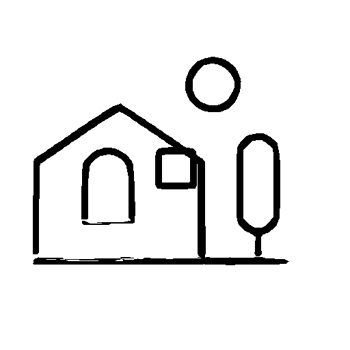} &
        \resultimg{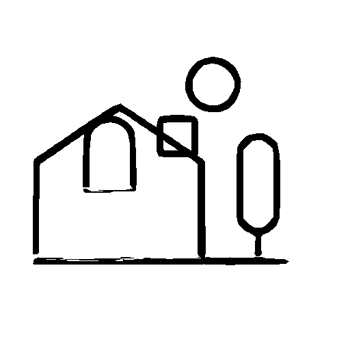} &
        \resultimg{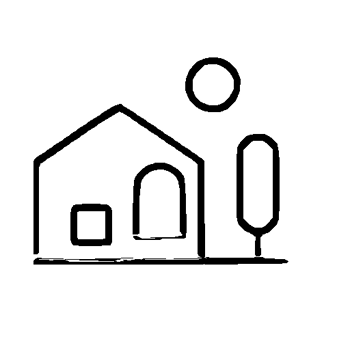} &
        \resultimg{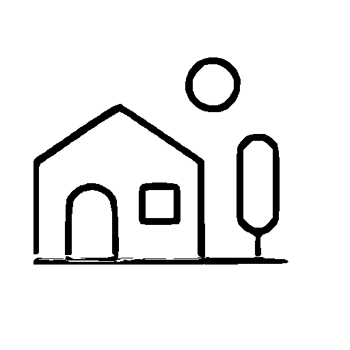}  & 
        \resultimg{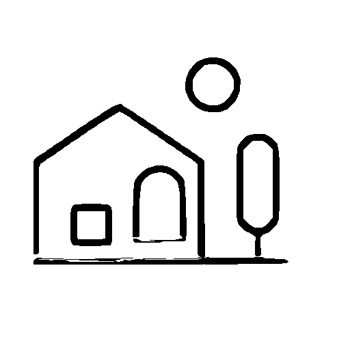}  
        \\
        \midrule
        \multicolumn{6}{l}{\parbox{7cm}{\textbf{"Move the middle star so that three stars form an equilateral triangle"}}}
        \\
        \resultimg{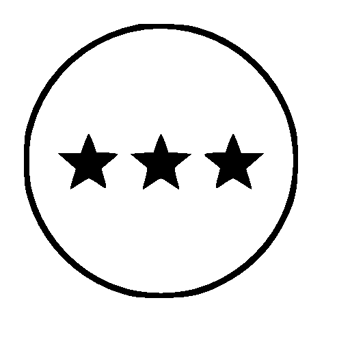} &
        \resultimg{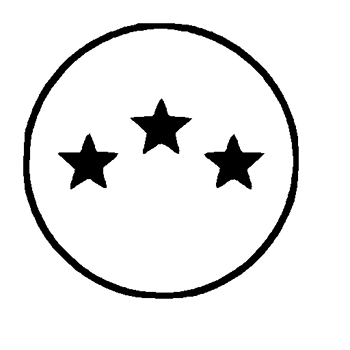} &
        \resultimg{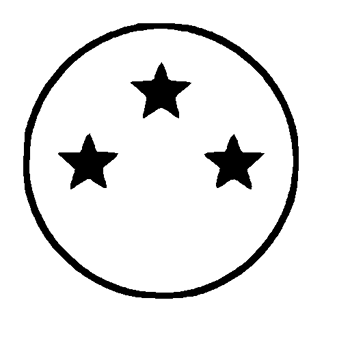} &
        \resultimg{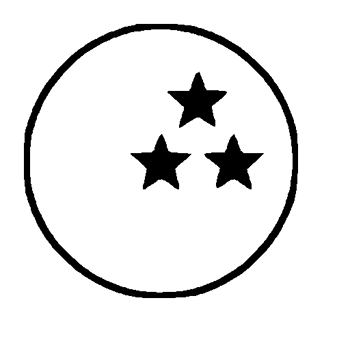} &
        \resultimg{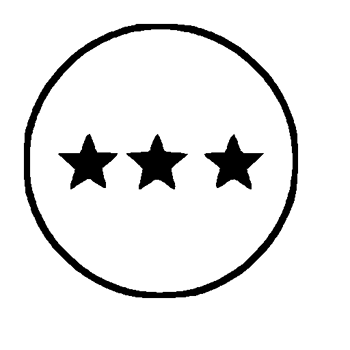}  & 
        \resultimg{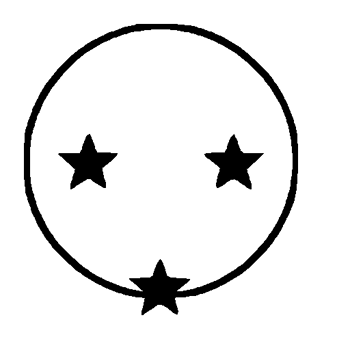}  
        \\
        \midrule
        \multicolumn{6}{l}{\parbox{7cm}{\textbf{"Move the left chair to be on the right side of the right chair"}}}
        \\
        \resultimg{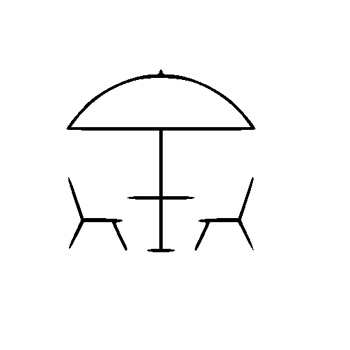} &
        \resultimg{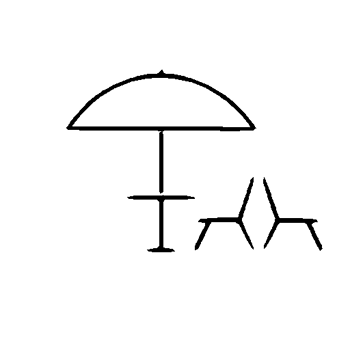} &
        \resultimg{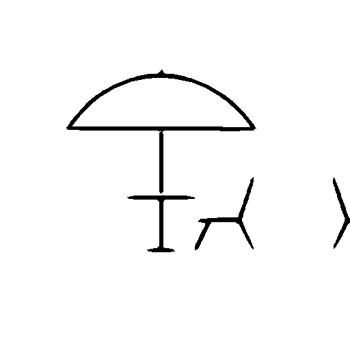} &
        \resultimg{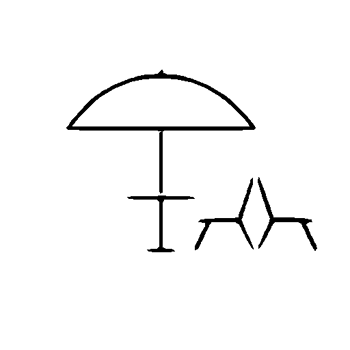} &
        \resultimg{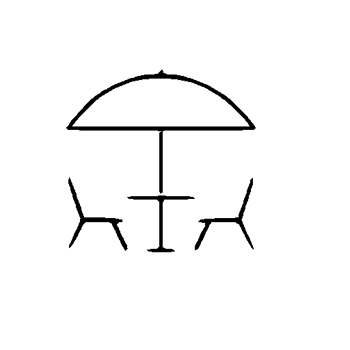}  & 
        \resultimg{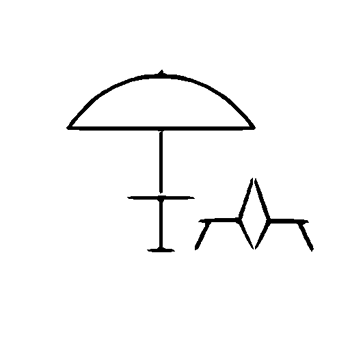}  
        \\
    \end{tabular}
    \caption{Additional LLM parametric editing comparison results}
    \label{fig:comp_gpt_2}
\end{figure}

\begin{figure}[htb!]
    \centering
    \small
    \setlength{\tabcolsep}{0pt}
    \renewcommand{\arraystretch}{0}
    \newcommand{\resultimg}[1]{\includegraphics[trim={0pt 0pt 0pt 0pt},clip,width=0.12\linewidth]{#1}}
    \begin{tabular}{c|c|c|c|c|c|c|c}
        \raisebox{0em}{Original} & \raisebox{0em}{sdi2i} & \raisebox{0em}{ip2p} & \raisebox{0em}{imagic} & \raisebox{0em}{sdi2iL} & \raisebox{0em}{ip2pL} & \raisebox{0em}{imagicL} & \raisebox{0em}{Ours} 
        \\
        \midrule
        \multicolumn{8}{l}{\parbox{7cm}{\textbf{"Reduce the vertical length of the cup by 50\%"}}}
        \\
        \resultimg{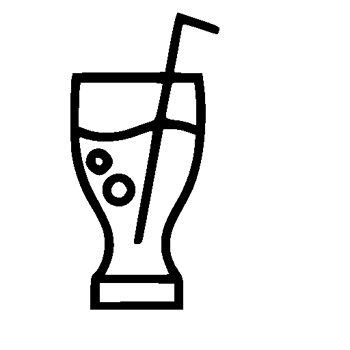} &
        \resultimg{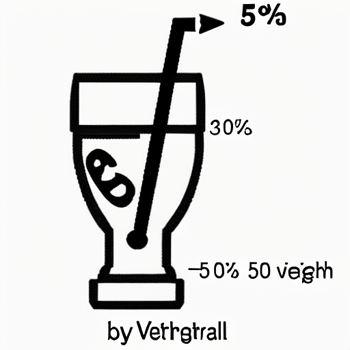} &
        \resultimg{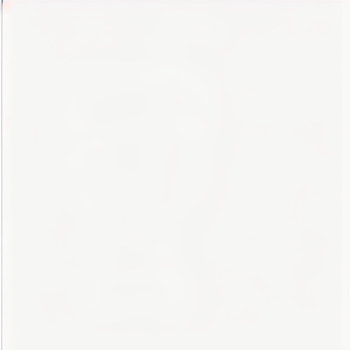} &
        \resultimg{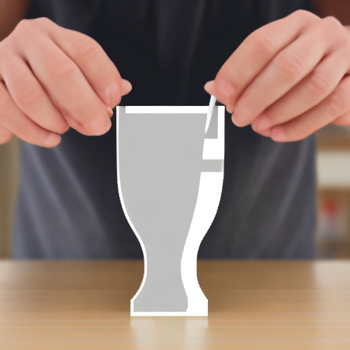} & 
        \resultimg{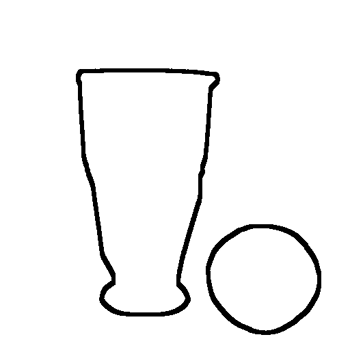} &
        \resultimg{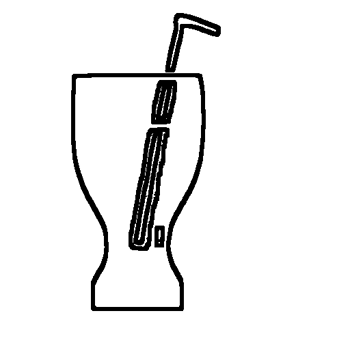} &
        \resultimg{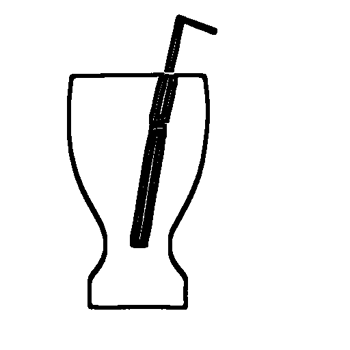} & 
        \resultimg{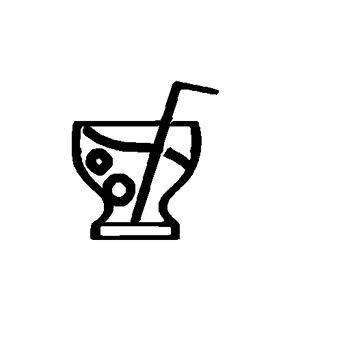}  
        \\
        \midrule
        \multicolumn{8}{l}{\parbox{7cm}{\textbf{"Move the water under the bottle spray"}}}
        \\
        \resultimg{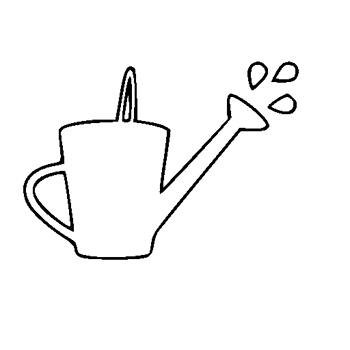} &
        \resultimg{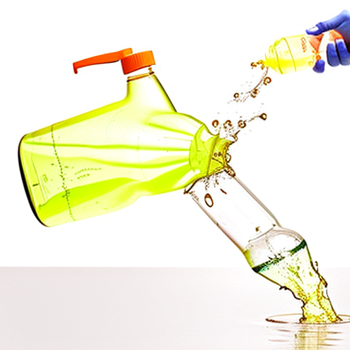} &
        \resultimg{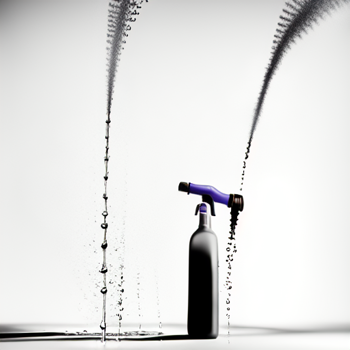} &
        \resultimg{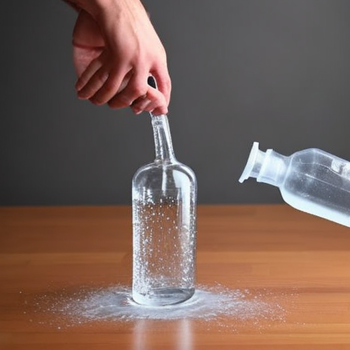} & 
        \resultimg{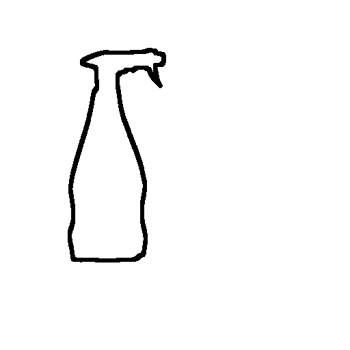} &
        \resultimg{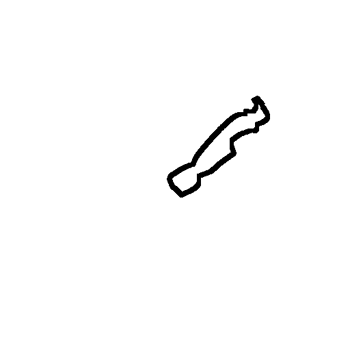} &
        \resultimg{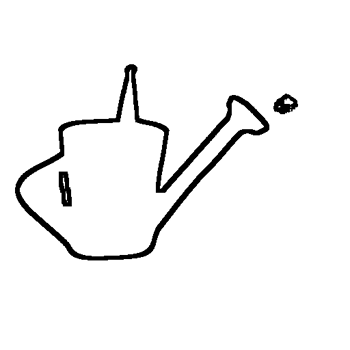} & 
        \resultimg{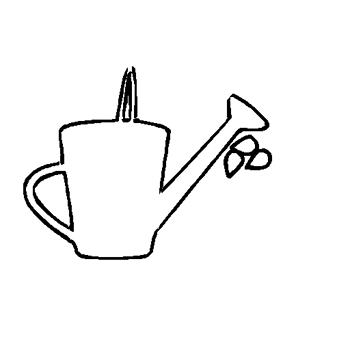}  
        \\
        \midrule
        \multicolumn{8}{l}{\parbox{7cm}{\textbf{"Reduce the vertical length of the sail column by 50\%"}}}
        \\
        \resultimg{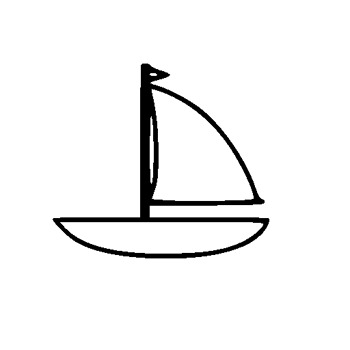} &
        \resultimg{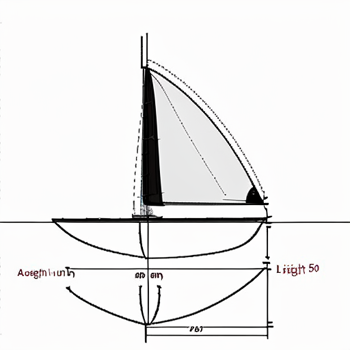} &
        \resultimg{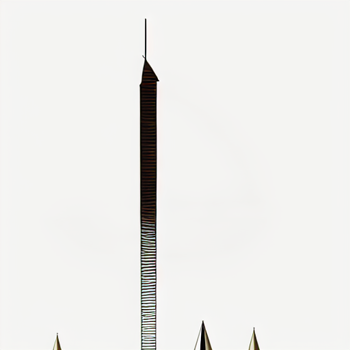} &
        \resultimg{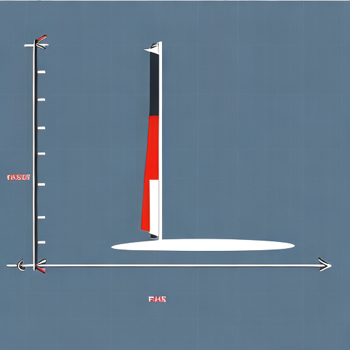} & 
        \resultimg{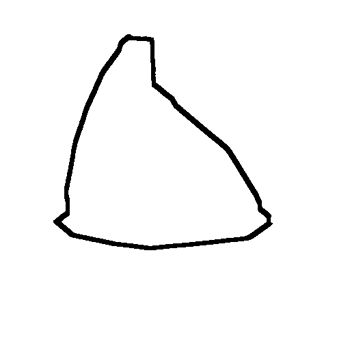} &
        \resultimg{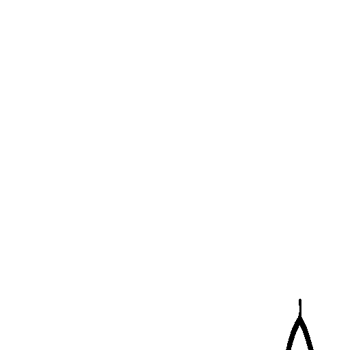} &
        \resultimg{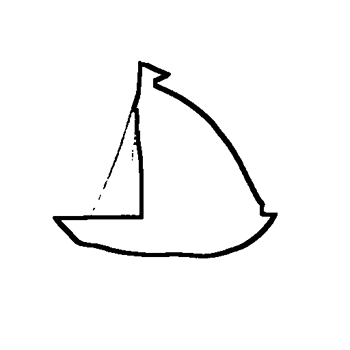} & 
        \resultimg{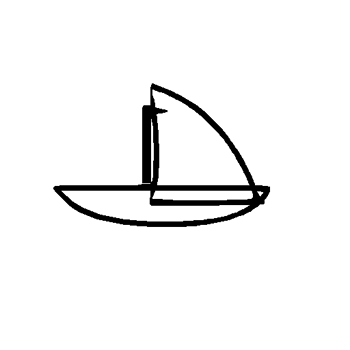}  
        \\
        \midrule
        \multicolumn{8}{l}{\parbox{7cm}{\textbf{"Move the bird, scale the swing, so that the bird is moved down by 25\%, keep the cage fixed"}}}
        \\
        \resultimg{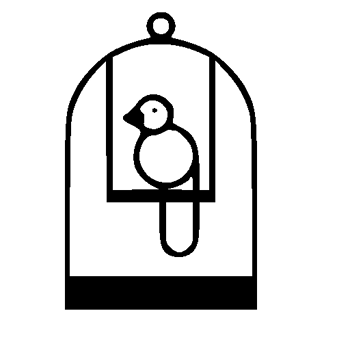} &
        \resultimg{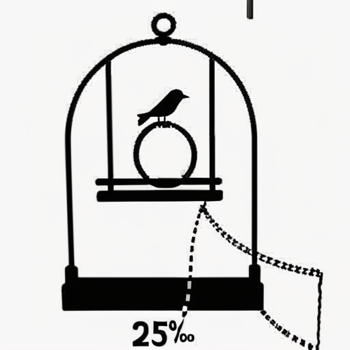} &
        \resultimg{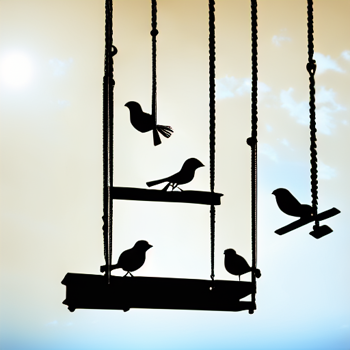} &
        \resultimg{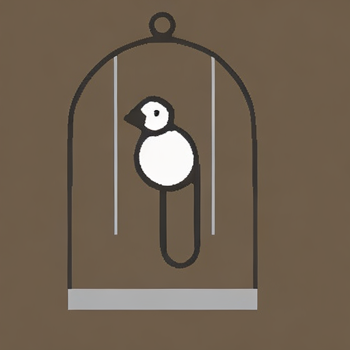} & 
        \resultimg{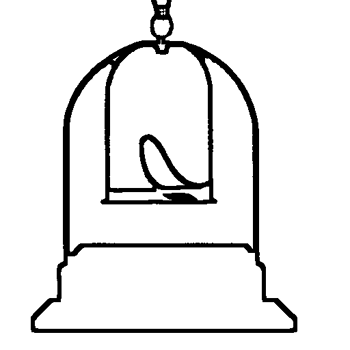} &
        \resultimg{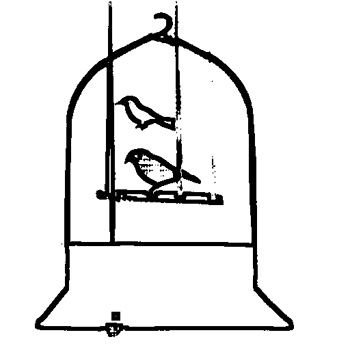} &
        \resultimg{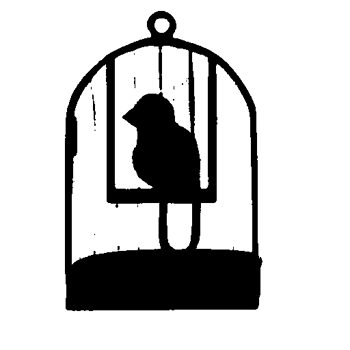} & 
        \resultimg{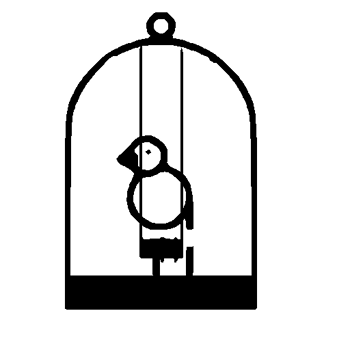}  
        \\
        \midrule
        \multicolumn{8}{l}{\parbox{7cm}{\textbf{"Move the clouds to touch their nearest trees"}}}
        \\
        \resultimg{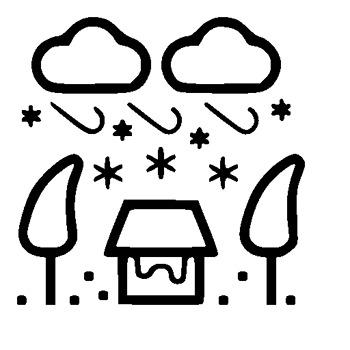} &
        \resultimg{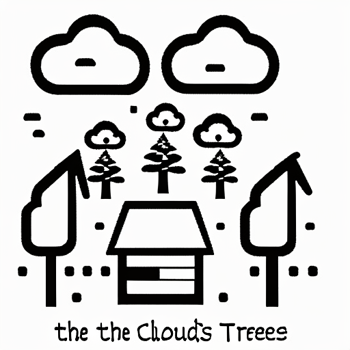} &
        \resultimg{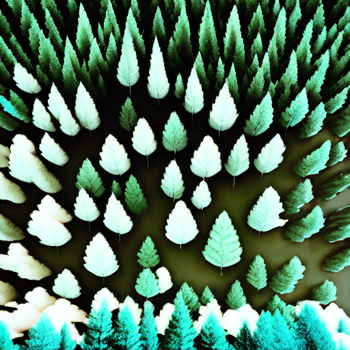} &
        \resultimg{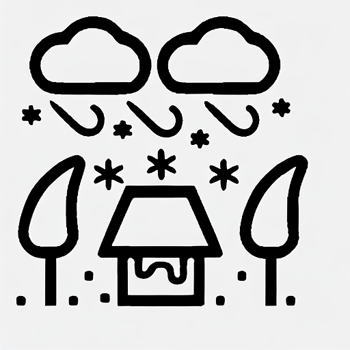} & 
        \resultimg{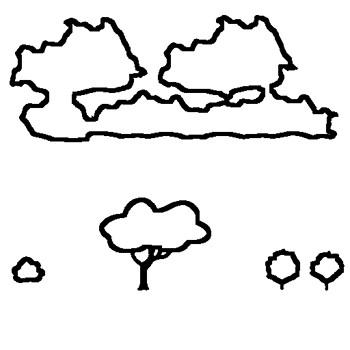} &
        \resultimg{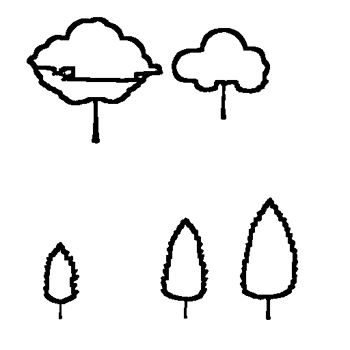} &
        \resultimg{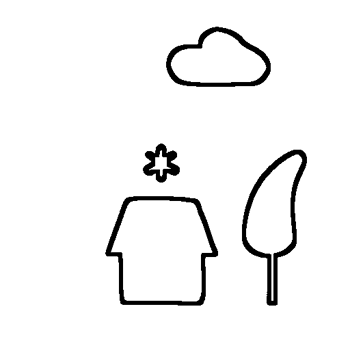} & 
        \resultimg{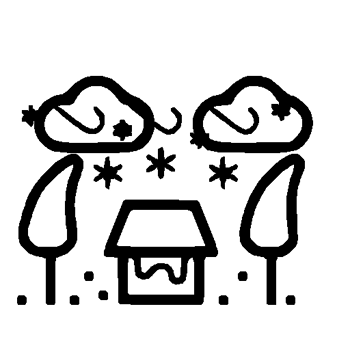}  
        \\
        \midrule
        \multicolumn{8}{l}{\parbox{7cm}{\textbf{"Scale all the grass so that their vertical length is equal to the vertical length of the box support column"}}}
        \\
        \resultimg{figs/results/original_white/53.png} &
        \resultimg{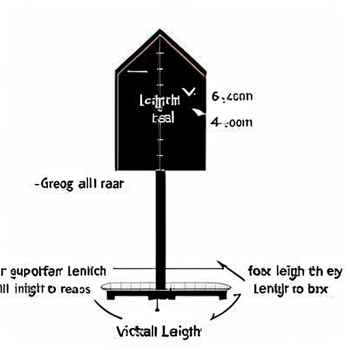} &
        \resultimg{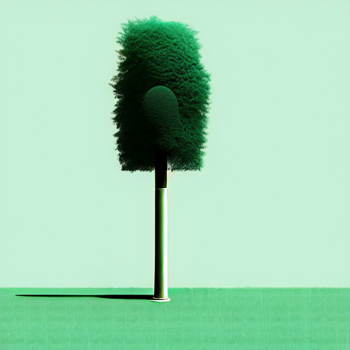} &
        \resultimg{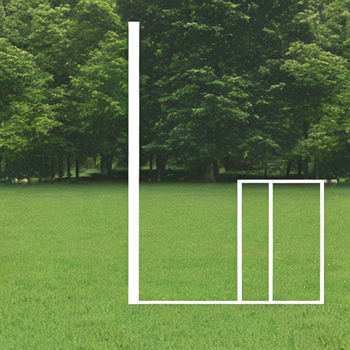} & 
        \resultimg{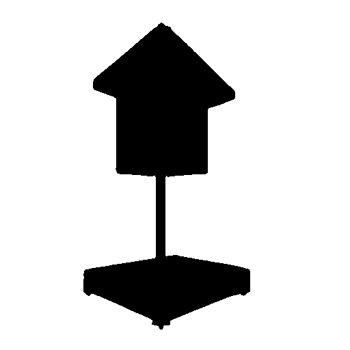} &
        \resultimg{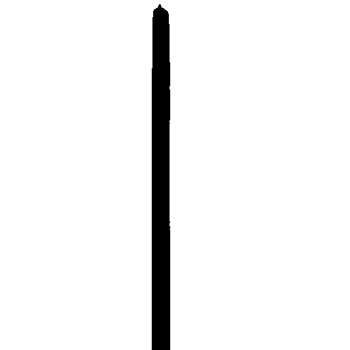} &
        \resultimg{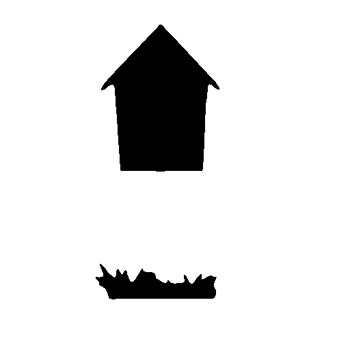} & 
        \resultimg{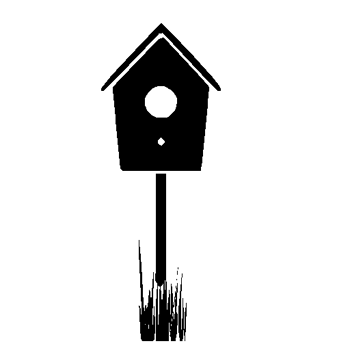}  
        \\
        \midrule
        \multicolumn{8}{l}{\parbox{7cm}{\textbf{"Grow the trees to the same vertical length as the house"}}}
        \\
        \resultimg{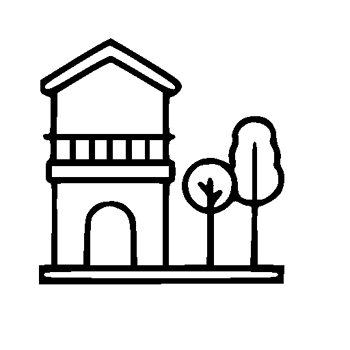} &
        \resultimg{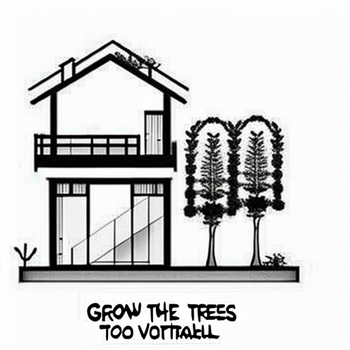} &
        \resultimg{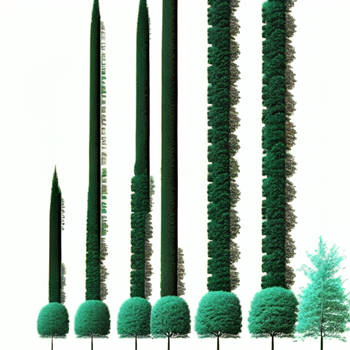} &
        \resultimg{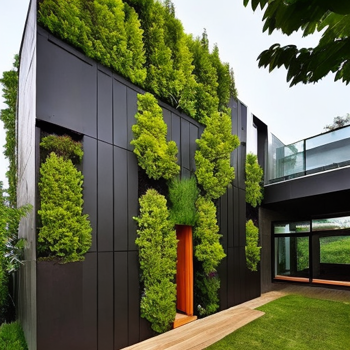} & 
        \resultimg{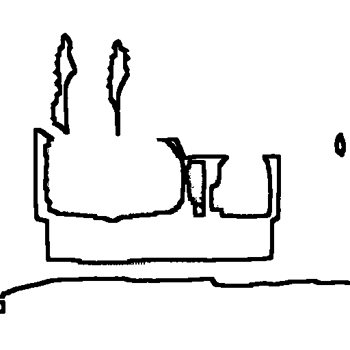} &
        \resultimg{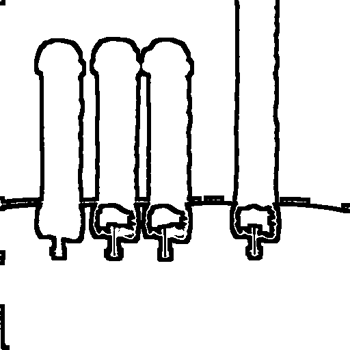} &
        \resultimg{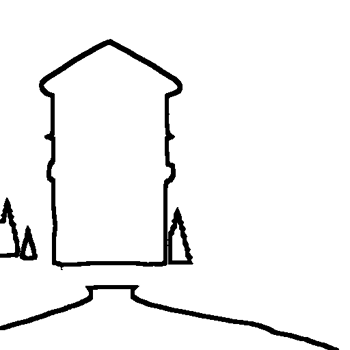} & 
        \resultimg{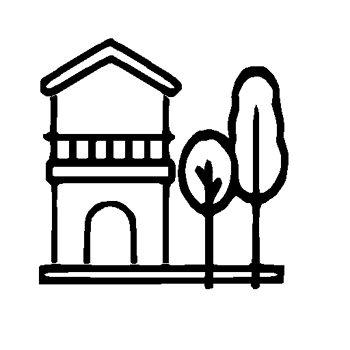}  
        \\
        \midrule
        \multicolumn{8}{l}{\parbox{7cm}{\textbf{"Make the ears 50\% smaller in horizontal length"}}}
        \\
        \resultimg{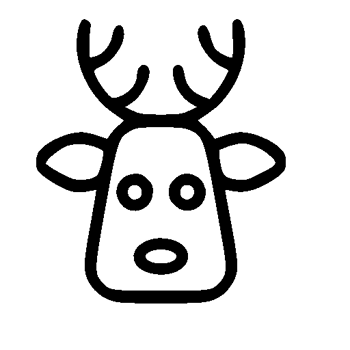} &
        \resultimg{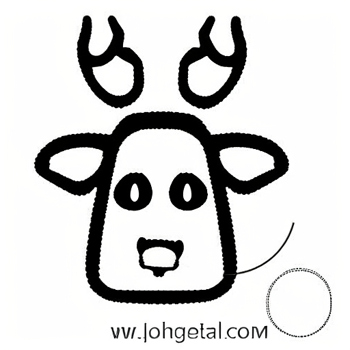} &
        \resultimg{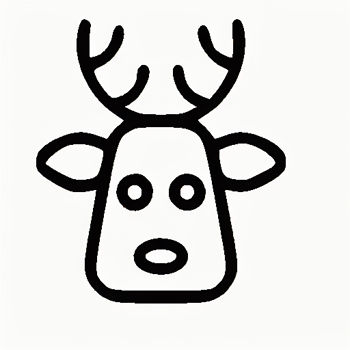} &
        \resultimg{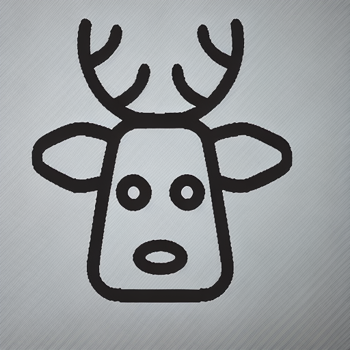} & 
        \resultimg{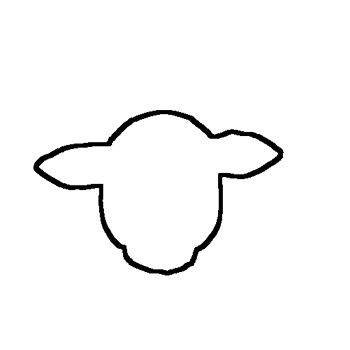} &
        \resultimg{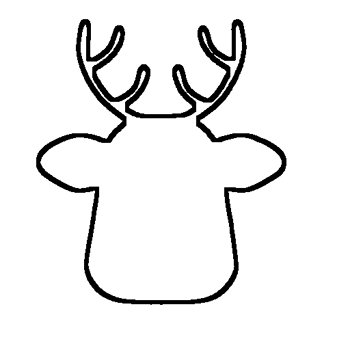} &
        \resultimg{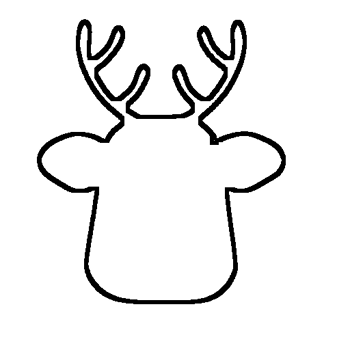} & 
        \resultimg{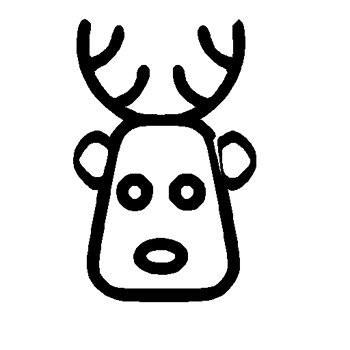}  
        \\
        \midrule
        \multicolumn{8}{l}{\parbox{7cm}{\textbf{"Halve the horizontal length of the roller head"}}}
        \\
        \resultimg{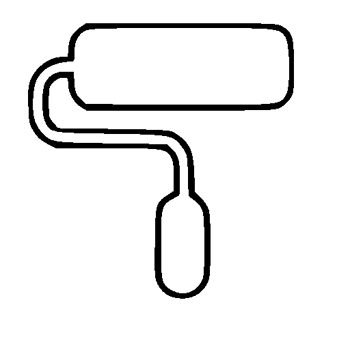} &
        \resultimg{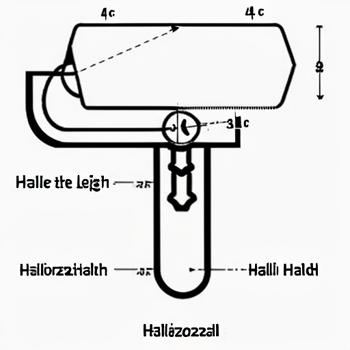} &
        \resultimg{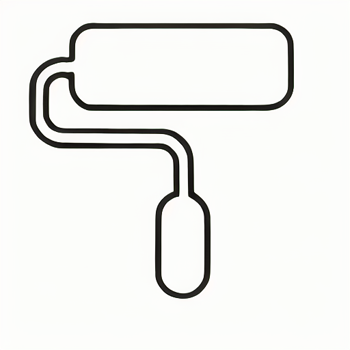} &
        \resultimg{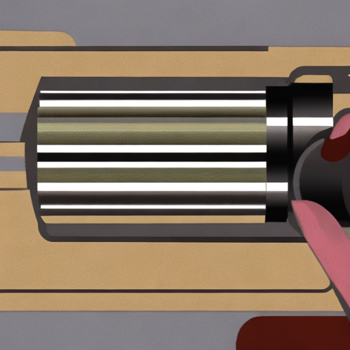} & 
        \resultimg{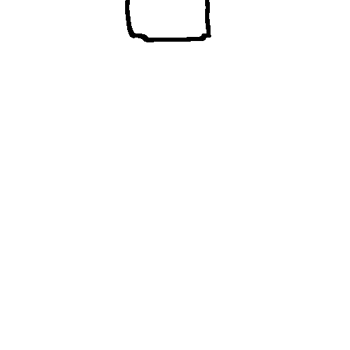} &
        \resultimg{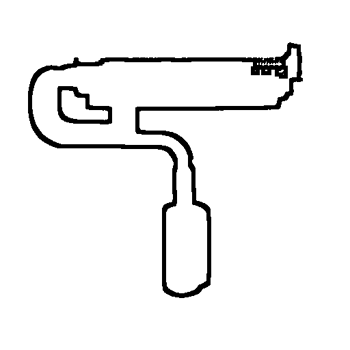} &
        \resultimg{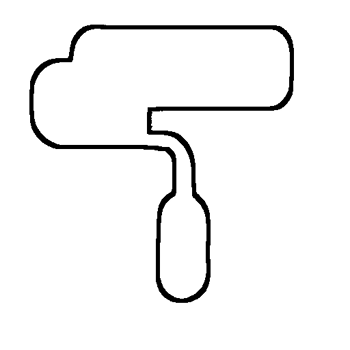} & 
        \resultimg{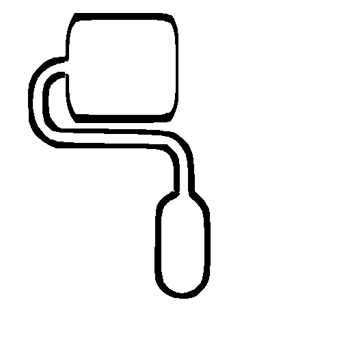}  
        \\
        \midrule
        \multicolumn{8}{l}{\parbox{7cm}{\textbf{"Increase the vertical length of the windows by 200\%"}}}
        \\
        \resultimg{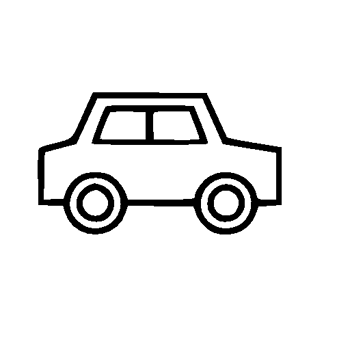} &
        \resultimg{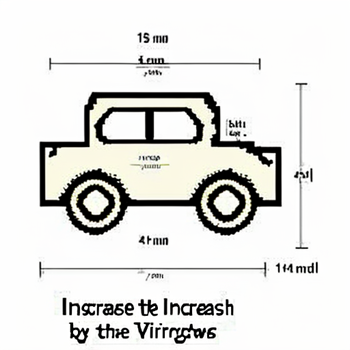} &
        \resultimg{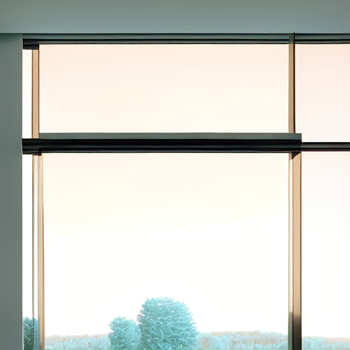} &
        \resultimg{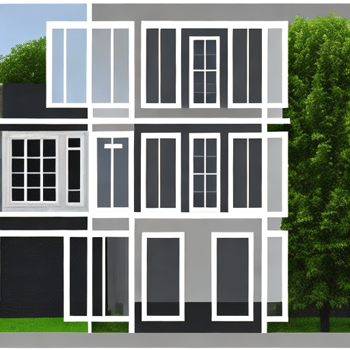} & 
        \resultimg{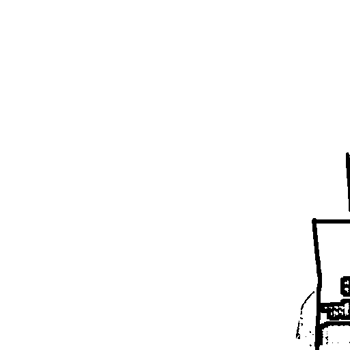} &
        \resultimg{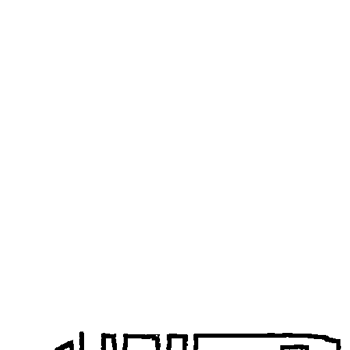} &
        \resultimg{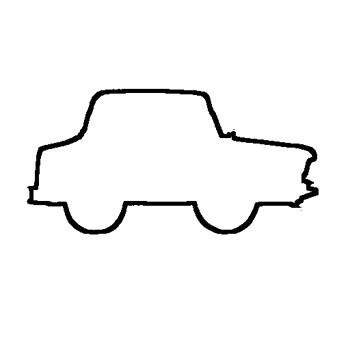} & 
        \resultimg{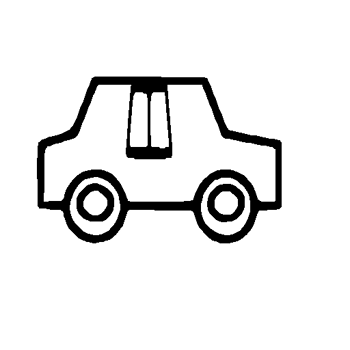}  
        \\
        \midrule
        \multicolumn{8}{l}{\parbox{7cm}{\textbf{"Make the door 50\% higher"}}}
        \\
        \resultimg{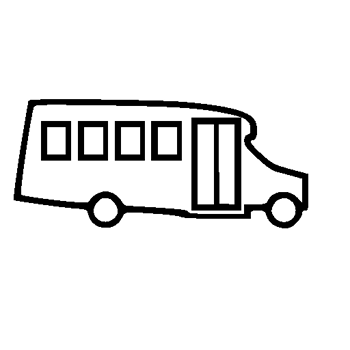} &
        \resultimg{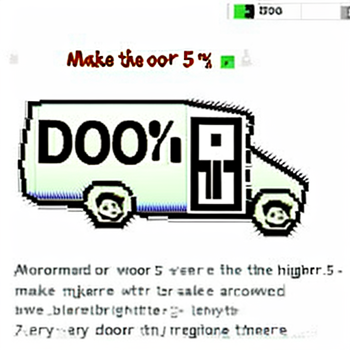} &
        \resultimg{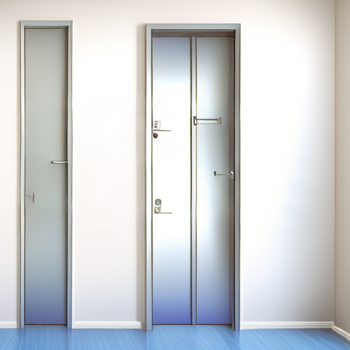} &
        \resultimg{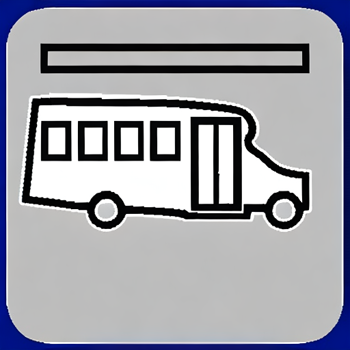} & 
        \resultimg{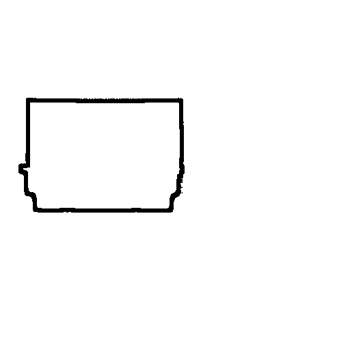} &
        \resultimg{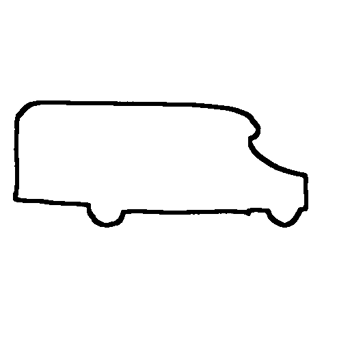} &
        \resultimg{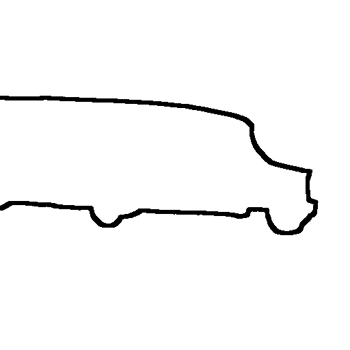} & 
        \resultimg{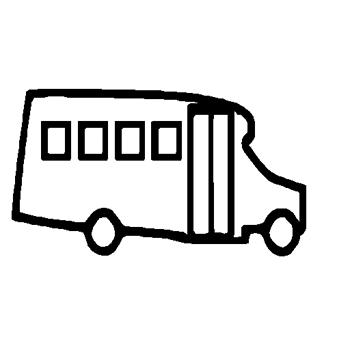}  
        \\
        \midrule
        \multicolumn{8}{l}{\parbox{7cm}{\textbf{"Shorten the flag handle and flag by 50\%"}}}
        \\
        \resultimg{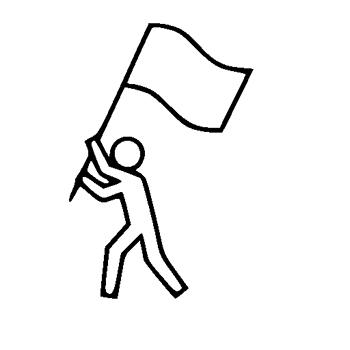} &
        \resultimg{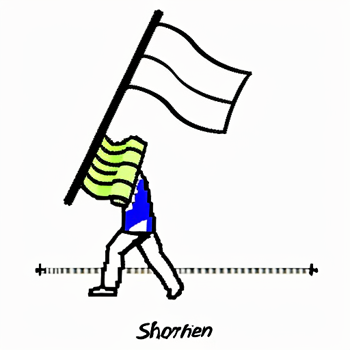} &
        \resultimg{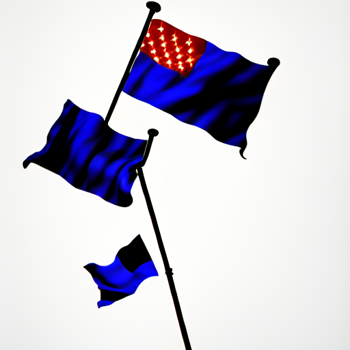} &
        \resultimg{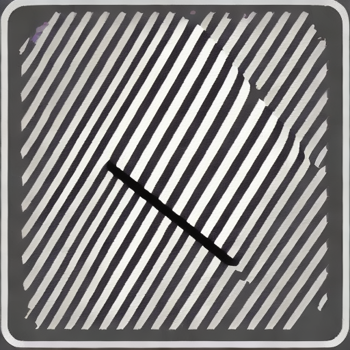} & 
        \resultimg{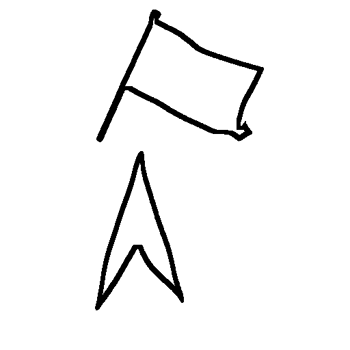} &
        \resultimg{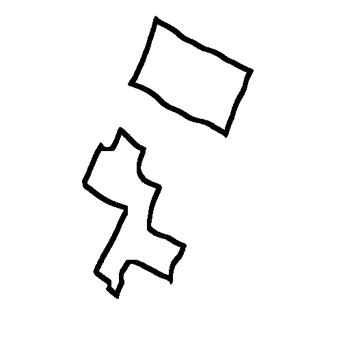} &
        \resultimg{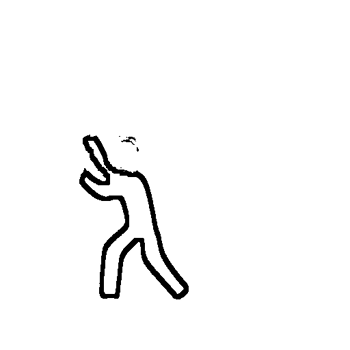} & 
        \resultimg{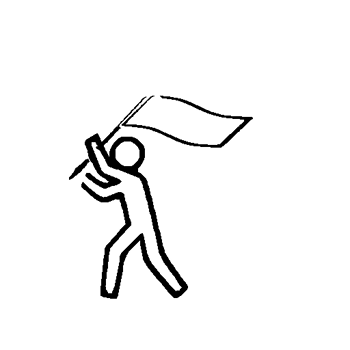}  
        \\
        \midrule
        \multicolumn{8}{l}{\parbox{7cm}{\textbf{"Move and scale the moon to touch the bed"}}}
        \\
        \resultimg{figs/results/original_white/32.png} &
        \resultimg{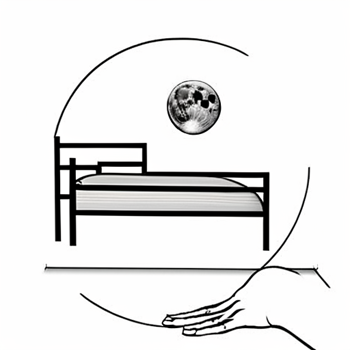} &
        \resultimg{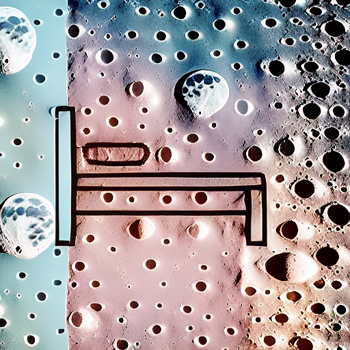} &
        \resultimg{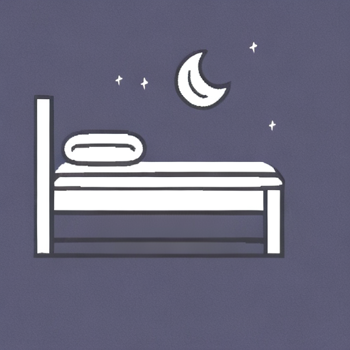} & 
        \resultimg{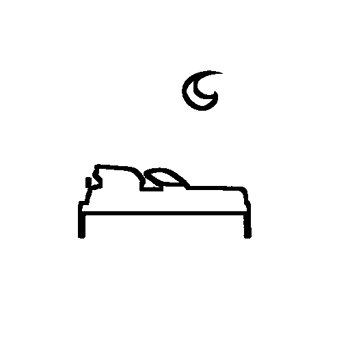} &
        \resultimg{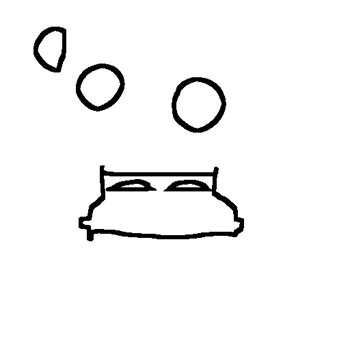} &
        \resultimg{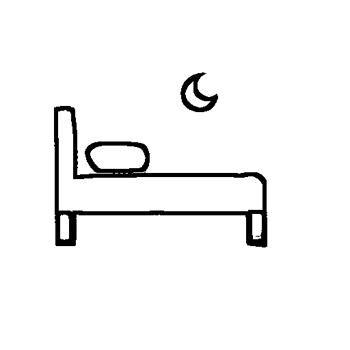} & 
        \resultimg{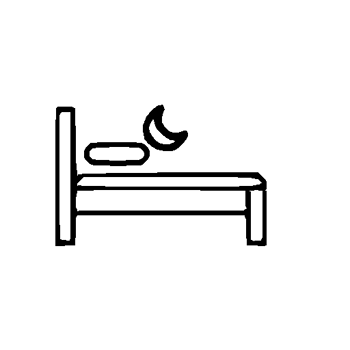}  
        \\
    \end{tabular}
    \caption{Additional diffusion image editing comparison results }
    \label{fig:comp_diffusion_2}
\end{figure}

\begin{figure}[htb!]
    \centering
    \small
    \setlength{\tabcolsep}{0pt}
    \renewcommand{\arraystretch}{0}
    \newcommand{\resultimg}[1]{\includegraphics[trim={10pt 10pt 10pt 10pt},clip,width=0.19\linewidth]{#1}}
    \begin{tabular}{ccccc}
        \raisebox{0em}{Original} & \raisebox{0em}{Seg} & \raisebox{0em}{Request1} & \raisebox{0em}{Request2} & \raisebox{0em}{Request3}
        \\
        \midrule
        \multicolumn{5}{l}{\parbox{7cm}{Request1 : Move and rotate the hand so that its bottom region is aligned with the top region of the 2nd cube from the left\\ Request2 : Move and rotate the hand so that its bottom region is aligned with the top region of the 2nd cube from the right\\ Request3 : Move and rotate the hand so that its bottom region is aligned with the top region of the leftmost cube}}
        \\
        \resultimg{figs/results/original_white/6.png} &
        \resultimg{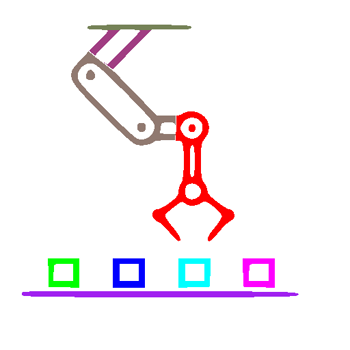} &
        \resultimg{figs/results/ours_c1m1/6.png} &
        \resultimg{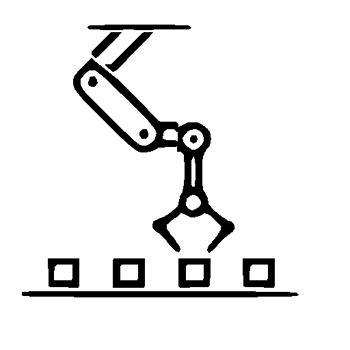} &
        \resultimg{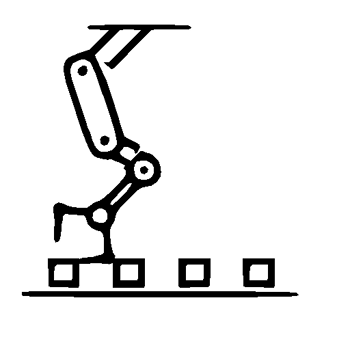} 
        \\
        \midrule
        \multicolumn{5}{l}{\parbox{7cm}{Request1 : Move the shade down to touch the sofa by shortening the lamp support \\ Request2 : Reduce the horizontal length of the sofa by 50\% \\ Request3 : Swap the positions of ing and the book together with bookshelf \\}}
        \\
        \resultimg{figs/results/original_white/25.png} &
        \resultimg{figs/results/seg_gt/25.png} &
        \resultimg{figs/results/ours_c1m1/25.png} &
        \resultimg{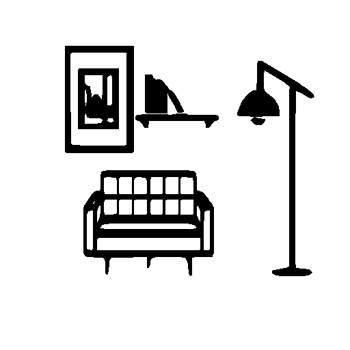} &
        \resultimg{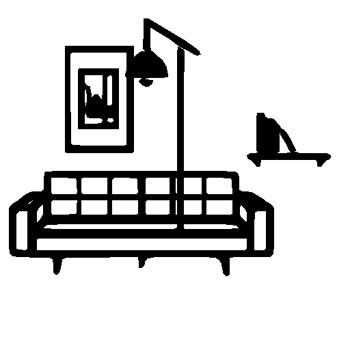} 
        \\
        \midrule
        \multicolumn{5}{l}{\parbox{7cm}{Request1 : Move the lamp shade to touch the base \\ Request2 : Halve both horizontal and vertical length of the shade \\ Request3 : Rotate the shade by 45 degree \\}}
        \\
        \resultimg{figs/results/original_white/5.png} &
        \resultimg{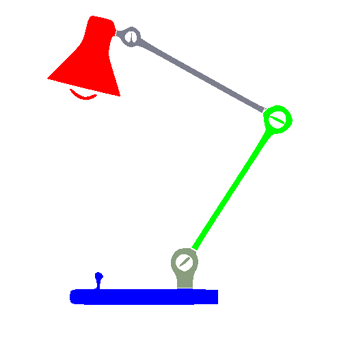} &
        \resultimg{figs/results/ours_c1m1/5.png} &
        \resultimg{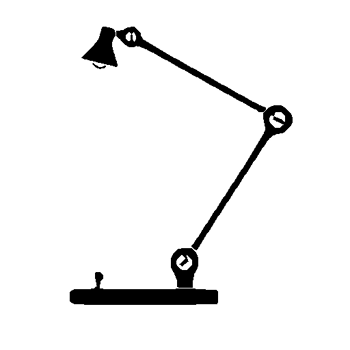} &
        \resultimg{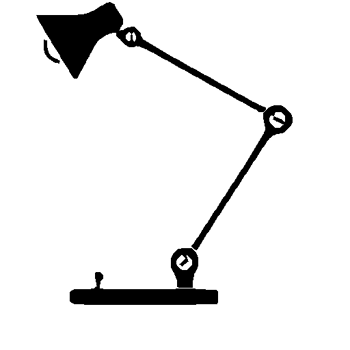} 
        \\
        \midrule
        \multicolumn{5}{l}{\parbox{7cm}{Request1 : Scale the vertical length of the right building and its windows as a whole by 50\% while keep the building on the ground \\ Request2 : Decrease the vertical length of all the windows of the right building by 50\% \\ Request3 : Scale the horizontal length of the left building by 50\% \\}}
        \\
        \resultimg{figs/results/original_white/20.png} &
        \resultimg{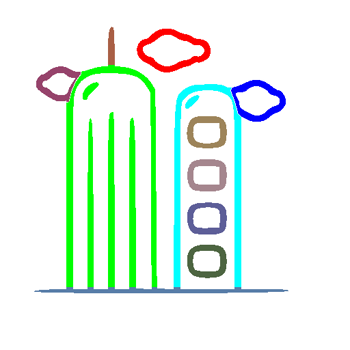} &
        \resultimg{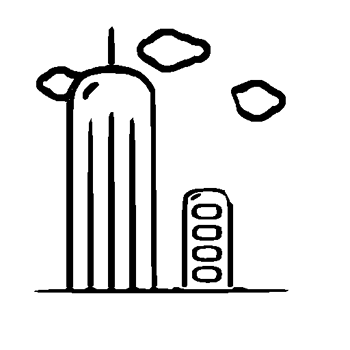} &
        \resultimg{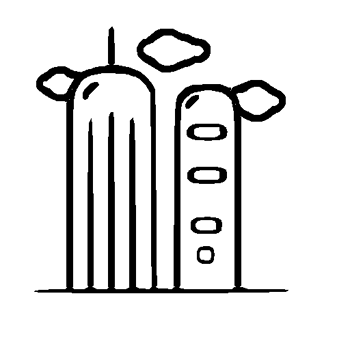} &
        \resultimg{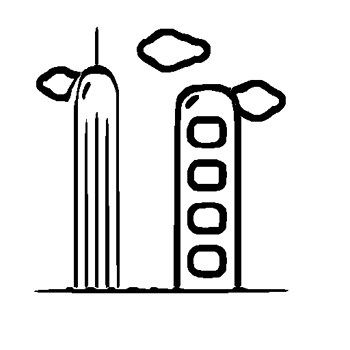} 
        \\
        \midrule
        \multicolumn{5}{l}{\parbox{7cm}{Request1 : Move the plant to touch and be on top of the sofa \\ Request2 : Move the plant together with the supporting table to touch and be on top of the sofa \\ Request3 : Scale the lamp to be the same height as the sofa \\}}
        \\
        \resultimg{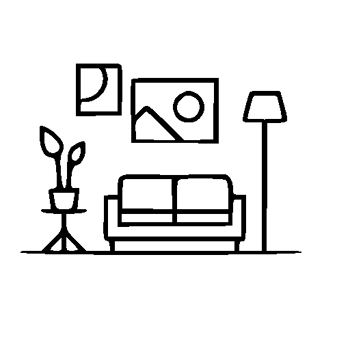} &
        \resultimg{figs/results/seg_gt/12.png} &
        \resultimg{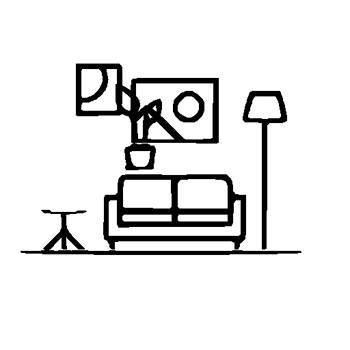} &
        \resultimg{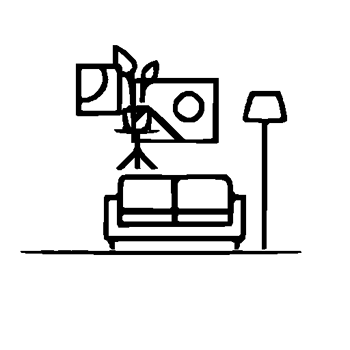} &
        \resultimg{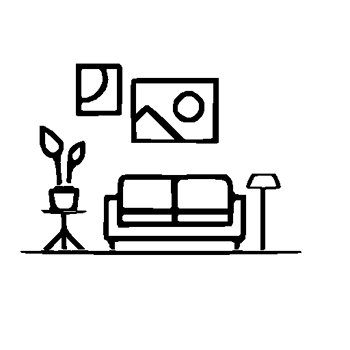} 
        \\
    \end{tabular}
    \caption{Single segmentation re-used for different editing requests}
    \label{fig:multiprompt}
\end{figure}

\begin{figure}[htb!]
    \centering
    \small
    \setlength{\tabcolsep}{0pt}
    \renewcommand{\arraystretch}{0}
    \newcommand{\resultimg}[1]{\includegraphics[trim={10pt 10pt 10pt 10pt},clip,width=0.2\linewidth]{#1}}
    \begin{tabular}{ccc}
        \raisebox{0em}{Original} & \raisebox{0em}{Seg-tool} & \raisebox{0em}{GT} 
        \\
        \resultimg{figs/results/original_white/2.png} &
        \resultimg{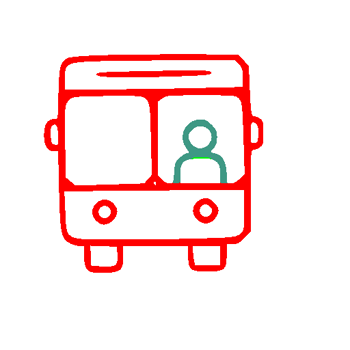} &
        \resultimg{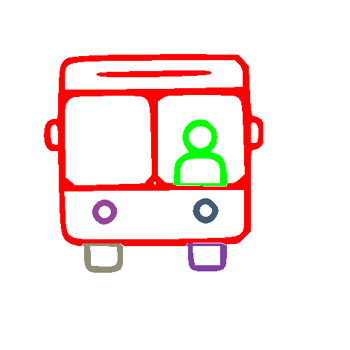} 
        \\
        \resultimg{figs/results/original_white/3.png} &
        \resultimg{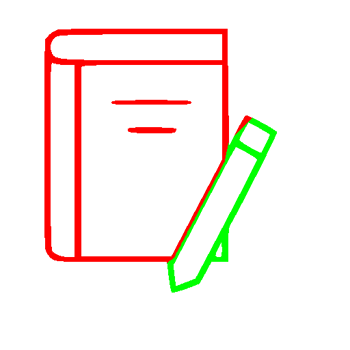} &
        \resultimg{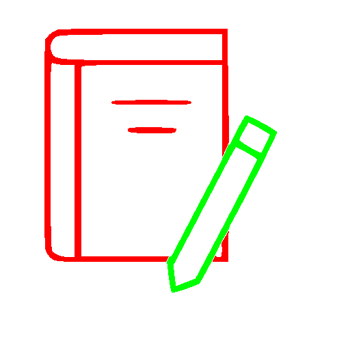} 
        \\
        \resultimg{figs/results/original_white/8.png} &
        \resultimg{figs/results/seg_pred/8.png} &
        \resultimg{figs/results/seg_gt/8.png} 
        \\
        \resultimg{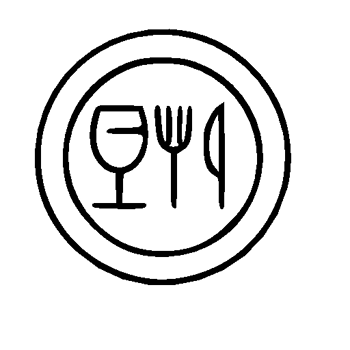} &
        \resultimg{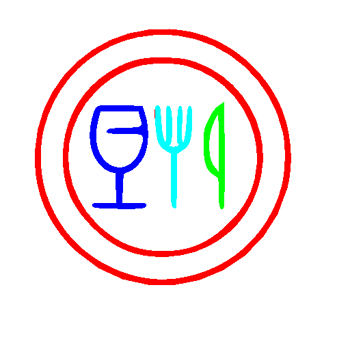} &
        \resultimg{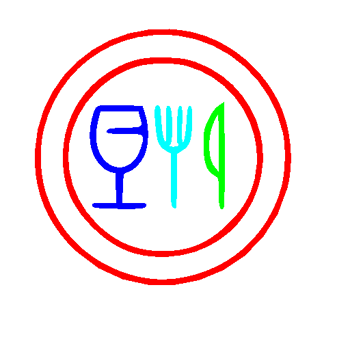} 
        \\
        \resultimg{figs/results/original_white/14.png} &
        \resultimg{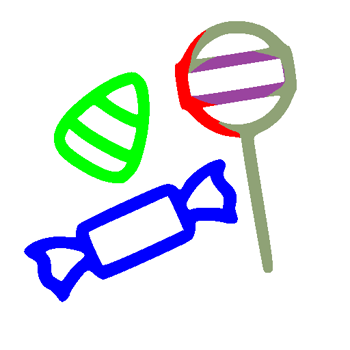} &
        \resultimg{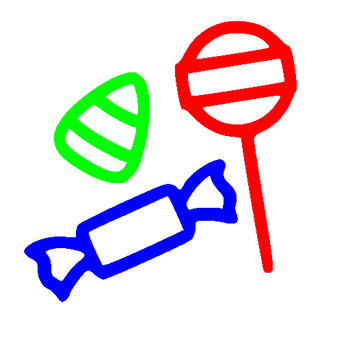} 
        \\
        \resultimg{figs/results/original_white/15.png} &
        \resultimg{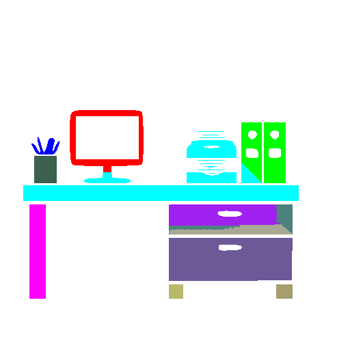} &
        \resultimg{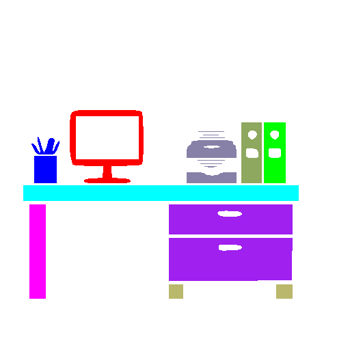} 
        \\
        \resultimg{figs/results/original_white/20.png} &
        \resultimg{figs/results/seg_pred/20.png} &
        \resultimg{figs/results/seg_gt/20.png} 
        \\
        \resultimg{figs/results/original_white/39.png} &
        \resultimg{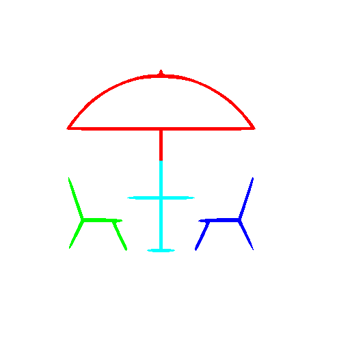} &
        \resultimg{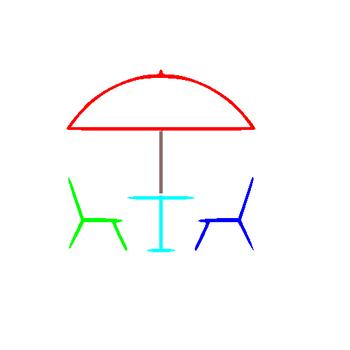} 
        \\
        \resultimg{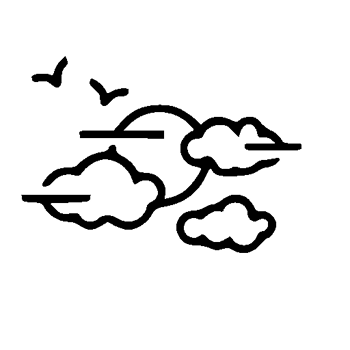} &
        \resultimg{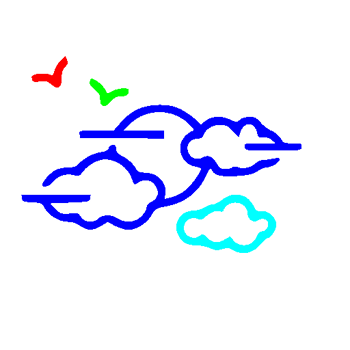} &
        \resultimg{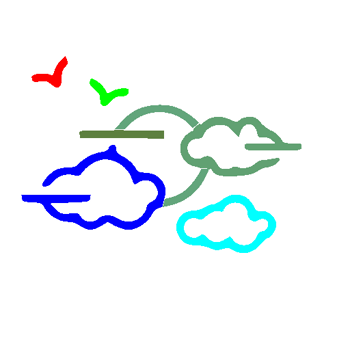} 
        \\
        \resultimg{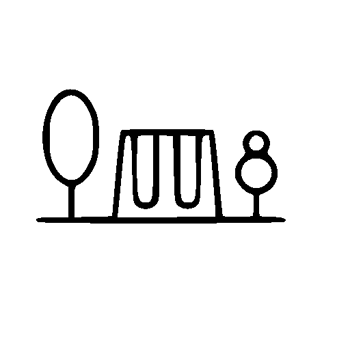} &
        \resultimg{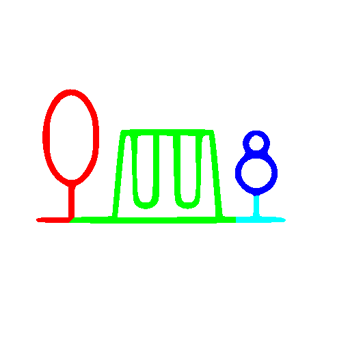} &
        \resultimg{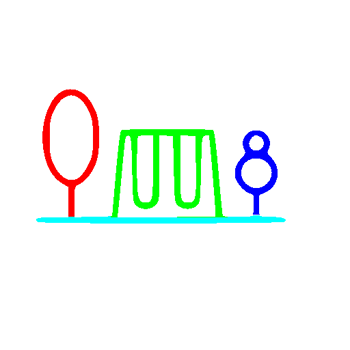} 
        \\
        \resultimg{figs/results/original_white/54.png} &
        \resultimg{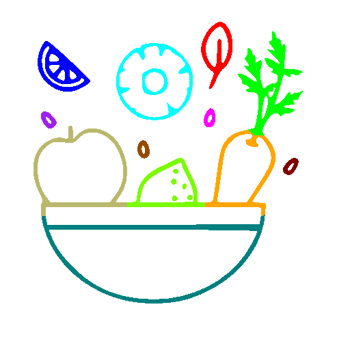} &
        \resultimg{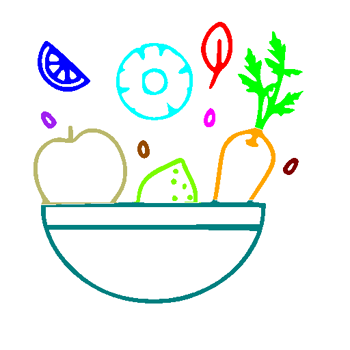} 
        \\
        \resultimg{figs/results/original_white/68.png} &
        \resultimg{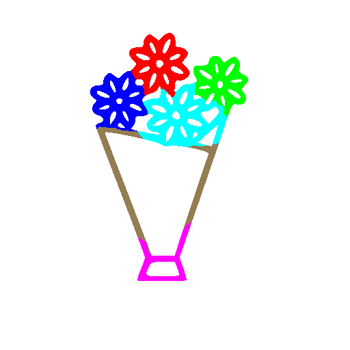} &
        \resultimg{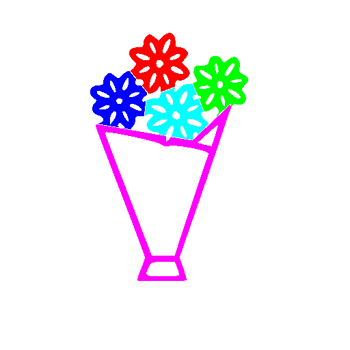} 
        \\
    \end{tabular}
    \caption{Predicted Segmentation vs. GT Segmentation}
    \label{fig:comp_seg_1}
\end{figure}

\begin{figure}[htb!]
    \centering
    \small
    \setlength{\tabcolsep}{0pt}
    \renewcommand{\arraystretch}{0}
    \newcommand{\resultimg}[1]{\includegraphics[trim={10pt 10pt 10pt 10pt},clip,width=0.2\linewidth]{#1}}
    \begin{tabular}{ccc}
        \raisebox{0em}{Original} & \raisebox{0em}{Seg-tool} & \raisebox{0em}{GT} 
        \\
        \resultimg{figs/results/original_white/60.png} &
        \resultimg{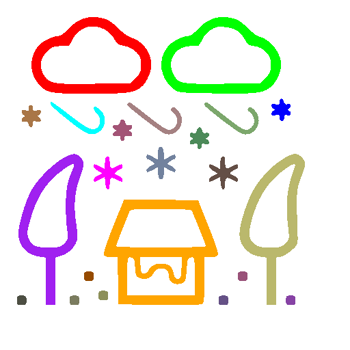} &
        \resultimg{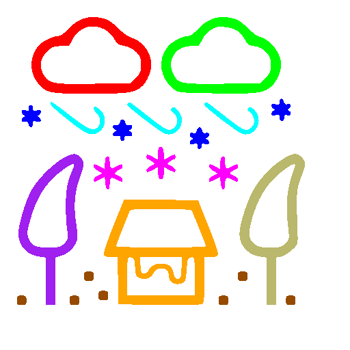} 
        \\
        \resultimg{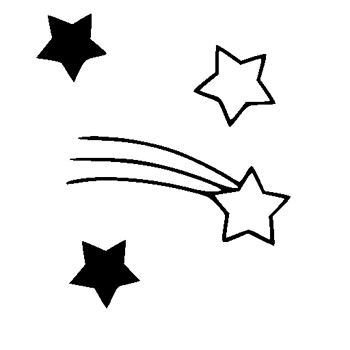} &
        \resultimg{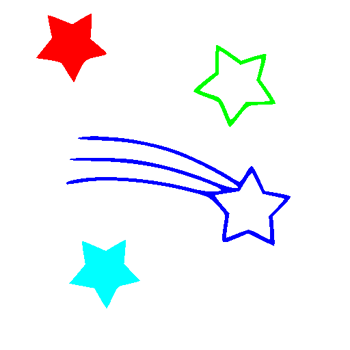} &
        \resultimg{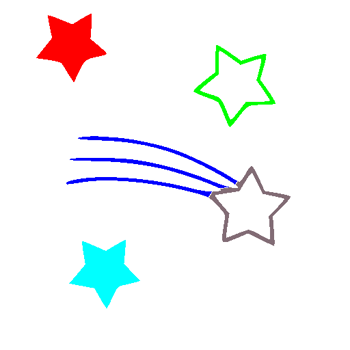} 
        \\
        \resultimg{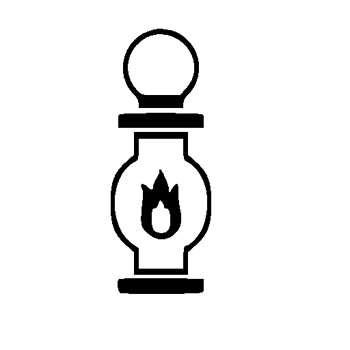} &
        \resultimg{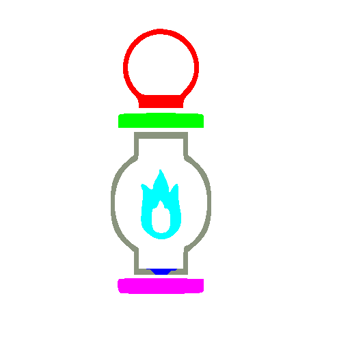} &
        \resultimg{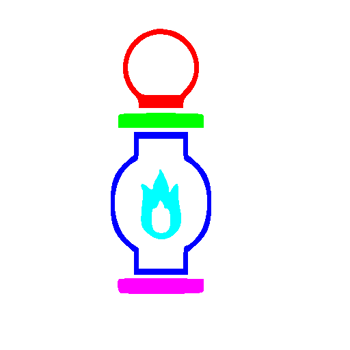} 
        \\
        \resultimg{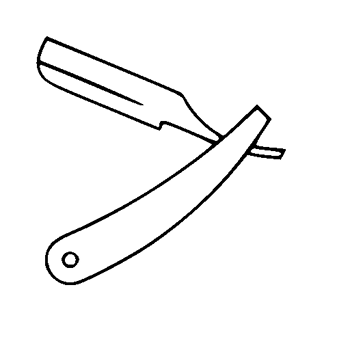} &
        \resultimg{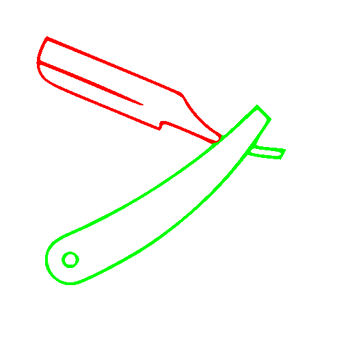} &
        \resultimg{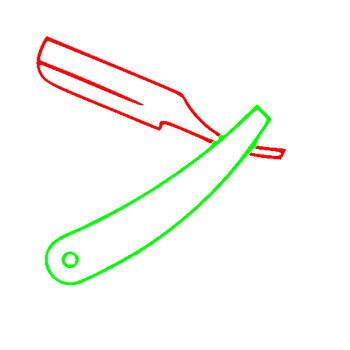} 
        \\
        \resultimg{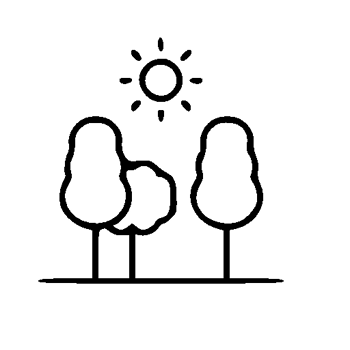} &
        \resultimg{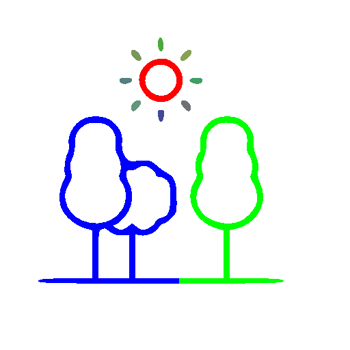} &
        \resultimg{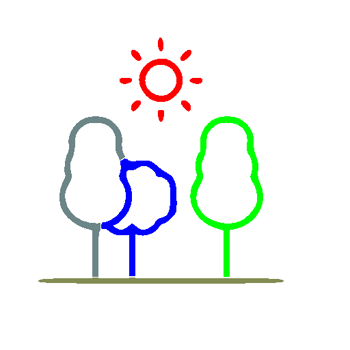} 
        \\
        \resultimg{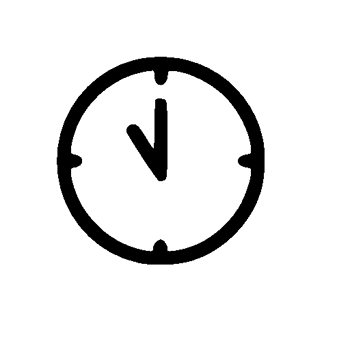} &
        \resultimg{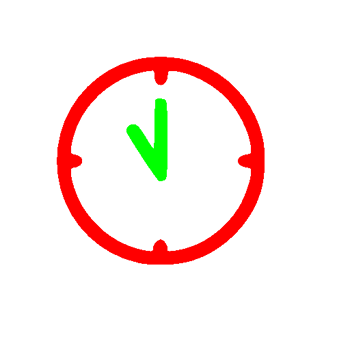} &
        \resultimg{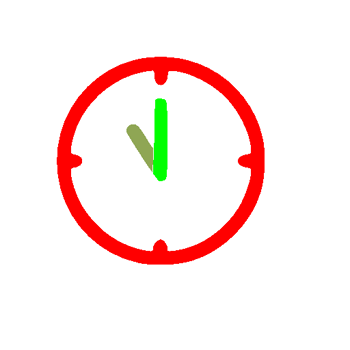} 
        \\
        \resultimg{figs/results/original_white/65.png} &
        \resultimg{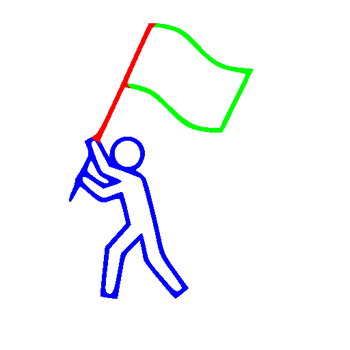} &
        \resultimg{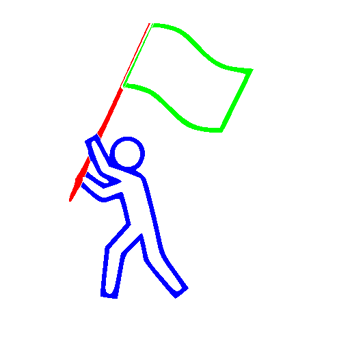} 
    \end{tabular}
    \caption{Predicted Segmentation vs. GT Segmentation}
    \label{fig:comp_seg_2}
\end{figure}

Please see fig \ref{fig:comp_gpt_2} for additional comparison results against different GPT based editing methods.  Please see fig \ref{fig:comp_diffusion_2} for additional comparison results against different diffusion based editing methods. Please see fig \ref{fig:comp_ablation_1} for additional ablation comparison results with different components removed from our editing system. Please see fig \ref{fig:multiprompt} for the editing results of how the same segmentation map can be re-used multiple times for different editing text prompt. Please see fig \ref{fig:comp_seg_1} and \ref{fig:comp_seg_2} for additional results for our segmentation tool. Please see table \ref{tab:program} for examples of what it looks like for the GPT to output constraints using our designed DSL. 

\subsection{Failure Cases}
\begin{figure}[htb!]
    \centering
    \includegraphics[width=0.9\linewidth]{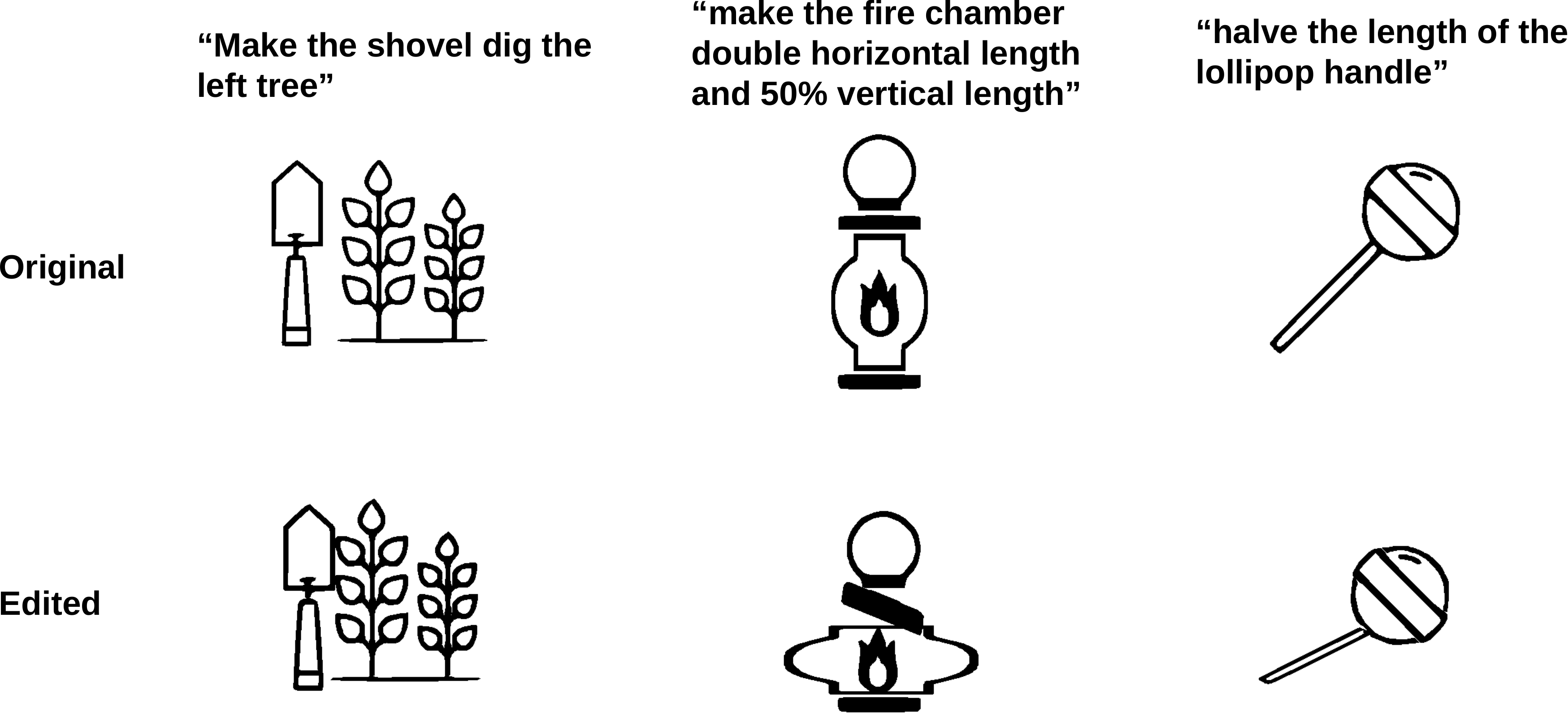}
    \caption{Failure Cases}
    \label{fig:failure}
\end{figure}

We observed several failure modes of our method. The first failure case occur when the editing request is too abstract so that the spatial relation is only implicitly specified. For example in the first editing task in fig \ref{fig:failure}, the LLM had a hard time associating the "dig" action with our DSL operators which then leads to inaccurate constraints. The second relative rare failure case is caused by sub-optimal hyper-parameters in the optimization method, and can lead to less desired result (the second task in fig \ref{fig:failure}). The third case is caused by inaccurate primary constraints generated by LLM, in the third task in fig \ref{fig:failure}, the LLM thinks the "length" in the editing request meant to be the vertical length of the handle rather than the principal length of the handle. We expect that as the performance of LLMs advanced in the future, some of these failure modes will be mitigated. Another type of failure modes is that when the input icon image contains many perspective rendered 3D objects so that the spatial relations of the objects can become quite complex and goes beyond the capability of our DSL and scene representation.

\clearpage

\begin{figure}[htb!]
    \centering
    \small
    \setlength{\tabcolsep}{0pt}
    \renewcommand{\arraystretch}{0}
    \newcommand{\resultimg}[1]{\includegraphics[trim={0pt 0pt 0pt 0pt},clip,width=0.19\linewidth]{#1}}
    \begin{tabular}{ccccc}
        \raisebox{0em}{Original} & \raisebox{0em}{-MS, -RS} & \raisebox{0em}{-RS} & \raisebox{0em}{-MS} & \raisebox{0em}{Ours} 
        \\

        \midrule
        \multicolumn{5}{l}{\parbox{7cm}{\textbf{"Reduce the vertical length of the sail column by 50\%"}}}
        \\
        \resultimg{figs/results/original_white/47.png} &
        \resultimg{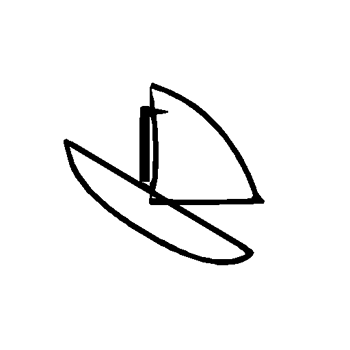} &
        \resultimg{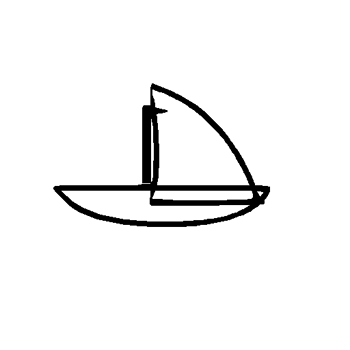} &
        \resultimg{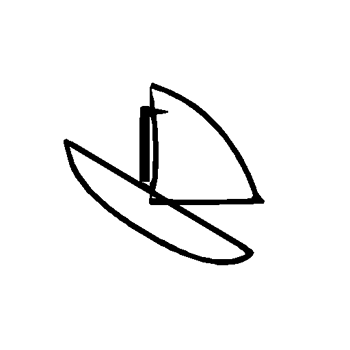} & 
        \resultimg{figs/results/ours_c1m1/47.png}  
        \\
        \midrule
        \multicolumn{5}{l}{\parbox{7cm}{\textbf{"Make the ears 50\% smaller in horizontal length"}}}
        \\
        \resultimg{figs/results/original_white/55.png} &
        \resultimg{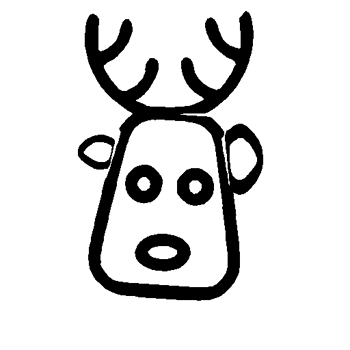} &
        \resultimg{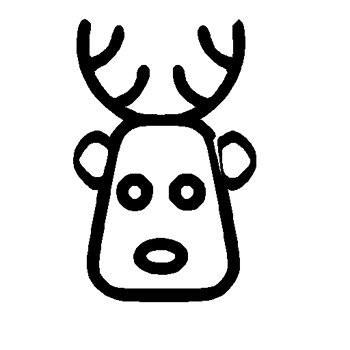} &
        \resultimg{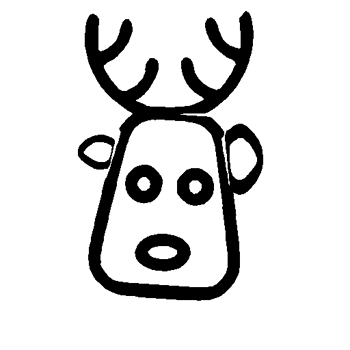} & 
        \resultimg{figs/results/ours_c1m1/55.png}  
        \\
        \midrule
        \multicolumn{5}{l}{\parbox{7cm}{\textbf{"Make the door 50\% higher"}}}
        \\
        \resultimg{figs/results/original_white/71.png} &
        \resultimg{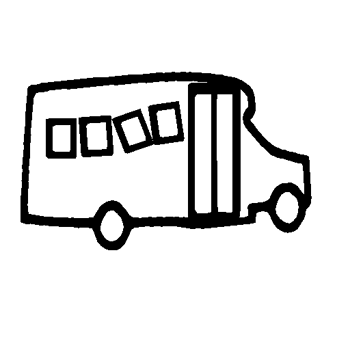} &
        \resultimg{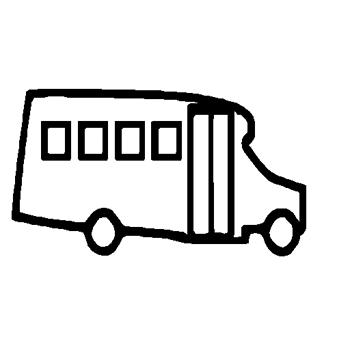} &
        \resultimg{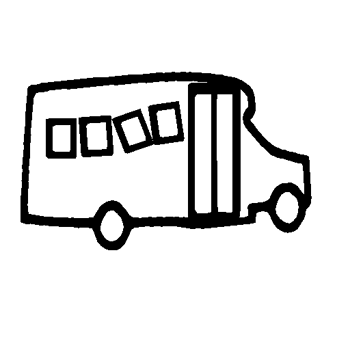} & 
        \resultimg{figs/results/ours_c1m1/71.png}  
        \\
        \midrule
        \multicolumn{5}{l}{\parbox{7cm}{\textbf{"Move the water under the bottle spray"}}}
        \\
        \resultimg{figs/results/original_white/44.png} &
        \resultimg{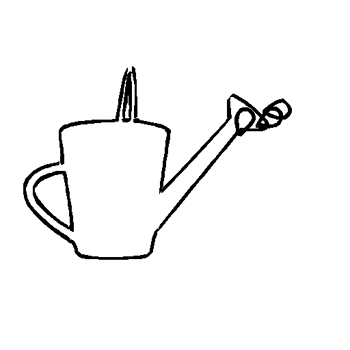} &
        \resultimg{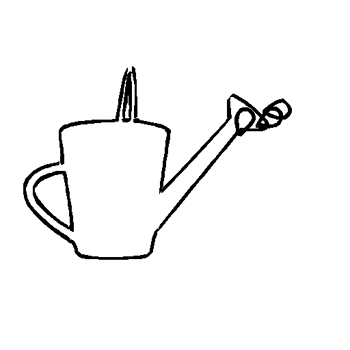} &
        \resultimg{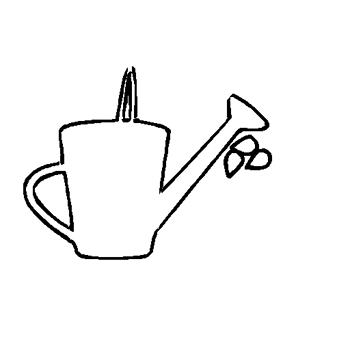} & 
        \resultimg{figs/results/ours_c1m1/44.png}  
        \\
        \midrule
        \multicolumn{5}{l}{\parbox{7cm}{\textbf{"Halve the horizontal length of the roller head"}}}
        \\
        \resultimg{figs/results/original_white/56.png} &
        \resultimg{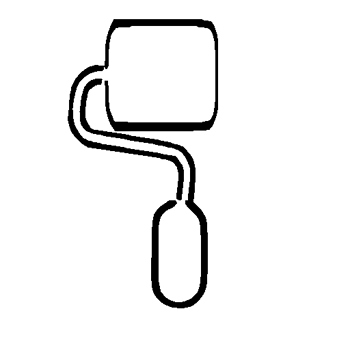} &
        \resultimg{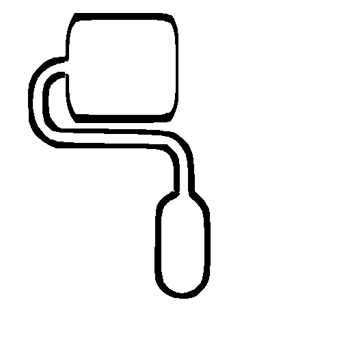} &
        \resultimg{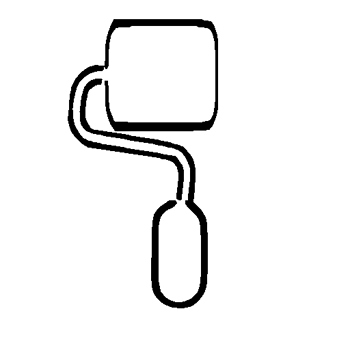} & 
        \resultimg{figs/results/ours_c1m1/56.png}  
        \\
        \midrule
        \multicolumn{5}{l}{\parbox{7cm}{\textbf{"Move the lamp to touch and to be on top of the sofa"}}}
        \\
        \resultimg{figs/results/original_white/17.png} &
        \resultimg{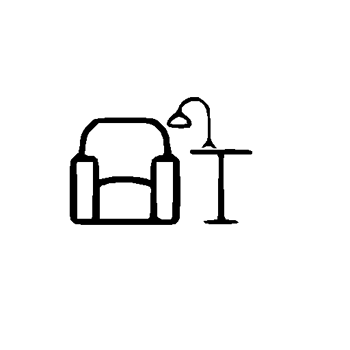} &
        \resultimg{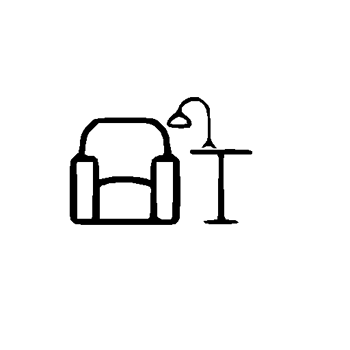} &
        \resultimg{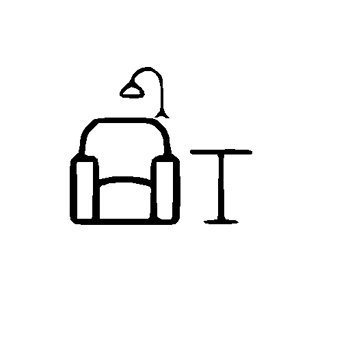} & 
        \resultimg{figs/results/ours_c1m1/17.png}  
        \\
        \midrule
        \multicolumn{5}{l}{\parbox{7cm}{\textbf{"Reduce the vertical length of the cup by 50\%"}}}
        \\
        \resultimg{figs/results/original_white/43.png} &
        \resultimg{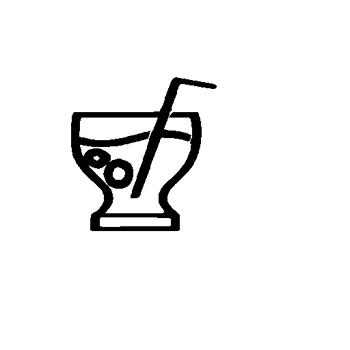} &
        \resultimg{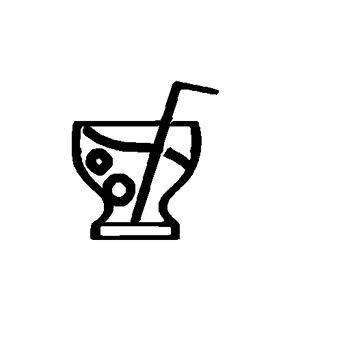} &
        \resultimg{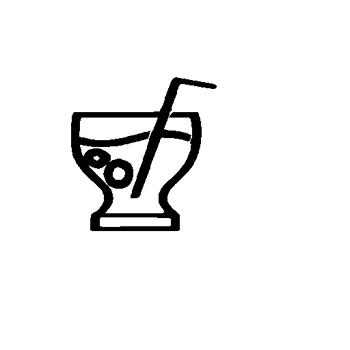} & 
        \resultimg{figs/results/ours_c1m1/43.png}  
        \\
        \midrule
        \multicolumn{5}{l}{\parbox{7cm}{\textbf{"Move the leftmost candle(with flame) and rightmost candle(with flame) to touch the 2nd layer of the cake"}}}
        \\
        \resultimg{figs/results/original_white/4.png} &
        \resultimg{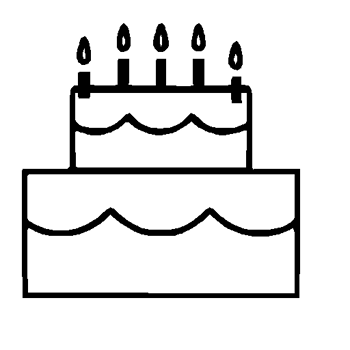} &
        \resultimg{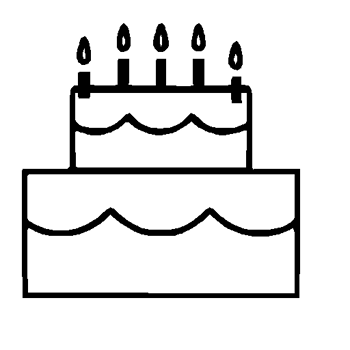} &
        \resultimg{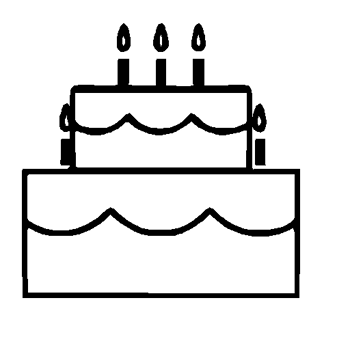} & 
        \resultimg{figs/results/ours_c1m1/4.png}  
        \\
        \midrule
        \multicolumn{5}{l}{\parbox{7cm}{\textbf{"Rotate the bike body so that the right side wheel's vertical position is raised by 50\%"}}}
        \\
        \resultimg{figs/results/original_white/8.png} &
        \resultimg{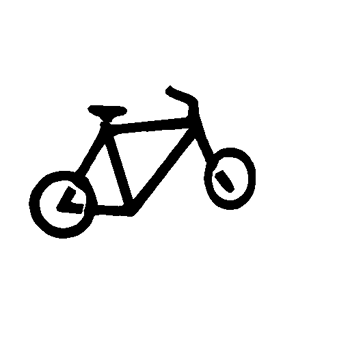} &
        \resultimg{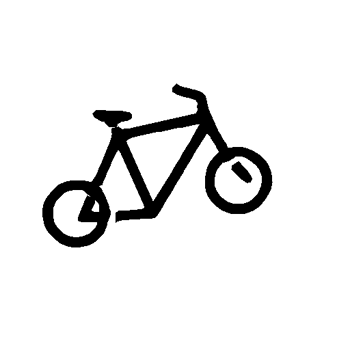} &
        \resultimg{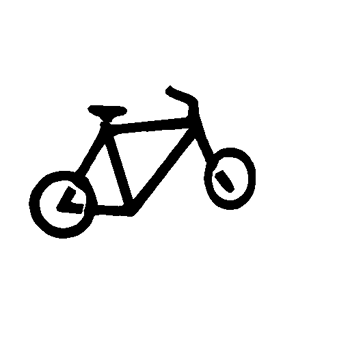} & 
        \resultimg{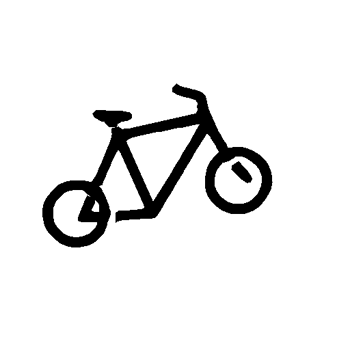}  
        \\
    \end{tabular}
    \caption{Additional ablation study results, MS: Motion Search, RS: Relation Search.}
    \label{fig:comp_ablation_1}
\end{figure}

\clearpage

\begin{table}[htb!]
\centering
\newcommand{\resultimg}[1]{\includegraphics[trim={10pt 10pt 10pt 10pt},clip,width=1.0\linewidth]{#1}}
\begin{tabular}{|p{2cm}|p{4cm}|p{10cm}|}
    \hline
    \raisebox{0em}{Original} & \raisebox{0em}{GT-Seg} & \raisebox{0em}{LLM primary constraints}
    \\
    \hline
    \resultimg{figs/results/seg_gt/0.png} 
    &
    \raisebox{3.5em}{\parbox{4cm}{seg0:(basket:handle)\\seg1:(basket:handle)\\seg2:(basket:top)\\seg3:(basket:body)}}
    &
    \raisebox{3.5em}{\parbox{10cm}{"\textbf{Rotate the handles apart so that the distance between their top regions is equal to half of the basket's horizontal length}" \\
    \\
    rotate(seg0), rotate(seg1) \\ equal(center\_distance(top\_inner\_region(seg0), top\_inner\_region(seg1)), 0.5 $*$ horizontal\_length(union(seg0, seg1, seg2, seg3)))}}\\ 
    \hline
    \resultimg{figs/results/seg_gt/1.png} &
    \raisebox{3.5em}{\parbox{4cm}{seg0:(bubble0:bubble)\\seg1:(fish:body)\\seg2:(fish:fin)\\seg3:(bubble2:bubble)\\seg4:(fish:eye)}}
    &
    \raisebox{3.5em}{\parbox{10cm}{"\textbf{Scale the fish body to reduce its horizontal length by 50\%}" \\
    \\scale(seg1), adjust(seg2), adjust(seg4) \\ equal(horizontal\_length(seg1), 0.5 $*$ horizontal\_length(old\_copy(seg1)))}}\\ 
     \hline
    \resultimg{figs/results/seg_gt/5.png} &
    \raisebox{3.5em}{\parbox{4cm}{seg0:(lamp:shade)\\seg1:(lamp:arm connector)\\seg2:(lamp:arm connector)\\seg3:(lamp:base connector)\\seg4:(lamp:base)}}
    &
    \raisebox{3.5em}{\parbox{10cm}{"\textbf{Move the lamp shade to touch the base}" \\
    \\{move(seg0)\\
    touch(seg0, seg4)} }}\\ 
     \hline
    \resultimg{figs/results/seg_gt/11.png} &
    \raisebox{3.5em}{\parbox{4cm}{seg0:(cloud:cloud)\\seg1:(umbrella:umbrella)\\seg2:(person:head)\\seg3:(cup:cup)\\seg4:(person:arm)\\seg5:(person:body)\\seg6:(chair:chair)\\seg7:(person:arm)}}
    &
    \raisebox{3.5em}{\parbox{11cm}{"\textbf{Raise the cup to be above the head}" \\
    \\{move(seg3), adjust(seg4),\\ on\_top(seg3, seg2)}}}\\ 
     \hline
    \resultimg{figs/results/seg_gt/25.png} &
    \raisebox{3.5em}{\parbox{4cm}{seg0:(painting:painting)\\seg1:(lamp:connector)\\seg2:(book:book)\\seg3:(lamp:support)\\seg4:(lamp:shade)\\seg5:(bookshelf:bookshelf)\\seg6:(sofa:sofa)}}
    &
    \raisebox{3.5em}{\parbox{10cm}{"\textbf{Move the shade down to touch the sofa by shortening the lamp support}" \\
    \\{move(seg4), scale(seg3), stay(seg6),\\ touch(seg4, seg6), \\smaller(vertical\_length(seg3), vertical\_length(old\_copy(seg3)))} }}\\ 
     \hline
    \resultimg{figs/results/seg_gt/13.png} &
    \raisebox{3.5em}{\parbox{4cm}{seg0:(airplane:fuselage)\\seg1:(airplane:engine)\\seg2:(airplane:engine)\\seg3:(airplane:wing)\\seg4:(airplane:wing)\\seg5:(airplane:engine)\\seg6:(airplane:engine)\\seg7:(airplane:tail)\\seg8:(airplane:tail)}}
    &
    \raisebox{3.5em}{\parbox{10cm}{"\textbf{Reduce the vertical length of the fuselage by 50\%}" \\
    \\
    \\scale(seg0), \\ equal(vertical\_length(seg0), 0.5 * vertical\_length(old\_copy(seg0)))}}\\ 
     \hline
\end{tabular}
\caption{Illustration of the primary constraints generated by GPT4 using our DSL as building blocks}
\label{tab:program}
\end{table}

\clearpage

{\small
\bibliographystyle{ieee_fullname}
\bibliography{main.bib}
}

\end{document}